# Nanoscale electrical measurements in liquids using AFM – progress and outlook

Liam Collins,*[1,2] Jason Kilpatrick,[3] Sergei V. Kalinin,[1,2] and Brian J. Rodriguez[3,4]

[1]Center for Nanophase Materials Sciences, Oak Ridge National Laboratory, Oak Ridge, Tennessee 37831, USA

[2]Institute for Functional Imaging of Materials, Oak Ridge National Laboratory, Oak Ridge, Tennessee 37831, USA

[3]Conway Institute of Biomolecular and Biomedical Research, University College Dublin, Belfield, Dublin 4, Ireland

[4]School of Physics, University College Dublin, Belfield, Dublin 4, Ireland

E-mail: collinslf@ornl.gov

## Contents

**Abbreviations**

| | |
|---|---|
| AC | Alternating current |
| AM | Amplitude modulated |
| AFM | Atomic force microscopy |
| BE | Band excitation |
| BDS | Broadband dielectric spectroscopy |
| CL | Closed loop |
| cKPFM | Contact Kelvin probe force microscopy |
| CPD | Contact potential difference |
| CR | Contact resonance |
| DC | Direct current |
| DF | Dual frequency |
| DH | Dual harmonic |
| EDL | Electric double layer |
| EcFM | Electrochemical force microscopy |
| EFGS | Electrostatic force gradient spectroscopy |
| EFM | Electrostatic force microscopy |
| EFS | Electrostatic force spectroscopy |
| FORC | First order reversal curve |
| FV | Force volume |
| FM | Frequency modulation |
| HBE | Half band excitation |
| HH | Half harmonic |
| H | Heterodyne |
| HOPG | Highly oriented pyrolytic graphite |
| IC | Intermittent contact |
| IM | Intermodulation |
| KFM | Kelvin force microscopy |
| KP | Kelvin probe |
| KPFM | Kelvin probe force microscopy |
| KPFS | Kelvin probe force spectroscopy |
| LCPD | Local CPD |
| LDLS | Local dielectric loss spectroscopy |
| LIA | Lock-in amplifier |
| MEMS | Micro electromechanical systems |
| MF | Multi-frequency |
| NIM | Nanoscale impedance microscopy |
| NC | Non-contact |
| OL | Open loop |
| OLBE | Open loop band excitation |
| OLEP | Open loop electric potential |
| PLL | Phase-locked loop |
| PFM | Piezoresponse force microscopy |
| SCM | Scanning capacitance microscopy |
| SECM | Scanning electrochemical microscopy |
| SKPM | Scanning Kelvin probe microscopy |
| SPFM | Scanning polarization force microscopy |
| SPM | Scanning probe microscopy |
| SSPM | Scanning surface potential microscopy |
| SHO | Simple harmonic oscillator |
| SF | Single frequency |
| SFA | Surface forces apparatus |
| 3D | Three-dimensional |
| TR | Time resolved |
| UHV | Ultra-high vacuum |
| UM | Ultramicroelectrode |
| USC | Ultra-small cantilevers |
| VM | Voltage-modulated |

## Abstract

Fundamental mechanisms of energy storage, corrosion, sensing, and multiple biological functionalities are directly coupled to electrical processes and ionic dynamics at solid-liquid interfaces. In many cases, these processes are spatially inhomogeneous taking place at grain boundaries, step edges, point defects, ion channels, etc. and possess complex time and voltage dependent dynamics. This necessitates time-resolved and real-space probing of these phenomena. In this review, we discuss the applications of force-sensitive voltage modulated scanning probe microscopy (SPM) for probing electrical phenomena at solid-liquid interfaces. We first describe the working principles behind electrostatic and Kelvin Probe Force microscopies (EFM & KPFM) at the gas-solid interface, review the state of the art in advanced KPFM methods and developments to (i) overcome limitations of classical KPFM, (i) expand the information accessible from KPFM, and (iii) extend KPFM operation to liquid environments. We briefly discuss the theoretical framework of the electrical double layer (EDL) forces and dynamics, the implications and breakdown of classical EDL models for highly charged interfaces, or under high ion concentrations, and briefly describe recent modifications of the classical EDL theory relevant for understanding nanoscale electrical measurements at the solid-liquid interface. We further review the latest achievements in mapping surface charge, dielectric constants, and electrodynamic and electrochemical processes in liquids. Finally, we outline the key challenges and opportunities that exist in the field of nanoscale electrical measurements in liquid as well as provide a roadmap for the future development of liquid KPFM.

# 1    Introduction

Knowledge of the local potential and ionic distributions at the solid-liquid interface is crucial for understanding material functionality and device performance across all fields of science and engineering spanning from material science, to physical chemistry and biological research. Importantly, local interface potentials and chemical and physical capacitances govern processes including corrosion,[1, 2] sensing,[3] energy conversion and storage,[4-6] and biology and biochemistry.[7, 8]

Although significant progress has been achieved in developing theoretical understanding of the complex diffuse charge dynamics at interfaces,[9-11] a complete understanding of their role in real systems necessitates techniques capable of probing 'local' electric double layer (EDL) characteristics and dynamics. While significant progress has been achieved for exploration of EDL *in situ* by emergent scattering measurements [12-14], local exploration of surface inhomogeneities as well as the complex electrodynamic phenomena taking place in the liquid adjacent to the solid surface is central to a vast array of electrochemical systems. To date, the most viable option for investigating the solid-liquid interface, electrochemical functionality, or material or device performance has been realized through macroscopic electrochemical methods. These methods are generally based on current detection of electrochemical responses in the time- or frequency-domains and are used to measure both equilibrium and dynamic processes. Electrochemical methods are well established for separation of electronic and ionic processes (e.g., impedance spectroscopy [1, 15, 16]), identifying potential thresholds between non-Faradaic and Faradaic regimes (e.g., cyclic voltammetry [17-19]), or for probing dynamic charge screening and storage processes (e.g., chronopotentiometric methods).[20] However, these traditional approaches have limited spatial resolution whereas the knowledge of the micro- and nano- structure of the interface is required for the quantitative interpretation of the data.

The necessity for local measurements has resulted in decades of research focused on miniaturizing conventional electrochemical probes, or ultramicroelectrode (UME), which spawned the development of a range of biased micro-probe electrochemical microscopies.[20-22] Undoubtedly the most widely adopted of these is scanning electrochemical microscopy (SECM), which was independently reported by the groups of Engstrom [23] and Bard.[24] SECM is a current sensitive technique combining the capillary electrode and distance feedback, which after 3 decades of development has enabled *in situ* investigations of electrochemical surface reactivity and interfacial properties. SECM has been used to study a host of electrochemical behaviors from ionic dissolution,[25] electrocatalysis,[26] biology [27-29] and can even be used for micro-patterning [30] of surfaces on the micron scale. The spatial resolution of SECM, however, is determined by the UME probe having precise geometry and strict fabrication requirements. To enable sufficient current signal detection, the probe size has to be relatively large, limiting the achievable resolution to 100s nm – μm. Despite many attempts at improvements,[31] the spatial resolution of SECM is usually orders of magnitude worse than that of other scanning probe microscopy (SPM) techniques, such as atomic force microscopy (AFM)[32] and scanning tunneling microscopy (STM).[33] Furthermore, SECM can be prone to crosstalk between topographical and electrochemical reactivity channels.[34, 35] For an in-depth description of the principles and operation governing SECM the reader is directed to the following papers,[34, 36, 37] while applications are described elsewhere.[27, 38, 39]

The alternative approach for probing interfacial functionalities is based on force detection. Researchers have probed the out of plane solid-liquid interface for decades using surface forces apparatus (SFA) combined with classical models of the EDL.[40-42] This body of work led to significant enhancement of our understanding of such systems and led to advances in related fields including colloid and interface science. However, a more complete understanding of non-spatially uniform and non-equilibrium phenomena requires approaches capable of providing reliable and achievable measurements with sufficient vertical and lateral resolution on complex material systems. Such techniques should be able to capture the characteristics of relevant electrochemical and transport phenomena both in and normal to the surface plane – across multiple length scales from a single structural defect, or a step edge to the diffuse EDL itself (e.g., Debye length [43]) and across multiple timescales comprising the dynamic charge relaxation and ion diffusion (e.g., ns – ms) processes which govern the operation of all electrochemical systems.

In this review, we will present a comprehensive overview of decades of research involving forced based detection of EDL structure, properties and dynamics specifically focusing our interest on recent application of force-based voltage-modulated (VM)-SPM for probing electronic,[44-50] ionic,[46, 51] dielectric,[52-55] and electrochemical [46, 51, 56] properties and processes in liquid. We begin by providing a background the underlying principles and theory governing both electrostatic force microscopy (EFM) and Kelvin probe force microscopy (KPFM) operation in air and ultra-high vacuum (UHV). We note that UHV-KPFM has become the gold standard for mapping electrochemical potentials at solid surfaces with atomic scale resolution demonstrated.[57] We will continue by presenting an overview of the state-of-art advanced modes of KPFM, including their advantages and disadvantages in terms of accuracy, lateral resolution, measurement speed, and overall level of information attainable. To better position the reader in understanding both the opportunities and complications associated with operating such measurements in aqueous environments and more generally in liquid electrolytes, we proceed by providing a brief background to the most common theoretical models used for describing the EDL under both equilibrium and dynamic regimes. We then discuss EDL dynamics at highly charged interfaces, highlight problems with the classical models for conditions relevant to liquid KPFM, and present modified theories of EDL dynamics which may prove to be important for understanding liquid KPFM imaging mechanisms. Beyond theoretical studies, crucial information on the equilibrium EDL forces has been derived from the successful implementation of SPM measurements *in-situ*. We will briefly review early comparisons between one-dimensional (1D) AFM force based measurements and classical theories describing the EDL forces (e.g., DVLO theory) as well as describing the state of the art in terms of atomically resolved three-dimensional (3D) force mapping of the solid-liquid interface using high resolution AFM. In the final sections of this review, we outline attempts at extending KPFM and KPFM-like measurements to the solid-liquid interface for mapping surface charges,[58] surface potentials,[44, 47, 49, 59] polarization forces,[52-55, 60] and dynamic electrochemical processes.[46, 51] We conclude by outlining the challenges and opportunities which exist for achieving liquid KPFM, as well as a providing a roadmap for successful implementation and adoption of these measurements.

## 2   Electrostatic force microscopy

This section provides a history and the theoretical background of EFM, the technique that formed the basis and impetus for KPFM. We begin by providing the necessary background theory on VM AFM as well as a description of long-range electrostatic interactions. The implementation and operational principles behind EFM for detection of electrostatic force and force gradient are outlined. Briefly, attempts to quantify local electronic properties, in particular, charge density, surface potential, and dielectric properties, using EFM are reviewed.

## 2.1 Voltage modulated AFM

The development of the AFM in 1986[32] was a significant boost to the emerging fields of nanoscience, nanotechnology, and bio-nanotechnology. Along with is predecessor, the STM,[61, 62] this new class of microscope helped to characterize the nanoworld, quickly becoming a mainstay research tool used in laboratories around the globe. The utilization of a sharp (1-100 nm radius) probe enabled structural mapping of materials with vastly superior spatial resolution than was previously available, ultimately down to the atomic scale.[63] Although early STM experiments,[64, 65] had allowed electronic properties to be measured and mapped down to the atomic level,[66] its operation was primarily limited to semiconducting [67, 68] and metallic materials.[64, 69, 70] AFM, which relies on force rather than electronic based detection mechanisms, opened up the possibility for broad applicability across all fields of science, as well as operation in all environments (e.g., UHV, ambient, and liquid). Soon after its initial development,[32] several functional modes of AFM operation were realized which went beyond topographical imaging and allowed simultaneous mapping of material properties with nanoscale resolution.

In particular, the combination of AFM with conductive probes and voltage modulation permitted imaging and manipulation of a wide range of electrical properties including electrostatic,[71-73] electronic,[74-77] electrochemical,[25, 35, 78-80] and electromechanical[81-85] functionalities. As the name suggests, common to most VM-AFM techniques is the application of a tip voltage given by:

$$V_{tip} = V_{dc} + V_{ac}\cos(\omega t) \tag{1}$$

where $V_{dc}$ and $V_{ac}$ are the applied direct current (dc) and alternating current (ac) voltage, respectively, and $\omega$ is the frequency. It is important to stress some critical distinctions amongst VM-AFM techniques. First, depending on the nature of the approach, in some modes of operation (e.g., EFM,[86] piezoresponse force microscopy (PFM) [81]), $V_{dc}$ is optional or constant, whereas it is variable and mandatory in others (e.g., closed loop (CL) KPFM[74]). Additionally, while traditional techniques discussed in this section utilize a single frequency of sinusoidal excitation, it will be seen later that almost any voltage waveform can be used as an excitation source provided an appropriate detection method is adopted (e.g., band excitation (BE),[87-89] intermodulation (IM) [90]). Finally, while most VM-AFM techniques are based solely on voltage modulation (e.g., amplitude modulated (AM)-KPFM) some do utilize a combination of voltage and mechanical modulation (e.g., EFM).

### 2.1.1   Contrast in VM-AFM

The principles and physical underpinnings of standard AFM are traditionally understood using force-distance curves, as shown in Figure 1(a). Typically, AFM imaging is performed in either contact, intermittent contact (IC), or non-contact (NC) regimes. Depending on the preferred image contrast mechanism, the operational regime of VM-AFM measurements can be sensitive to; elastic interactions (e.g. bias induced softening in contact resonance (CR)-AFM[91, 92]), van der Waals, (e.g., electrostatic actuation of a AFM cantilever for NC-AFM[50, 93, 94]), electromechanical effects (e.g., PFM, contact KPFM[95-97]) or long-range interactions (e.g., electrostatic or magnetic).[98] Measurements of electrostatic tip-sample forces are often, but not always, performed far from the sample surface using lift mode, sometimes referred to as interleave, dual pass, or constant height mode. Mapping interactions with the tip positioned away from the sample surface ensures that the long-range forces are dominant and eliminates the influences of short-range chemical, electromechanical, and van der Waals forces.

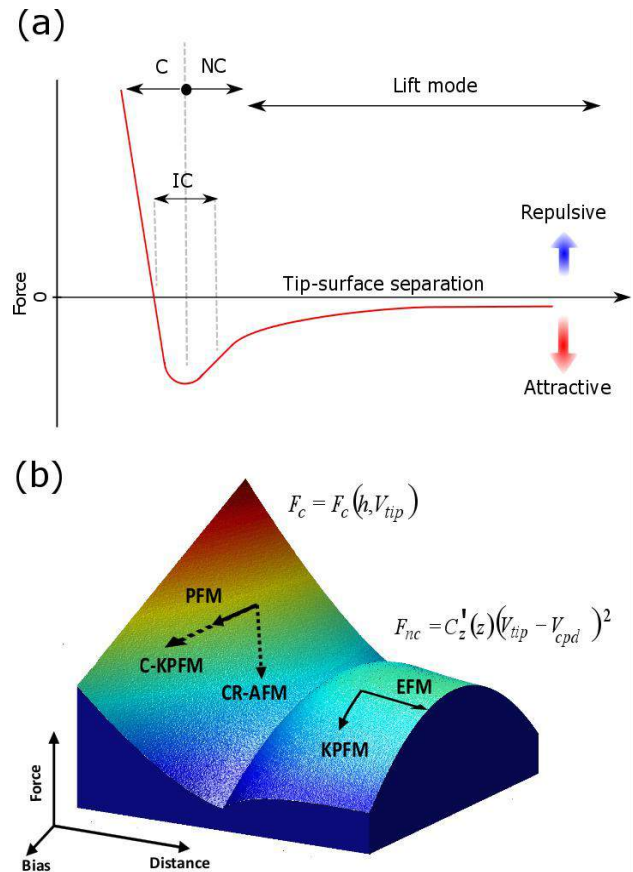

Figure 1. (a) A schematic of the dependence of force on sample tip separation showing the regimes in which contact (C), non-contact (NC), intermittent contact (IC), and lift mode imaging are performed. (b) VM-AFM can be described using a force-distance-bias map. In the small signal limit, the signal is directly related to the derivative in bias (e.g., piezoresponse force microscopy (PFM) and KPFM) or distance (e.g., contact resonance (CR)-AFM, EFM) direction. Adapted from [99]. *Seeking permissons.*

Instead of a single force distance relation, VM-AFM is described by two independent variables, tip-sample separation and tip bias, giving rise to force-distance-bias surface, as depicted in Figure 1(b).[99] The fundamental factors underpinning the contrast formation mechanism is the nature or functional form of dominant electrical interactions and the signal dependence on tip-sample separation or contact radius for contact based measurements.[100] These factors govern the potential for quantitative measurements as well determining the best strategies for instrumentation and technique development.

When held in contact with the material, a conducting probe interacts with the sample through short range interactions. In contact, concentrated electric fields in the small volume (e.g. ~10s nm) of the tip-sample junction can be extremely large ($> 10^8$ V/m)[96] and can be used for both sample modifications or probing material functionality. Examples of sample modification using VM-AFM in contact include ferroelectric polarization switching,[82, 97] controlled nanoscale melting, [92, 101] and lubrication,[102] as well as controlled local electrochemical processes.[103, 104] PFM is probably one of the most well-known VM-AFM contact characterization technique. In PFM the electric field at the tip can induce strains in piezoelectric samples which are generated by linear electromechanical coupling. The modulated electric field induces periodic deformation of the material volume, which is detected by the coupled cantilever motion using a lock-in amplifier(LIA) (see section 3.2.1.1). In the contact regime, the imaging mechanism is ultimately controlled by the shape of the force-distance-bias surface, i.e., $F_c = F_c(h, V_{tip})$, where $h$ is the indentation depth. In the small signal approximation, the PFM signal is given by $(\partial h / \partial V_{tip})_{F=const}$. [99] PFM will be discussed in more detail in the context of measurements in liquid environments in a subsequent section. A recently developed, but related approach, cKPFM, allows measurement of electrostatic forces while the tip is held in contact with the sample, which is useful for differentiating between ferroelectric and non-ferroelectric hysteresis and for exploring triboelectric effects in contact.[97, 105] CR-AFM is another contact mode technique with contrast related to mechanical properties. CR-AFM is highly sensitive to the tip-sample contact area. In contrast to PFM or cKPFM, in CR-AFM,[106-109] the ac voltage is applied to a mechanical actuator (e.g. piezoelectric actuation, photothermal actuator). This modulation voltage vibrates the tip into the sample, and the amplitude and phase of the resulting lever oscillations are decoupled, much in the same way as in PFM. Combining CR-AFM with voltage spectroscopy has shown it is possible to explore local bias-induced phenomena ranging from purely bias induced elastic softening to surface electrochemical processes.[91, 110]

In the non-contact regime, for techniques such as KPFM and EFM, the electrostatic force has a parabolic dependence on the tip voltage, described by:

$$F_{cap} = -\frac{1}{2} C_z{'} (V_{tip} - V_{cpd})^2 \qquad (2)$$

where $C_z{'}$ is tip-sample capacitance gradient with respect to separation distance ($z$) and $V_{cpd}$ is the contact potential difference (CPD). In similar fashion the force gradient acting between the probe apex and sample can be described by,

$$F_{cap}{'} = \frac{\partial F_{cap}}{\partial z} = -\frac{1}{2} C_z{''} (V_{tip} - V_{cpd})^2 \qquad (3)$$

where $C_z''$ is the derivative of the capacitance gradient.

KPFM and EFM are both sensitive to the voltage derivative of the force, $\partial F_{nc}/\partial V_{tip}$. This effect will form the basis behind the nulling technique and CL bias feedback, and the known functional form of this dependence in KPFM makes the approach readily interpretable and insensitive to topographic artifacts. Without the force vs. bias relationship, the signal mechanism is a convolution of electronic surface properties and the capacitance gradient, which is the case in EFM.

## 2.2 Basics of electrostatic force microscopy

In the late 1980s, Martin *et al.*[71] along with Stern[72] and Terris *et al.*,[73] independently adapted the concept of the century old Kelvin method[111] to the nanoscale by combining it with AFM. Their combined approaches would become known collectively as EFM, although operated by fundamentally different detection methodologies (e.g., electrostatic force[71] vs. force gradient detection[72]). EFM presented the first opportunity for nanoscale electrical characterization of all materials (e.g., conductors, semiconductors, insulators etc.). This is because, unlike the current sensitive scanning Kelvin probe (KP) or potentiometric measurements using STM,[64] EFM detects the long-range electrostatic forces acting on an AFM cantilever, while offering far superior spatial resolution to KP.

These attributes resulted in EFM being applied to mapping electric field distributions in devices,[112] imaging of self-assembled monolayers on surfaces, [113, 114] potential and polarization mapping on DNA and proteins, [115] surface potential variations in oxide bicrystals,[116, 117] static and dynamic properties of ferroelectric materials,[118] as well as observation of charge storage and leakage in various materials. [119-121] Although the nature of the interaction is always the same, several possible technical implementations of EFM exist today.[86, 122]

### 2.2.1 Force gradient detection

As in most AFM approaches, the dynamic response a cantilever is more easily detected than static forces due to the removal of the $1/f$ component. The most common implementation of EFM involves mechanically driving the cantilever close to its resonance frequency ($\omega_0$) by applying a voltage, $V_{drive} = V_{ac}\sin(\omega_0 t)$, to an actuator such as a piezo element (i.e., mechanical coupling) or a laser driver (i.e., photothermal coupling [123, 124]).[125] The dynamics of a freely oscillating cantilever can be modelled by using a damped simple harmonic oscillator (SHO) model.

$$m\frac{d^2z}{dt^2} + \frac{m\omega_0}{Q}\frac{dz}{dt} + kz = F(z,t) \qquad (4)$$

where $m$ is the effective mass, $\omega_0$ is the mechanical resonance frequency, and $z$ is the tip deviation from the equilibrium position ($z_0$). The viscous damping coefficient is $m\omega_0/Q$ defined the energy lost per oscillation and the dimensionless quality factor, $Q$, is a measure of how fast the cantilever responds to perturbation. $Q$ is defined as the full width at half maximum of the resonance peak and in practical terms, it determines the cantilever bandwidth, $\omega_0/2Q$. The external forces $F(z,t)$ are composed of the time-dependent excitation force, $F_{exc}$, and the distance-dependent tip sample force, $F_{ts}$.

$$F(z,t) = F_{exc}(t) + F_{ts}(z) \qquad (5)$$

The sinusoidal excitation of the actuator induces cantilever oscillations and the time dependent tip-sample separation is $z(t)$:

$$z(t) = z_0 + \text{A}(\omega)\sin(\omega t - \varphi) \qquad (6)$$

where $\text{A}(\omega)$ is the frequency dependent oscillation amplitude and $\varphi$ is the phase shift between the driving voltage on the actuator and the resultant cantilever oscillation. The following expressions for the oscillation amplitude $\text{A}(\omega)$ and phase $\varphi(\omega)$ are given by:

$$A(\omega) = \frac{A_{max}\omega_0{}^2}{\sqrt{(\omega^2 - \omega_0{}^2)^2 + (\omega\omega_0/Q)^2}} \qquad (7)$$

$$\varphi(\omega) = \tan^{-1}\left(\frac{\omega\omega_0/Q}{\omega^2 - \omega_0{}^2}\right) \qquad (8)$$

If during non-contact operation the conductive tip is then biased the resulting electrostatic force gradient produces a shift in the resonance frequency ($\omega_0$), governed by the following equation:

$$\Delta\omega \approx \frac{\omega_0}{2k} - F_{cap}{}' \qquad (9)$$

Figure 2, the electrostatic force gradient results in a shift in the fundamental resonance peak which can be considered as an effective softening of the cantilever spring constant. Notice that this frequency shift results in corresponding changes to the phase and amplitude of the cantilever oscillation.

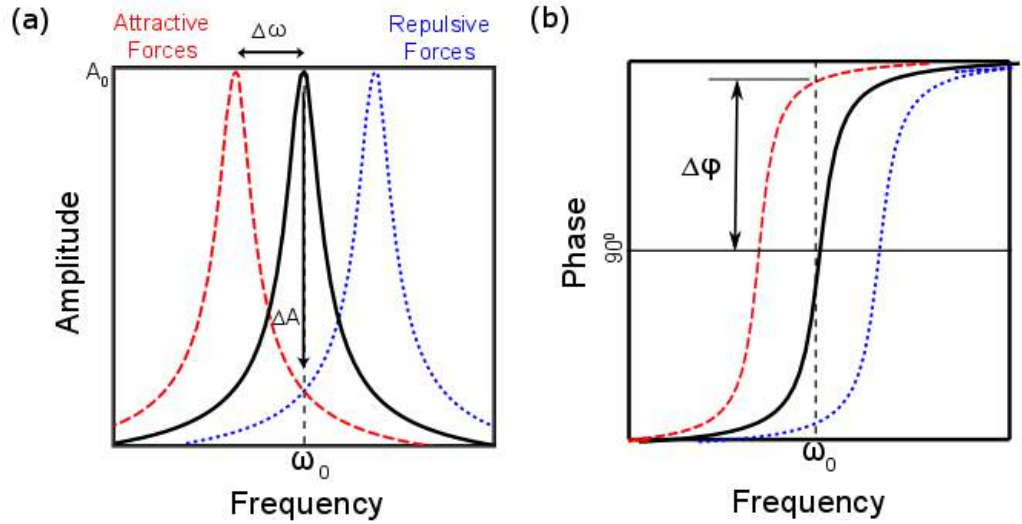

Figure 2. The harmonic potential of the cantilever spring is modified by an attractive or repulsive force gradient. This modification leads to a shift of the cantilever resonance (a) frequency and an associated (b) phase shift, which is detected by the AFM control electronics.

Therefore, the phase of a mechanically modulated (biased) cantilever at a constant driving frequency allows an indirect method of force gradient detection. The phase shift at resonance can detected using heterodyne detection and related to the capacitive force gradient described by:

$$\Delta\varphi \approx \frac{Q}{k} - F_{cap}{'} \qquad (10)$$

For large force gradients, when the resonance shift exceeds the full width half maximum of the resonance equation (10) will break down and the phase imaging method becomes impractical. Normally this is not the case for large tip-sample separation used, however it precludes operation in single pass mode where mechanical tip-sample interactions can further induce phase shifts and produce parasitic signals or topographic artifacts on the EFM signal.

In 1991 Albrecht *et al.*[126] developed non-contact AFM based on the dynamic AFM developed previously by Martin *et al.*[127] and soon after further improvements were made by introduction of the phase-locked loop (PLL) technique.[128] The PLL presented an alternative method to determine the electrostatic force gradient. Instead of phase detection using LIA, the PLL allowed frequency shifts to be detected directly, and an additional feedback loop was introduced to adjust the driving frequency to match the cantilever resonance. In this frequency-detection regime, the relationship between the resonant frequency shift and the force gradient is given by equation(9).

### 2.2.2 Force detection

EFM force gradient detection has been described using combined mechanical and electrical excitation. However, it is also possible to perform electrostatic force sensitive EFM. By combining $V_{tip}$ with equation (2), and splitting the forces acting on the cantilever into their frequency dependence or spectral components including static ($F_{dc}$) and dynamic ($F_\omega$ and $F_{2\omega}$) interactions, the forces on the cantilever can described by:

$$F_{dc} = -\frac{1}{2}C_z'\left[(V_{dc} - V_{cpd})^2 + \frac{1}{2}V_{ac}^2\right] \tag{11}$$

$$F_\omega = -C_z'(V_{dc} - V_{cpd})V_{ac}\sin(\omega t) \tag{12}$$

$$F_{2\omega} = C_z'\frac{1}{4}V_{ac}^2\cos(2\omega t) \tag{13}$$

Variations in the capacitance terms arise from variations in the dielectric constant and geometry of the tip-sample interaction. The latter arises from non-electrostatic forces acting on the tip, imperfect feedback control, as well as sample topography such as step edges. Practically, the dc component can be determined by directly monitoring the raw static deflection of the cantilever, whereas the amplitude (and phase) of the dynamic cantilever response at the harmonic frequencies are normally captured using traditional LIA detection. However, the convolution of $V_{CPD}$ and capacitance terms in the electrostatic forces makes techniques such as EFM extremely difficult to quantify directly and requires precise knowledge of tip-sample geometry.

## 2.3 Quantifying EFM

Over the past decade, several groups have focused on extracting quantitative information on surface potential,[118, 129, 130] dielectric constant,[131-135] trapped charges[136, 137] and polarizability[138, 139] from EFM measurements. Most applications of quantitative EFM have focused on mapping variation in surface charge or dielectric constant, often by modeling the electric field at the biased tip and/or calibrating the precise tip-shape. Another focus has been on mapping the variation in surface potential which can be deduced from the functional force-bias dependence. However, it is important to consider, particularly for insulating materials, that the surface potential does not uniquely define the materials electronic properties. Equally, the tip-sample forces in EFM are strongly coupled to variations in dielectric properties, as well as surface bound and volume trapped charges.[138, 140] Quantifying such properties, however, requires a precise understanding of the EFM force or force gradient contrast. Such analysis is complicated due to an unknown capacitance term determined by the complex tip-sample geometry, as well as the dielectric properties of the tip-sample gap in which electric field propagates. Indeed, quantitative analysis of EFM data calls for detailed knowledge of the tip geometry as well as the exact nature of the tip-sample interaction.[141]

### 2.3.1 Electrostatic force spectroscopy

Spectroscopic techniques can be used to determine the functional form of the force-bias or distance) relationship in EFM, which in turn can be used to extract information on the potential and capacitive sample properties. Indeed, in the macroscopic KP community, "off-null" bias spectroscopy measurements have been shown to be beneficial over classical closed loop (CL) nulling techniques. [142] In electrostatic force spectroscopy (EFS), the $V_{dc}$ is varied linearly as the electrostatic interaction is detected, allowing the functional form of the force (or force gradient) to be measured and verified against equations (11)-(12). In Figure 3, an example EFS measurement is provided in which the tip was positioned (50 nm) above a freshly cleaved highly oriented pyrolytic graphite (HOPG) surface, as an AC voltage having a frequency of 15 kHz and amplitude ranging from 500 mV to 7 V is superimposed on a linear dc bias ramp (-3 V to +3 V) applied to the tip. The static bending of the cantilever has a parabolic bias dependence and demonstrates a squared ac voltage dependence (Figure 3(a,b)). The second term at the fundamental drive frequency has linear bias dependence (equation (12)) and demonstrates a linear capacitive coupling with $V_{ac}$ (Figure 3(c,d)). Note, as described in equation (12), the amplitude at the fundamental drive drops to zero when the tip bias $V_{dc} = V_{CPD}$ (see Figure 3(c)). In this way, if the functional form of the force is known then the response of the cantilever can be used to determine the CPD (a basic premise in the development of KPFM). For this reason, the approach is sometimes refered to as Kelvin probe force spectroscopy, Typically, EFM measurements are performed at $V_{dc} = 0$, or at some constant $V_{dc}$ used to enhance the electrostatic force between tip and sample. This aspect makes quantification of surface potential impractical. However, combined with single point EFS or by recording EFM images at different dc offsets, surface potential properties can be inferred.[129, 130] From Figure 3(e,f), the third term at $2\omega$ is shown to depend only on the capacitive coupling between tip and sample and is not influenced by the tip bias, having a squared dependence on the ac voltage. In the following sections, we will describe how this response can be useful in measuring polarization forces and quantifying dielectric properties.

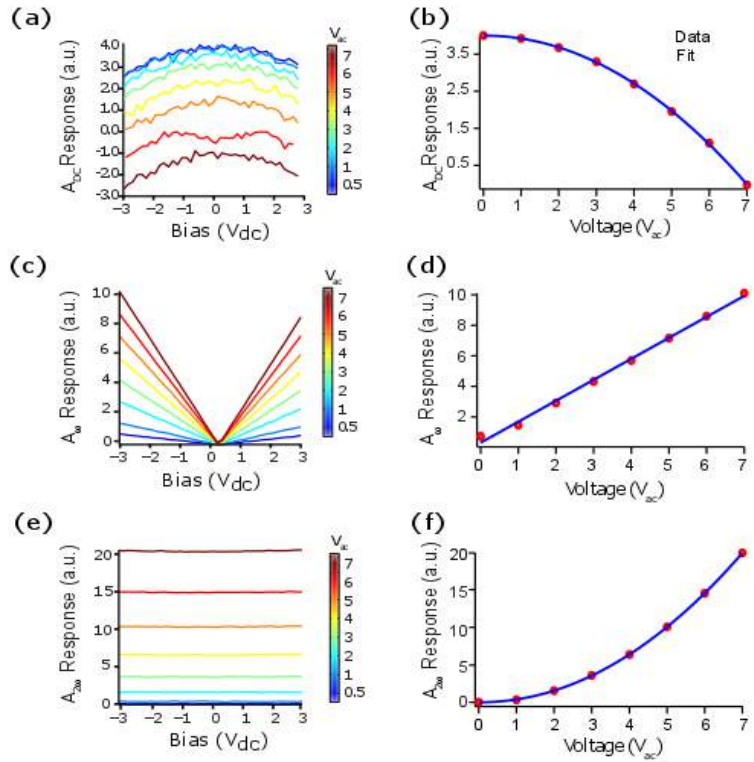

Figure 3. (a,c,e) DC bias and (b,d,f) AC voltage dependence of the (a,b) static deflection, (c,d) first and (e,f) second harmonic amplitude response. Reprinted with permission from [143]. Copyright 2016, Institute of Physics.

### 2.3.2   Stray capacitance

In EFM (and KPFM) capacitive tip-sample interactions are heavily influenced by the geometry of the tip-sample capacitor (i.e., lift height, probe architecture, surface topography) as well as the dielectric properties of the tip-sample interaction. A major difficulty in interpreting EFM/KPFM is due to the lack of precise descriptions of the electrostatic force and capacitance acting on the tip in the presence of a polarized object. This complexity arises from the irregular shape of the AFM probe: a conical or pyramidal tip ending in a spherical apex, attached to a beam-shaped or triangular cantilever. In stark contrast to contact mode techniques, where the tip-sample interaction is dominated by interaction at the tip apex (e.g., PFM), in non-contact techniques, the contribution of the entire probe can contribute significantly. Consider that equation (2) implies that the electrostatic force is a function of the tip-sample geometry, as well as dielectric properties in the gap, coupled through the capacitance gradient term. Consequently, $F_{cap}(z)$ and $F_{cap}'(z)$ are rapidly decaying functions of tip-sample separation.

Several models describing tip-sample interactions have been proposed. Convenient analytical expressions are often used, where the total capacitance $C(z)$ is approximated as the sum of the contributions due to the tip apex, bulk tip, and the cantilever, correspondingly:

$$C(z) = C_{apex}(z) + C_{bulk}(z) + C_{cant}(z) \qquad (14)$$

Approximate geometric models of the tip-sample interaction based on isolated point charges [73, 144] and parallel-plate geometry [127, 145, 146] have been proposed and extended by considering tip shape, sample surface geometry, and image charge effects in the substrate [119, 141, 147]. Descriptions of the tip as a sphere,[73] hyperboloid,[148-150] and cone [151] have been proposed to find analytical descriptions of the electrostatic force (Figure 4(a)).

Hudlet *et al.*[152] developed one such analytical model for a cone with spherical apex which has been widely adopted for describing the interaction between an AFM probe and metallic substrate, where the cone contribution is written as:

$$F_{cone} = \frac{\pi\varepsilon_0 V_{tip}{}^2}{(ln\tan(\theta/2))^2} ln\frac{H}{z} \qquad (15)$$

where $z$ is the distance from the surface, $H$ is the cone height, and $\theta$ is the cone angle. The apex contribution of the tip for a sphere-plane geometry is given by:

$$F_{apex} = \pi\varepsilon_0 R^2 \frac{(1 - \sin(\theta))}{z[z + R(1 - \sin(\theta))]} V_{tip}{}^2 \qquad (16)$$

where $R$ is the effective radius of the tip apex. In addition, Hochwitz *et al.*[153] and others [141, 154, 155] showed that the electrostatic force acting over the cantilever supporting the tip has a non-negligible influence on the overall force. The role of the cantilever cannot be ignored in the quantitative analysis of EFM data, especially where claims about the lateral resolution or attempts at quantitative surface potential are made.[118] Belaidi *et al.*[141] compared an equivalent charge model used to determine the electrostatic force acting on a tip with existing analytical models. They showed experimentally that the force acting on the cantilever becomes dominant for distances $d > 10$ nm in the case of sharp ($R = 10$ nm). They concluded that this effect can be minimized by using tips with larger angles, apex radii or lengths. For a simple description of the cantilever force, an analytical expression using a parallel plate [71] model can be used:

$$F_{cant} = -\frac{1}{2}\frac{\varepsilon_0 A}{(z+H)^2}V_{tip}{}^2 \tag{17}$$

Wagner *et al.*[156] calculated the total contribution to the tip apex, bulk tip, and cantilever beam of an AFM probe to the measured capacitive force (i.e. amplitude modulation (AM)) and force gradient (i.e. frequency modulation (FM)) as a function of distance for a mechanical oscillation amplitude of 5 nm (shaded region shows range of amplitude ($0.1 - 50$ nm). The influence of different aspects of the AFM probe on electrostatic force and force gradients are summarized in Figure 4(b). For intermediate and large tip-sample separation the cantilever provides the largest contribution to the electrostatic force. The major contribution to the force gradient is due to the bulk tip; the cantilever providing a distance-independent offset.

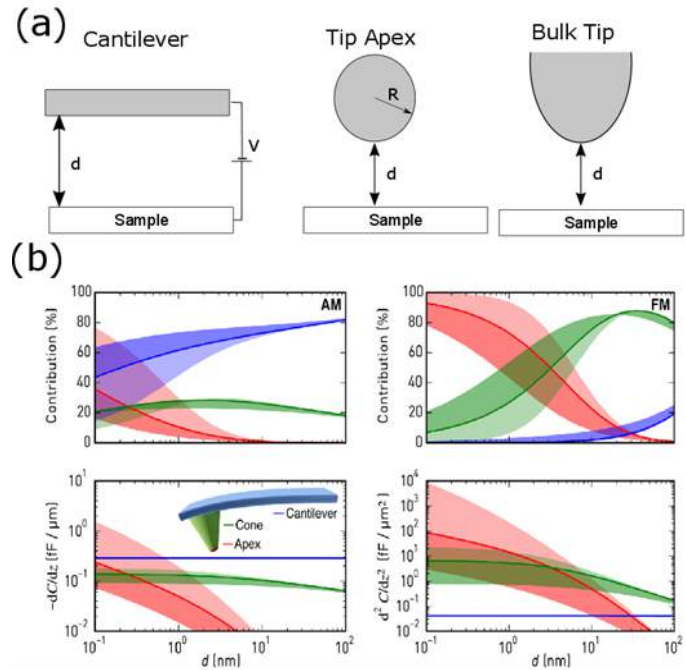

Figure 4. (a) Simple schematics of the geometric approximation of the cantilever (parallel plate), tip apex (sphere) and probe bulk (hyperboloid). (b) Calculated contributions of apex, cone, and cantilever to the first (AM, left) and second (FM, right) order capacitance gradient as a function of distance for a mechanical oscillation amplitude of 5 nm (solid line) and an Olympus AC160 cantilever (Rtip = 5 nm, Htip = 14 μm, θ = 17.5°, and Alever = 160 μm × 40 μm). The light (dark) shaded regions indicate the range up to A = 0.1 nm (50 nm). Reprinted with permission from [156]. Copyright 2015, Beilstein-Institut.

While simple approximate models provide convenient analytic expressions of the tip-sample interaction, for studying the lateral variation of the sample surface properties (e.g., topography and trapped charge distribution) precise calculation of tip-sample capacitance $C(z)$ is required, which can only be obtained only by numerical methods. One such approach is to use molecular-dynamics [157, 158] to track the motion of the atoms of the probe tip and the sample surface, while modeling the cantilever as a point mass to calculate the resulting the electrostatic force-field. Barth *et al.*,[159] highlighted that although these theories could correctly predict trends in contrast changes when varying the tip-sample distance, quantitative agreement with experiment is limited due to the non-ideal atomic structure of real tips. As is typical of molecular-dynamics systems, artificial numerical constraints had to be introduced to these models (e.g., periodic boundary conditions).[160] Several authors modelled the tip-sample system as a series of point charges and/or line charges and their image charges.[141, 161] Similar analysis has been performed for conductive and dielectric samples with topographic and/or dielectric inhomogeneities [131, 162, 163]. The electrostatic boundary value problem can also be solved using continuum level numerical schemes such as finite element methods,[164] self-consistent integral equations,[165] and the boundary element methods.[160, 166]

### 2.3.2.1   Dielectric polarizability

Equations (11)-(13) are commonly cited to describe contrast in EFM/KPFM. This is despite the equations being derived for the ideal conditions of a metal tip and metal sample in vacuum. Quantification of the electrostatic tip-sample interaction is significantly more complicated for lossless dielectric materials. Unlike conductors, dielectric materials can sustain both surface charges and volume trapped charges, which contribute to the total electrostatic force, as well as capacitive interactions due to tip-induced image charges.[167] Consider the interaction between the tip and a point charge on the surface of the insulting film. Image or mirror charges will be induced on the probe apex ($Q'_s$) and the back electrode. In EFM, the dominant forces can be separated into two parts: Coulombic forces due to static charges and multipoles and capacitive forces due to surface potential and dielectric screening.[138]

Following Sarid,[168] for a charge $Q_s$ on a dielectric surface the electrostatic interaction between a biased tip and the sample includes three distinct components: effective surface charge on the sample, $Q_s$, constant dc bias applied between the tip and the sample, $V_{dc}$, and periodic ac voltage applied between the tip and the sample, $V_{ac}\cos(\omega t)$. The charge on the tip can be approximated as:

$$Q_{tip} = Q'_s + Q_{dc} + Q_{ac} \tag{18}$$

where $Q'_s$ is the image charge and $Q_{dc} = C(z,k)V_{dc}$ and $Q_{ac} = C(z,k)V_{ac}\sin(\omega t)$ where $C(z,k)$ is the effective capacitance of the tip-sample system, $k$ is the dielectric constant of the material. In the limit of high dielectric constant, $C(z,k) \approx C(z)$, the total expression for the force can be written down as:

$$F = \frac{Q_s Q_t}{4\pi\varepsilon_0 z^2} + \frac{1}{2}C'(V_{dc} + V_{ac})^2 \tag{19}$$

Substituting the expression for $Q_t$, the spectral components $F_{dc}$ becomes:

$$F_{dc} = \frac{Q_s{}^2}{4\pi\varepsilon_0 z^2} + \frac{Q_s V_{dc}}{4\pi\varepsilon_0}\frac{C}{z^2} + \frac{1}{2}C'\left(V_{dc}{}^2 + \frac{1}{2}V_{ac}{}^2\right) \tag{20}$$

and $F_\omega$ becomes:

$$F_\omega = \left(\frac{Q_s V_{dc}}{4\pi\varepsilon_0}\frac{C}{z^2} + C'V_{dc}\right)V_{ac} \tag{21}$$

and $F_{2\omega}$ remains unchanged (equation (13)). Thus, static dielectric charges and tip-induced charges contribute to the harmonic components of the force differently and can thus be distinguished. In practice however, the lack of the information on tip-sample capacitance and the charge state of the dielectric significantly complicate analysis.

### 2.3.3   Quantifying surface charge

Determination of surface or volume charge density in dielectric is often the purpose of the EFM experiment. This, however, requires a detailed understanding of the tip-sample capacitance, and hence precise knowledge of the tip-sample geometry. Several different approaches have been explored. One effective way of extracting quantitative information on charge dynamics using EFM is to exploit time-varying phenomena. Transient responses such as charge decay,[169, 170]  blinking,[144] or charge transport under the influence of an applied voltage have been used to quantitatively measure local trapped charges.[119, 144, 145, 171]

However, the measurement of static properties, such as local dielectric constant, precludes the use of these approaches. Tip-sample simulations of EFM have been shown to be useful in the determination of physical quantities from EFM images, particularly for flat samples, as complex sample topography considerably complicates the situation.[172] One popular approach is to use simplified analytical models of probe geometry (apex, bulk, cantilever) and compare these with individually calibrated tips by comparing C′ (z). This approach can be used to determine the influence of geometry on the measured signal. For more complex systems the influence of sample geometry should also be considered. Several methods have been described for quantitative analysis of charge and dielectric properties on chemically synthesized nanocrystals (e.g., CdSe,[144] PbSe,[139] nanorods,[173] and Au [138] and SiGe [172] nanostructures).

Although good correlation between model and experimental data has been demonstrated, additional correction terms are often necessary for the model to capture the relevant parameter dependences.[138] For measurements on complex samples, having nanostructured topography, simulations must be extended to consider 3D electrostatic contributions. Qi *et al.*[174] demonstrated quantitative measurement of charge density using EFM combined with a millimeter-sized conductive sphere as a charge reference. The reference sample has a well-defined charge density dependence on applied voltage and hence EFS measurements could be used to calibrate against this reference. In this way, the charge density of an unknown sample could then be determined by comparing curves measured on sample and the reference. This approach normalizes the influence of the complex geometry of EFM tips, providing a facile approach for quantitative analysis of the charge density on sample surfaces at the nanometer scale. However, it does not take into account sample topography. It was later applied to the determination of surface charge density and temperature dependence of purple membrane.[175]

### 2.3.4   Quantifying dielectric properties

Determining the local dielectric properties of materials and devices is important for dielectric materials, particularly in the fields of polymer or biological research. Classically, dielectric properties are measured macroscopically using broadband dielectric spectroscopy (BDS),*[176]* which is based on the measurement of the capacitance between parallel plates. BDS, however, lacks spatial information and hence is not suitable for investigating local properties or nanoscale devices. Several groups attempted to gather similar information comparable with BDS at the nanoscale by reducing the size of the electrode towards the dimensions of an AFM probe.[177] Other approaches for extracting related material properties include nanoscale impedance microscopy (NIM) [178] [179, 180] and scanning capacitance microscopy (SCM)[181] which are current sensitive AFM modes, as such, these modes typically suffer from poor sensitivity due to extremely small currents, corresponding to a low capacitance of the tip-sample capacitor. These techniques often require bespoke electronics and great care to avoid and characterize stray capacitances present. Also should be mentioned the applications of the scanning impedance microscopy[75, 182-187] and its non-linear analogs[188, 189] that can be applied for probing lateral ac transport in lateral devices, proving the characteristics of corresponding lumped elements or parameters of transmission line models[190].

The ability to extract dielectric properties directly from the electrostatic forces detected by VM-AFM occurred in 1995 with the development of scanning polarization force microscopy (SPFM)[191] This was the first technique to qualitatively map dielectric polarization, particularly for liquid droplets and other weakly adsorbed materials. In the SPFM mode, a bias voltage (dc or ac) is applied to a conductive cantilever tens of nanometers from the surface while constant force feedback is employed. Figure 5 exemplifies a utilization of SPFM to map the water film structures (average thickness of about 2 Å) covering a mica surface. The two phases present in the figure are due to a difference in dielectric constant related with different water layer ordering.[192] While SPFM is a useful imaging technique, it remains difficult to interpret quantitatively.

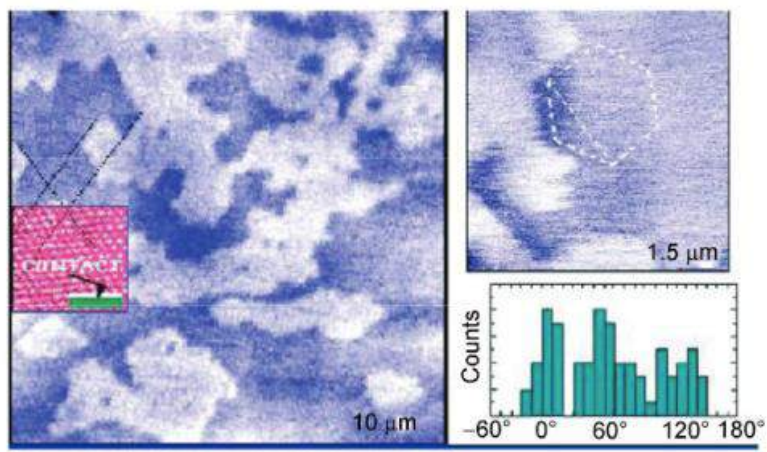

Figure 5. Scanning polarization force microscopy (SPFM) images of ordered (ice-like) water structures formed on mica surface. Dark and bright areas represent first and second water layer, respectively. Reprinted with permission from [192]. Copyright 1995, AAAS.

Krayev *et al.* demonstrated that variations of relative dielectric permittivity could be mapped from the force gradient signal in EFM (phase detection). This was realized on thick (several microns) heterogeneous polymer blend.[193, 194] The authors quantified the value of $\varepsilon$ by using a simple spherical capacitor model and a calibration with two known polymers. This approach has some caveats, including the model is not universally appropriate especially for thin samples when the thickness of the sample is very small compared to the tip geometry. [193, 194] It also requires two reference polymers, so that the ratio of the phase measured between them can be used to measure a third polymer of unknown dielectric constant. Using different approaches, EFM in combination with analytical descriptions have been used to quantifying dielectric properties of a range of materials including polymer films,[193-195] nanotubes,[196, 197] and nanoparticles [138, 139, 144]

The group of Gomilla have shown remarkable success in the extraction of physical values directly from the electrostatic force in EFM for both thick[198] and thin [133] insulators. Gramse et al.,[133] presented a simple method to measure dielectric constant of thin films by recording the static bending of a cantilever combined with a simple analytical expression for capacitance. The method was validated on thin silicon dioxide films[133] and purple membrane monolayers.[132]

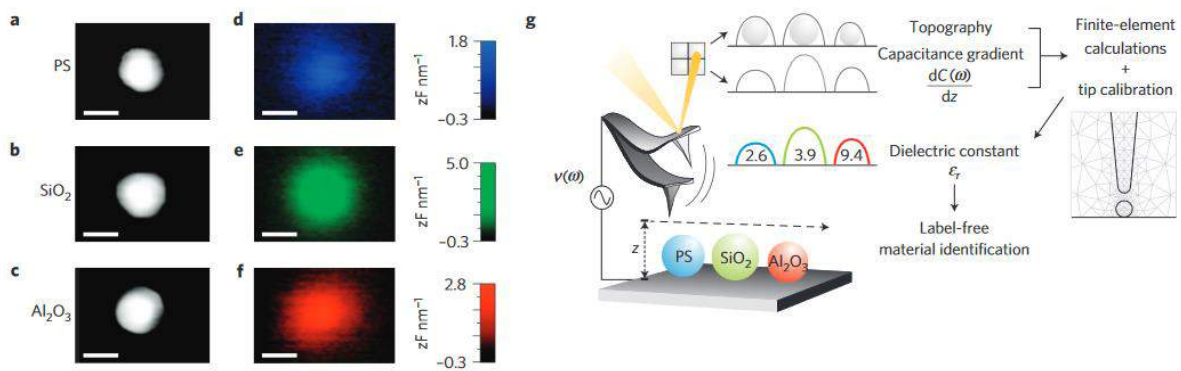

Figure 6. (a–f) Dielectric constant measurement and material identification of individual dielectric nanoparticles in multiple experiments. (a–c) Topography and (d–f) dielectric (capacitance gradient, C) images of polystyrene (blue), $SiO_2$ (green) and $Al_2O_3$ (red) nanoparticles measured on different graphite substrates. Scale bars, 50 nm. (g) Schematic representation of the material identification procedure used by the for determining dielectric constant.[134] Reprinted with permission from [134]. Copyright 2012, Nature publishing group.

The same authors also developed an approach which allowed for the quantification of the dielectric constant of thick insulators from the amplitude of the second harmonic component of the electrostatic force.[198] This dynamic approach is generally more sensitive than static dc methods as LIAs effectively attenuate noise that occurs at frequencies other than that of interest, thereby excluding, e.g., 1/f noise. The dynamic approach precludes the establishment of a universal analytic expression and finite element calculation are required. Additionally, fitting the second harmonic force component as a function of tip-sample separation is required for the quantification of the dielectric constant. The procedure used for extracting quantitative dielectric measurements is shown schematically in Figure 6 along with its application to the quantification of individual nanoparticles composed of different materials. Of particular significance, much attention in the field has been given to application of these newly developed methods in EFM for studying dielectric properties of biological systems including biomembranes,[53] DNA,[199] viruses[134] and bacteria.[200, 201]

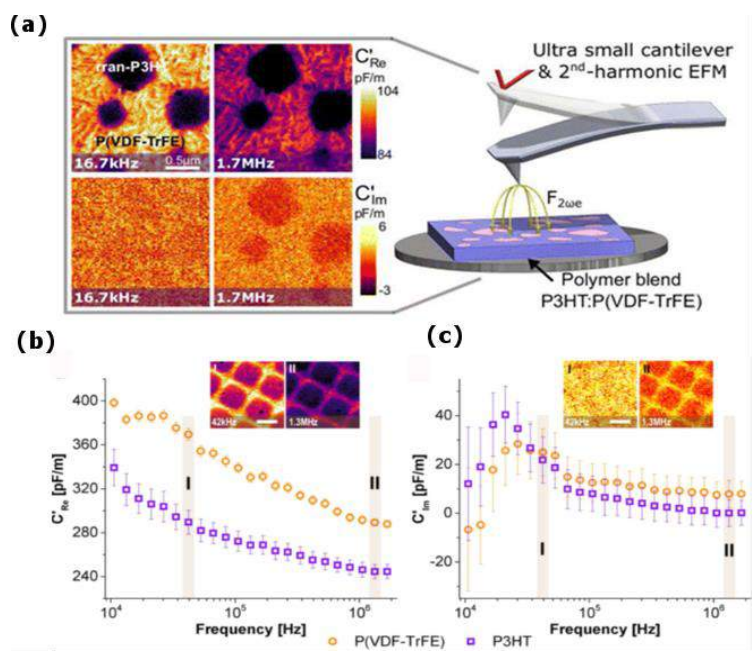

Figure 7. Frequency dependence plots of real ($C_{re}$) and imaginary ($C_{Im}$) capacitance for the nanostructure device of P3HT:P(VDF-TrFE). The squared pattern corresponds to P(VDF-TrFE) and the inner holes to P3HT. Average values of (a) $C_{re}$ and $C_{Im}$ as a function of frequency. Experimental parameters: $V_{ac}$= 5 V, $\omega_m$= 4.8 MHz, $A_0$ = 8.3 nm, $A_{sp}$ = 80%, scale bar 1 μm. Adapted from [202]. Copyright 2016, ACS publications.

Another significant advance in quantifying dielectric properties using EFM was in the development of local dielectric loss spectroscopy (LDS) by Crider *et al.*[203, 204] and Riedel *et al.*[205] These approaches consist of detection of the low frequency EFM response (FM-EFM)[203, 204] or the second harmonic component of the force[205] as a function of temperature and relating the response to dielectric losses in the sample.[205] Schwartz *et al.*[135] developed a method to map the local dielectric frequency response by measuring EFM as a function of frequency, allowing complex dielectric properties to be measured. However, the frequency range of EFM-based dielectric spectroscopy has been limited to a few kilohertz by the resonance frequency of conventional cantilevers with sufficient force sensitivity. Recently, Cadena *et al.*[202] successfully performed LDS measurement over 3 decades of frequency (kHz to MHz) by exploiting the high resonance frequency and low thermal noise of ultra-small cantilevers (USCs).

# 3 Kelvin probe force microscopy

The following sections focus on the historical development and operating principles behind KPFM at the gas-solid interface. We begin by describing the Kelvin method and the development of scanning KP. Next, the working principles behind AM and FM-KPFM are outlined. Followed this, a brief overview of KPFM applications, including citation analysis of published literature, from the first implementation of KPFM to modern applications spanning important areas of science and technology. Finally, a review of artifacts and limitations of the various implementations is given.

## 3.1 The Kelvin method

In 1898,[111] Lord William Thompson Kelvin originally from Belfast, Ireland, outlined a method to measure the "contact electricity" between two metal electrodes. Nowadays this approach is referred to as the Kelvin method, although Lord Kelvin himself did acknowledge that a similar measurement had been described previously by Pellat in 1881.[206] More precisely, the Kelvin method measures an "electrochemical" or contact potential between a metal probe and a metal sample of different work functions ($\Phi_1$ and $\Phi_2$), shown schematically in Figure 8. The metals are placed in close proximity but not electrically connected. The work function is defined as the amount of energy required to eject a bound electron in the Fermi level ($\varepsilon$) to the vacuum energy level ($\varepsilon_{vac}$). When an electrical connection is made between the dissimilar metals, as shown in Figure 8(b), electrons will flow from the metal with the lower work function to the one with the higher work function. If the electrodes are in a parallel-plate capacitor configuration, then equal and opposite surface charges will form on the capacitor plates. The contact potential developed across the plates (i.e., CPD) is given by $V_{CPD} = (\Phi_1 - \Phi_2)/\text{-e}$. The CPD can be measured by applying an external "backing" potential ($V_{dc}$) to the probe (or sample), until the point at which the surface charges between the probe and sample are nullified. In this equilibrated state, the applied backing potential is exactly equally to the CPD or the work function difference ($\Delta\Phi = \Phi_1 - \Phi_2$) between the metals, $V_{dc} = \pm V_{CPD}$. Note that the $\pm$ sign depends whether $V_{dc}$ is applied to the sample (+) or the probe (−).[207] For the remainder of this section, we frame the discussion with respect to application of $V_{dc}$ to the probe.

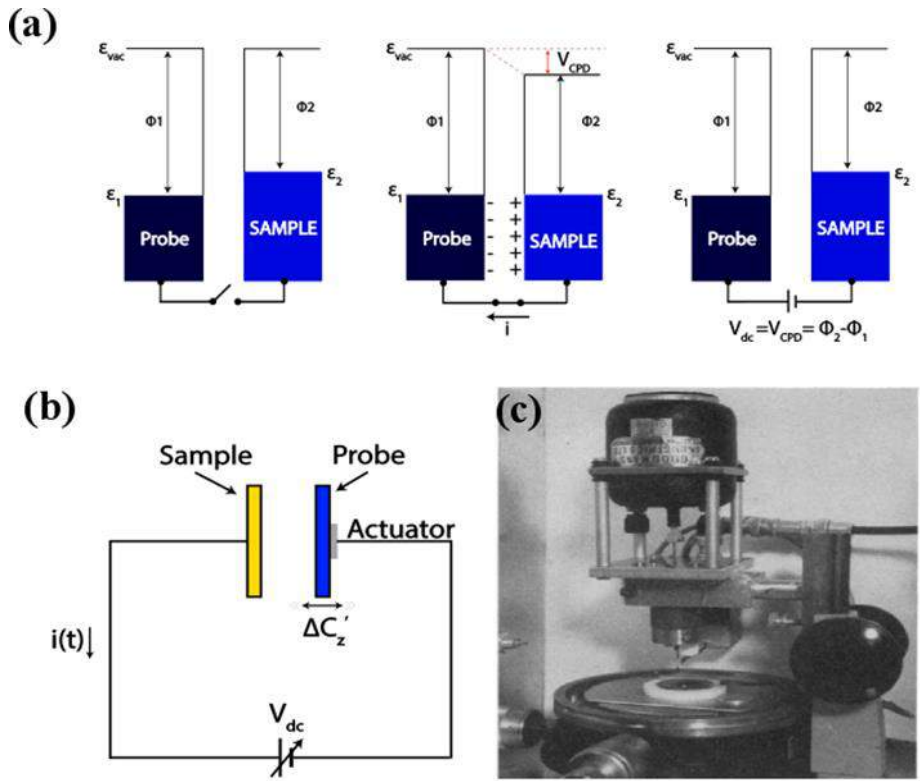

Figure 8. Concept of the Kelvin probe. (a-c) Band gap diagrams depicting the principle behind the Kelvin method. (b) Simple schematic of the KP apparatus and (b) an image of an early vibrating Kelvin probe system developed by Shockley et al. Adapted with permission from [208]. Copyright 1964, Elsevier.

### 3.1.1 The scanning Kelvin probe

The basic operational principles of the KP are shown in Figure 8(b). In the KP two conductors are arranged in parallel plate configuration with a small spacing in between. In 1932, Zissman *et al.*[209] implemented the alternating current mode of the KP by modulating the distance between the probe and sample using a vibrating piano wire as an actuator. The actuation of the metal electrodes results in a time varying capacitance making the current flow back and forth between the plates. The periodic vibration (e.g., where $\omega$ is the angular frequency of vibration) results in an observable current described by $i(t) = C_z'(V_{dc} - V_{CPD})\cos(\omega t)$, where $C_z'$ is the capacitance gradient as a function of distance. Notice that once $V_{dc}$ nullifies the CPD between the capacitor plates, $i(t) = 0$. This detection and corresponding compensation of the current using a backing potential is the fundamental mechanism behind the KP and is realized using a bias feedback loop, which continually adjusts $V_{dc}$ to nullify the measured current.

Decades of research have gone into the continued improvement in accuracy and sensitivity of KP.[142, 210-212] Innovations in UHV-compatible driver design including external tuning fork,[213] voice coil [210] or piezoelectric actuators [214, 215] as well as utilizing data acquisition systems have made for a more versatile approach.[212] Adoption of scanning KP using XYZ stage drivers as well as miniaturization of probes [216] allowed for spatially resolved CPDs to be recorded. Generally, a trade-off between sensitivity and lateral resolution is to be expected with probe diameter, with large electrodes allowing changes as low as 0.005 meV to be detected, whereas miniaturized probes have decreased accuracy (~20 meV) but with improved lateral resolution (< 40 µm).[142]

Since the topmost ~1 – 5 layers of atoms or molecules typically define the CPD of a surface,[217] the approach is very sensitive to surface condition. This makes KP a widely applicable surface analysis technique for the investigation of corrosion,[218, 219] metallurgy,[220] semiconductors,[221] and even for fingerprint detection,[222-224] as demonstrated in Figure 8(c). In addition, KP has been used in the study of thin films,[225, 226] adsorption kinetics,[227-229] surface photovoltage spectroscopy,[230, 231] and work function topographies[232].

Relevant to the topic of this review, several authors have noted that work function measurements using KP are prone to difficulties caused by parasitic capacitance and poor signal to noise levels offered by traditional LIA and self-nulling detection, leading to the implementation of methods that aim to overcome these difficulties, including improved shielding and off null operation.[142] Decades later, similar issues were identified and tackled by the KPFM community and is the topic of subsequent discussions (see Section 3.2.1.2).

## 3.2    Kelvin probe force microscopy

### 3.2.1    Introduction

As described in the previous section, a major drawback of EFM is that the image mechanism is governed by an unknown capacitance gradient, rendering quantitative interpretation of the EFM signal more challenging. Towards quantitative surface potentials using EFM, Weaver *et al.*[233] explored the possibility of measuring potentials for EFM by introducing bias feedback into the EFM approach. In 1991 Nonnemacher *et al.*[74] coined the name KPFM to describe a similar approach used for determination of CPD. Since then, KPFM, sometimes called scanning surface potential microscopy (SSPM), scanning Kelvin probe microscopy (SKPM), or Kelvin force microscopy (KFM), has found diverse applications in several fields, due to the direct and quantitative measurement principle. The approach has been widely adopted by physicists, materials scientists, and physical chemists to measure 'contact potentials', 'electrochemical potentials', 'work functions' or 'surface potentials', a somewhat confusing variety of misnomers for the same measurement. Care should be taken by researchers to identify whether true work functions (i.e. in the case of metallic conductors) or surface potentials (i.e. in the case of insulators) are being measured and reported or if in fact only the CPD between the probe, of nominally unknown work function, and the sample are measured.

The success of KPFM is demonstrated through the increasing number of publications and citations it has garnered in the short time since its inception as seen in Figure 9(a). As of January 2017, the web of science database includes over 1590 published articles have KPFM (or SSPM, SKPM, KFM) including in the topic of the research. Its broad applicability to conducting and non-conducting samples alike makes it a versatile approach, having widespread applicability across all fields of science including physics,[234],[235] biology,[236-242] electrochemistry,[243-245] and material sciences.[207] Finally, the unparalleled spatial resolution of the approach makes it useful for understanding the role of inhomogenetics on the length scales of nanostructures,[246, 247] grain boundaries,[248-250] step edges,[251-253] point defects, [254] or even individual atoms.[255, 256] KPFM has found applications for mapping local electrostatic potential profiles across working devices,[112] semiconductor junctions,[234, 257-260] photovoltaics materials,[261-263] as well as investigating charge dynamics in ferroelectrics [264-266] and dielectrics.[267, 268] For biological materials KPFM has been demonstrated useful for; the label-free detection of biomolecules,[236, 237, 269] surface photo-voltage measurements in optically active proteins,[270-274] for investigation DNA [275-277] and for understanding the relationship between structure and functionality in model lipid membranes.[238, 278-281] Note, however, that nearly all of the success achieved by KPFM to date has been realized at the solid-gas interface, whereas the study of biological systems and energy materials requires measurements in physiologically relevant conditions, which necessitates the application of KPFM or KPFM-like techniques in ionically-active liquids. The progress and development within the field of KPFM is well captured through analysis of citation networks, using the CiteSpace program, and is illustrated in Figure 9(b). This analysis makes it possible to identify the major keywords in papers relating to KPFM, and observe their evolution in time. The citation analysis identified a total of 51 clusters (first 25 shown) of common terms. It is evident that the KPFM technique services a need for measurements in areas of importance including "organic" and inorganic (e.g. "Gallium nitride", "GaN heterostructures") semiconductor research, investigation of oxide materials (e.g. "$NiO_2$"), "alloys", as well as in chemical (e.g. "self-assembled monolayers") and biological (e.g. "pulmonary surfactant") systems. In general, however, the large number of individual clusters indicates that the field of KPFM is extremely diverse, lacks a single common thread, likely as a result of the broad applications in which the measurement has been applied too.

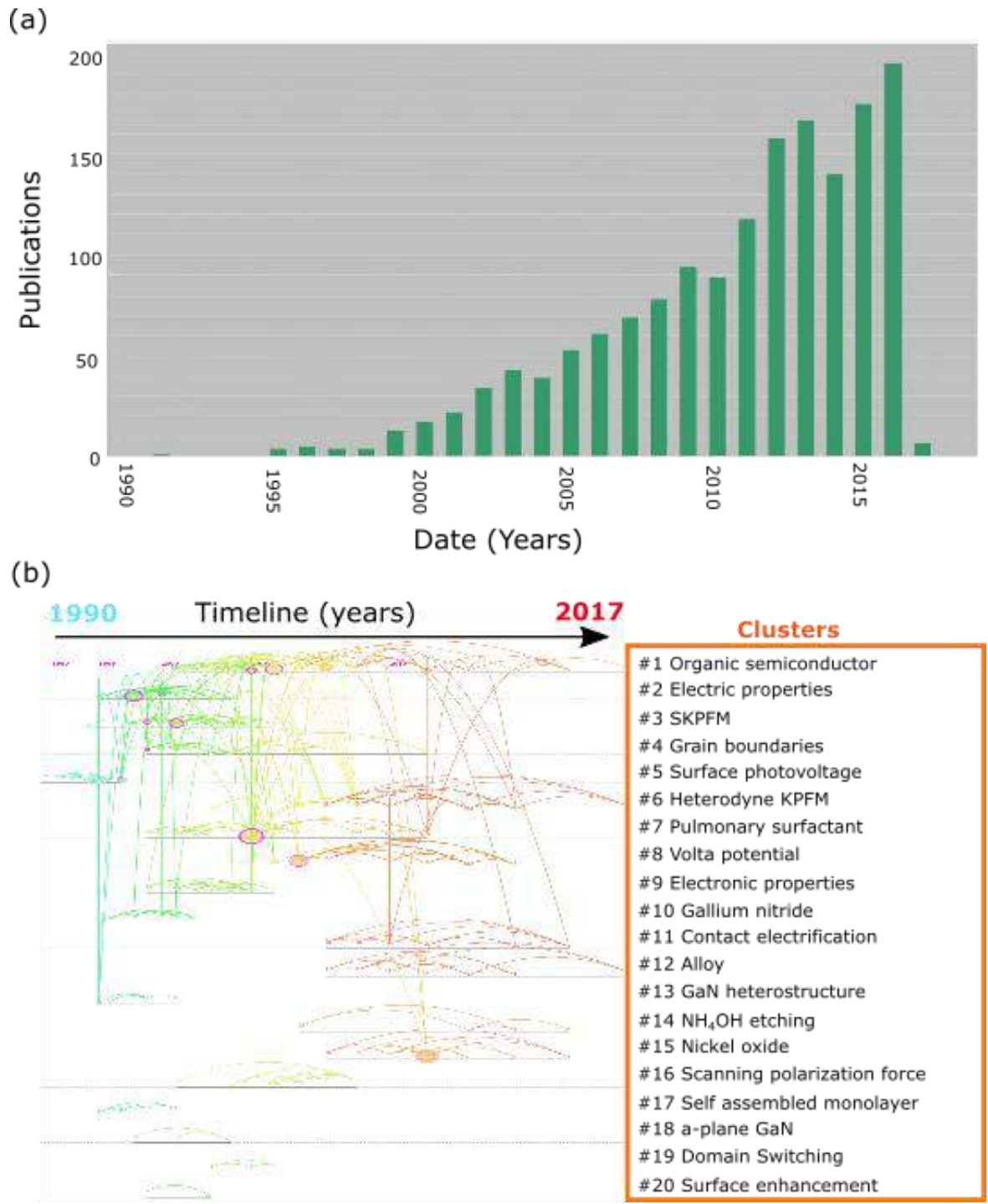

Figure 9. Graphical representation of the success of KPFM evidenced through the (a) number of publications (c) CiteSpace analysis of the literature to date. Data was generated from a Web of science search for publication which include any of the terms "Kelvin probe force microscopy", "Kelvin force microscopy" or "Scanning potential probe microscopy".

### 3.2.2 Amplitude modulated-KPFM

The first demonstration of KPFM was realized using AM detection. A simplified schematic for a typical AM-KPFM setup is shown in Figure 10(a). During the AM-KPFM measurement the cantilever is not driven mechanically; rather, the tip is excited directly by $V_{tip}$. For detection purposes the driving frequency, $\omega$, is typically selected close to the resonant frequency ($\omega_0$) of the cantilever to ensure a strong response of the cantilever. The general principles for the quantification of the CPD in AM-KPFM rely on the fact that the first harmonic response ($F_\omega$) described by equation (12) demonstrates a linear bias dependence. Note, the amplitude at the fundamental drive is minimized when $V_{dc} = V_{CPD}$ (see Figure 3(c)). KPFM utilizes this principle by using a LIA to detect the amplitude first harmonic amplitude $A_\omega$ which is related to the force component by $A_\omega = G(F_\omega/k)$, where $G$ is the cantilever transfer function gain. Next, a feedback loop is used to adjust the constant component of the tip bias $V_{dc}$ until $A_\omega$ and hence the electrostatic force is nulled, a map of the nulling potential ($V_{dc}$) yields a map of the CPD. In fact, the input to the feedback loop is the in-phase amplitude, $X_\omega$ (or quadrature, $Y_\omega$), component rather than the amplitude as it contains information on the phase and hence polarity of the response.

Normally AM-KPFM is operated in life mode, in which the first line trace in an image is used to measure the sample topography (e.g., in either contact, intermittent, or non-contact modes) with the tip grounded. During the second pass, the measured topography is retraced but this time at a specified distance above the surface (usually $20 - 100$ nm) in a highly-controlled manner. At the same time, the topography information and short-range forces are available through the topography line scan. However, as an alternative to lift mode, AM-KPFM is sometimes performed simultaneously with surface topography acquisition in a single pass.[282, 283] Topography is captured using AM-AFM operated on the first cantilever resonance frequency, while the KPFM electrical excitation is performed at a frequency far from the cantilever resonance frequency. Using an off-resonance frequency will result in a reduction in sensitivity compared to lift mode (which benefits from resonant amplification), so often a higher eigenmode frequency is chosen for the electrical measurement. The single pass method offers several advantages over the lift-up mode, including higher spatial resolution and more localized measurement of CPD, as well as minimization of topographical errors due to electrostatic forces (i.e., compensation of electrostatic forces).[282-284] At the same time, spontaneous excitation of the higher eigenmodes can make this approach prone to indirect topographical crosstalk[285, 286] as the resonance peak can shift in response to changes in the sample topography, indirectly effecting the measured CPD.[286] Recently Li *et al.*,[287] performed a study on the influence of crosstalk in single pass KPFM. They concluded that crosstalk between topography and CPD image can be minimized if the electrical driving frequency is carefully tuned such that it is neither a multiple nor a factor of the topography driving frequency. However, even then, finite crosstalk in either the topography channel (due to electrostatic forces) or CPD channel (due to abrupt topographical changes) could be observed.[287] Regardless, these measurements demonstrate the importance of careful consideration into the effects of crosstalk and development of measurements and protocols to minimize crosstalk for reliable application of ambient single pass KPFM.

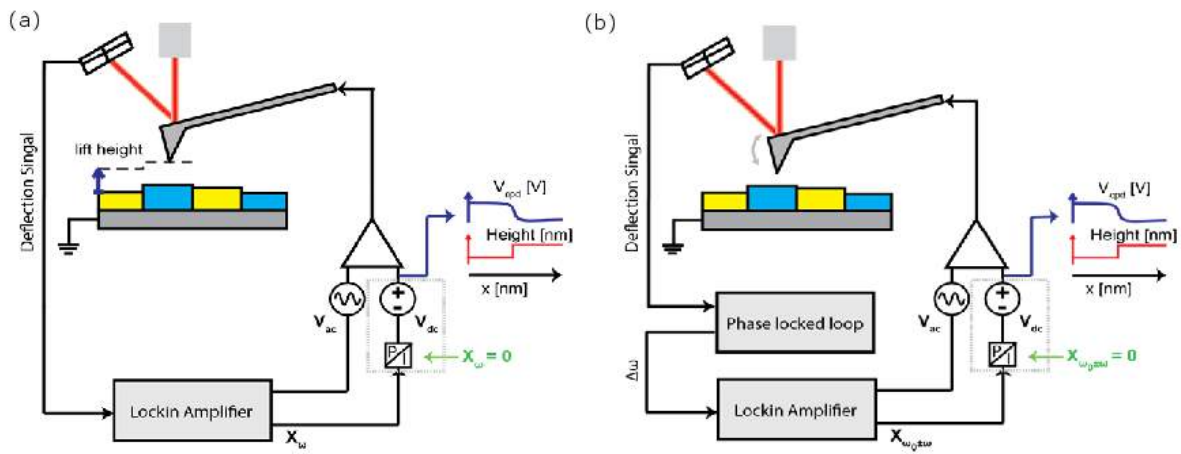

Figure 10. Simple Schematic of an (a) AM- and (b) FM-KPFM experimental setup.

### 3.2.3 Frequency modulated-KPFM

A major improvement in terms of KPFM lateral resolution was the development of frequency modulated (FM)-KPFM, first proposed by Kitamura *et al.*[288] as shown schematically in Figure 10(b). This approach combines aspects of both EFM and AM-KPFM. In FM-KPFM, the tip is mechanically actuated at its resonance frequency while a low frequency electrical excitation is applied to the tip. Figure 11 shows a schematic of the frequency spectrum of the cantilever oscillation. Besides peaks at $\omega_{mod}$ and $2\omega_{mod}$ resulting directly from the electrostatic force modulation, sidebands appear at $\omega_0 \pm \omega_{mod}$ and $\omega_0 \pm \omega_{mod}$ adjacent to the fundamental resonance peak at $\omega_0$. These sidebands originate from the modulated force gradient, $F_\omega'$ and $F_{2\omega}'$, found by substituting the derivative of the electrostatic force components described in equations $(11) - (13)$ into equation $(9)$.

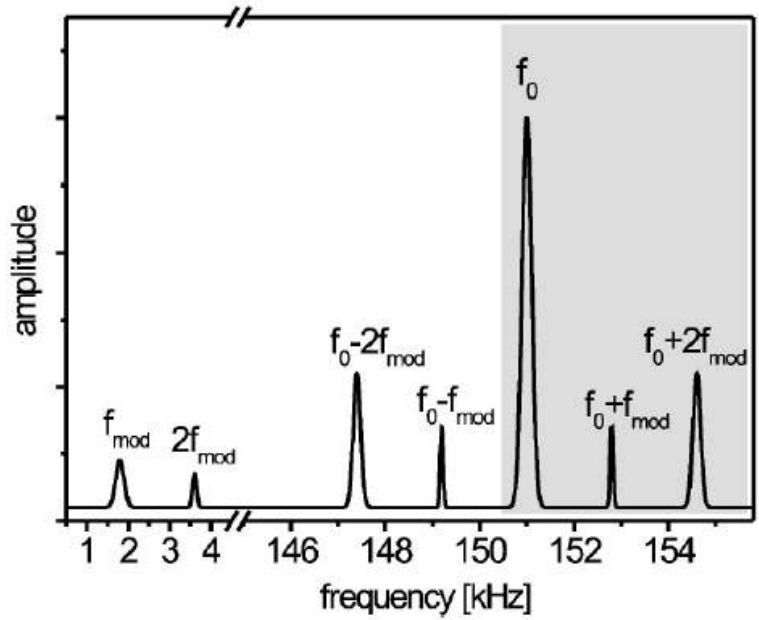

Figure 11. Schematic frequency spectrum of the tip oscillation in FM-KPFM. The peaks at $f_{mod}$ and $2f_{mod}$ originate from the electrostatic force, whereas peaks at $f_0 \pm f_{mod}$ and $f_0 \pm 2f_{mod}$ show the frequency modulation of $f_0$, produced by the oscillating electrostatic force gradient. *Seeking permissions.*

In this way, FM-KPFM, like EFM is sensitive to electrostatic force gradient, as opposed to electrostatic force. A PLL is used to determine the bias-induced frequency shift. If the frequency of the electrical excitation is within the PLL bandwidth then a LIA can be used to measure the dynamic changes in the resonance frequency. FM-KPFM nullifies the signal at $\omega_0 \pm \omega_{mod}$ by applying an appropriate $V_{dc}$. This again, like AM-KPFM, allows direct detection of CPD. In UHV, where FM is used for distance control in the attractive force regime,[126] FM-KPFM is normally operated simultaneously with the topography measurement in a single scan.[289] However, in ambient or on rough or contaminated samples stable non-contact operation is not trivial. As a result, Ziegler *et al.*[290] developed force gradient-sensitive detection in lift mode KPFM for operation of FM-KPFM in ambient.

### 3.2.4   Comparison of AM- and FM-KPFM detection

Although in principle both AM- and FM-KPFM on the same sample should provide a measure of the true work function, however this is never the case, primarily due to the fundamentally different imaging mechanisms involved. From Figure 4(b), we expect the electrostatic force gradient to be predominantly confined to the interaction between tip apex and sample whereas the electrostatic force is expected to be more sensitive to influences from the bulk cone and the cantilever shank.

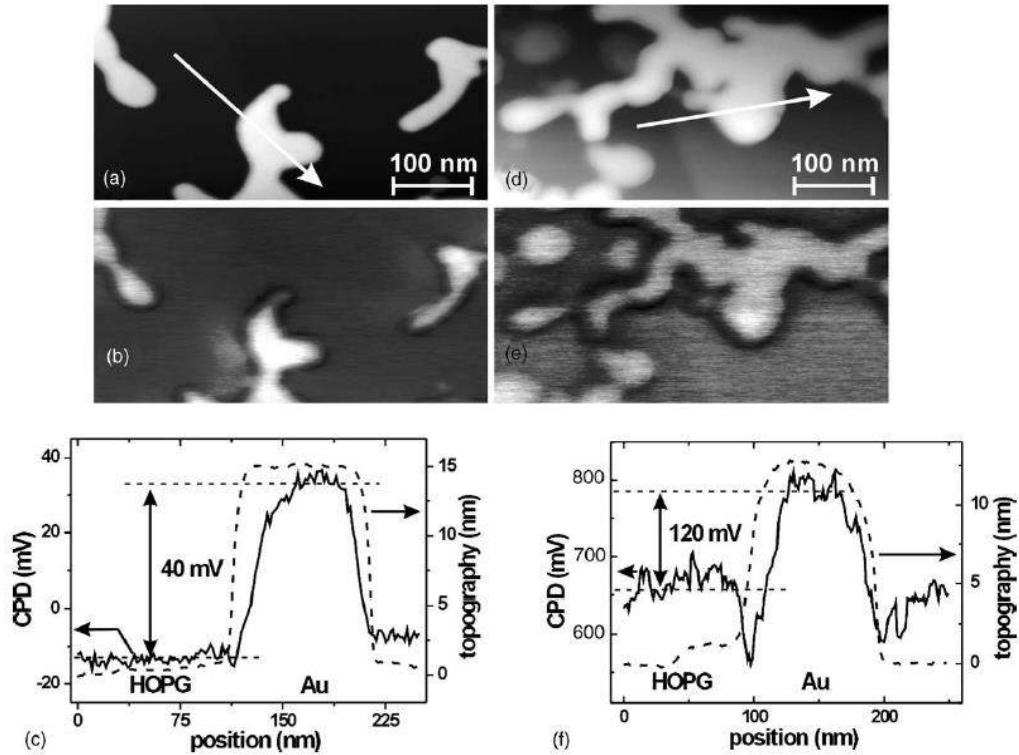

Figure 12. Comparison of AM- and FM-KPFM measurements on HOPG with Au islands. (a) Topography ($\Delta z = 17$ nm) and (b) CPD ($\Delta CPD = 50$ meV) collected using AM-KPFM with a PSI cantilever. (d) Topography ($\Delta z = 15$ nm) and (e) CPD ($\Delta CPD = 290$ meV) collected using FM-KPFM. Line scans through the images presented in (c) and (f) for AM- and FM-modes, respectively. Reprinted with permission from [291]. Copyright 2013, Elsevier.

Glatzel *et al.*[291] compared AM- and FM-KPFM measurements by performing both methods on gold islands formed on a highly oriented pyrolytic graphite (HOPG) substrate, shown in Figure 12. AM-KPFM was shown to suffer from a strong averaging effect of the cantilever itself, decreasing the resolution of the measurement compared to FM-KPFM. They concluded that FM-KPFM provides a higher lateral resolution and more accurate representation of the local CPD compared to AM-KPFM. They further performed measurements using different tips which indicated that both detection modes have a strong dependence of the measured signal on the tip geometry, with a long tip being preferably for KPFM measurements in agreement with previous simulations performed by Jacobs *et al.* [292] Importantly, Glatzel *et al.*[291] concluded that FM-KPFM suffers from lower sensitivity than AM-KPFM and therefore necessitates the application of higher ac voltages. A major drawback of this is that high ac voltages can contribute to the topography signal and result in tip induced band bending on the surface of semiconducting samples.[257, 293, 294] In terms of excitation frequency, in FM-KPFM the frequency is restricted to a narrow band of frequencies (~1-3 kHz), where the lower frequency is determined by the bandwidth of topography feedback (i.e., required to avoid crosstalk between channels), and the high frequency limited by the bandwidth of the PLL. In comparison the limiting factor in AM-KPFM mode is the bandwidth of the photodiode used for the detection of the cantilever oscillation, normally 1-10 MHz, and tip selection (e.g., eigenmode frequencies). Zerweck *et al.* [295] performed further comparisons between detection modes by performing a combined experimental and theoretical study of the accuracy and resolution of both methods. They used a 3D numerical model of a truncated cone (i.e., disk, cone, and half sphere) and located over a boundary between two regions of different surface potential, which good agreement being found between experimental and theoretical results.[295] In agreement with the study by Glatzel *et al.*,[291] and corroborated by ultraviolet photoelectron spectroscopy (UPS) measurements they found that FM-KPFM provides a more accurate measurement of CPD than AM-KPFM due to localization of the interaction. From modelling they concluded that for accurate FM-KPFM measurements of small objects, tips having a radius smaller than the minimum feature are required, even suggesting that combining FM-KPFM with very sharp tips might even allow quantitative atomic contrast to be obtained. By exploring the distance dependence in AM- and FM-KPFM, as shown in Figure 13, they found that in their setup the FM method did not vary significantly for small distances (< 30 nm), on the contrary, the CPD measured with the AM method strongly deviated as a function of tip-sample distance. In FM-KPFM, for distances larger than ~30 nm, the voltage induced frequency shift decreases and as a result the Kelvin bias feedback controller gets unstable. In comparison to previous studies,[291] they found that their setup allows the use of small modulation voltages (> 100 mV).[295]

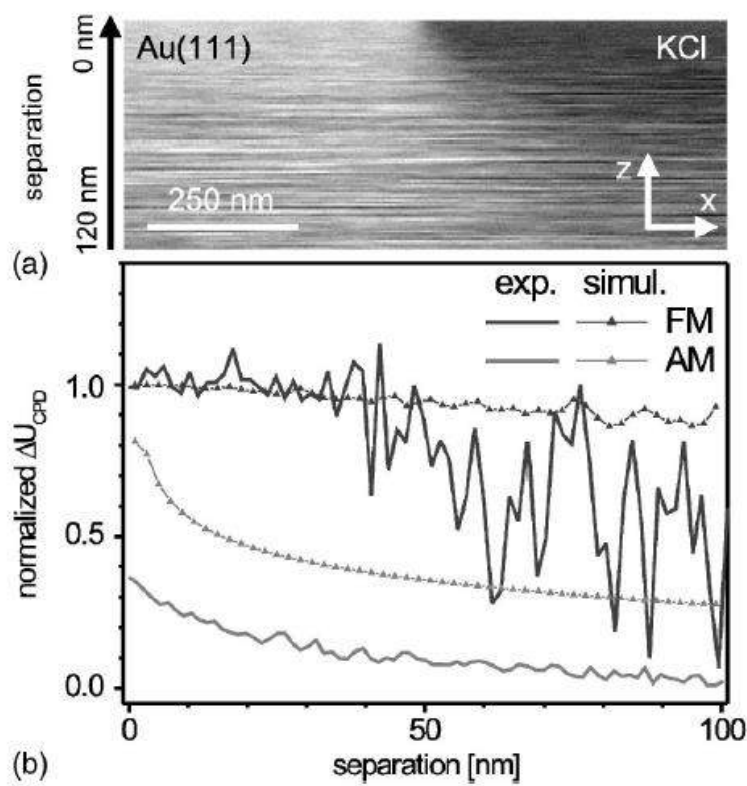

Figure 13. Distance dependence of normalized values of CPD between KCl/Au(111) and Au(111). (a) FM-KPFM image on an *xz* scan over the boundary. (b) Tip-sample separation profiles for the experimental (solid) and simulated curves (scattered) for both force and force gradient detection methods. Reprinted with permission from [295]. Copyright 2005, American Physical Society.

Moores *et al.* [238] compared the resolution of lift mode, and single pass AM- and FM-KPFM for imaging biological samples. Unsurprisingly, they found that lift mode implementation was shown to sacrifice resolution and bias sensitivity in exchange for ensuring only electrostatic interaction between tip and sample.[238] Their results comparing AM- and FM-KPFM resolution were in agreement with previous studies,[291, 295] with FM-KPFM providing a more localized measure of surface potential compared to AM-KPFM on pulmonary surfactant film, shown in Figure 14. They reported that FM-KPFM, unlike AM-KPFM, nm-scale variations in surface potential, corresponding to excess cholesterol, to be spatially resolved.[238]

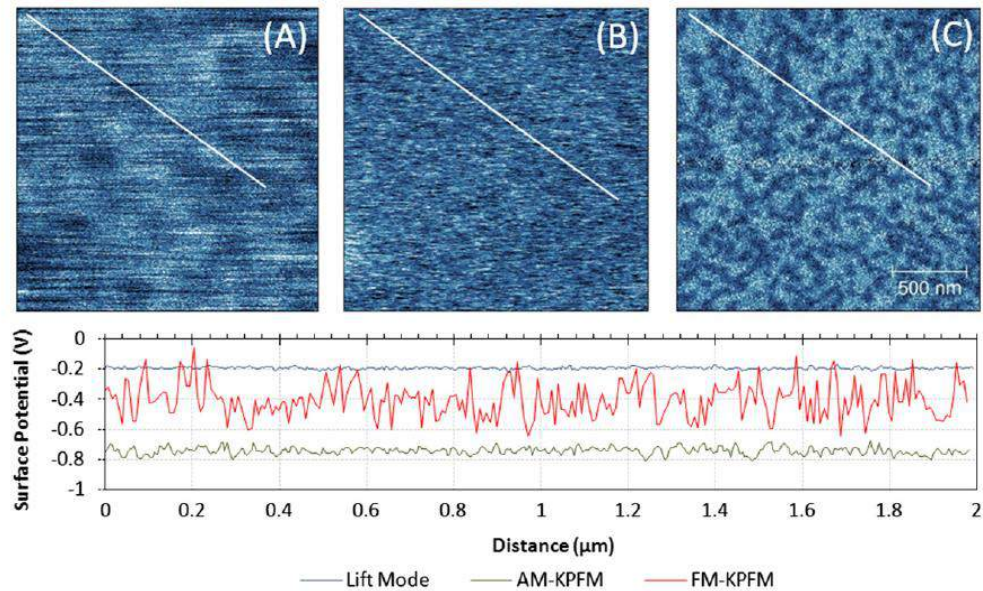

Figure 14. 2 x 2 µm KPFM images of bovine lipid extract surfactant with 20% cholesterol in (a) lift mode, (b) AM-KPFM and (c) FM-KPFM. Reprinted with permission from [238]. Copyright 2010, Elsevier.

## 3.3    Artefacts and limitations of KPFM

An implicit assumption of KPFM is that if the oscillation of the cantilever at the excitation frequency, $\omega$, is minimized, the applied bias, $V_{dc}$, effectively compensates the electrostatic force (*i.e.*, $F_\omega = 0$) and, therefore, equals the local CPD ($V_{dc} = V_{CPD}$). Under the condition of $F_\omega = 0$, the measured CPD can then be determined independently from measurement parameters, equation (12).[286] Although quantitative measurements should be the goal of any KPFM measurement, in practice KPFM measurements of CPD are known to be strongly dependent on the measurement environment,[296] tip geometry,[292] instrumentation effects,[297] as well as chosen experimental parameters.[118] Such deviations in recorded CPD values make comparison between measurements from different groups, or with theory, extremely difficult. In practice, comparing KPFM measurements which have been performed on different instruments, on different days, or even using different tips, requires that influences from instrumentation or chosen experimental parameters is known. In this section, we provide a review of artefacts and/or crosstalks present in KPFM, as well as highlighting attempts to eliminate, or minimize, such effects.

### 3.3.1   Influence of the tip in KPFM

In the seminal work by Jacobs *et al.*,[297] they discussed the importance of proper experimental practices in KPFM for stable and accurate imaging of surface potential distribution on conducting and non-conducting samples. In particular, they emphasized the importance of tip choice and preparation for optimal KPFM measurements. Ideally in KPFM the tip acts as a passive sensor, however, this is not always that case, particularly in ambient conditions were the tip can easily be affected by contaminants and adsorbates in addition some metal tips exhibit poor stability and even breaking away during scanning.[298] As shown in Figure 15(a-f), damage to metal-coated tips can greatly reduce the contrast as well as introducing significant deviations and noise into the KPFM measurement compared to similar non-coated highly doped silicon probes.[297] For this reason, significant efforts in improving tip coating design and fabrication such as using gold particle-coated silicon probes,[299] Ga needle probes,[269] carbon nanotube probes,[300-303] or focused ion beam micro-machined probes,[304-307] among others (solid Pt wire probes, diamond probes, etc.).

In addition to tip wear, tip (and sample) cleanliness is an extremely important consideration in KPFM which is very sensitive to surface chemistry. In Figure 15(g,h) we present unpublished results of changes in the work function of tip (Multi 75-G Pt/Ir coating, Budget sensors), measured in ambient as a function of time. The work function of a Pt/Ir-coated tip was measured against freshly cleaved HOPG with a known work function ($\Phi_{HOPG}$= $(4.475 \pm 0.005)$ eV[308]). The $\Phi_{tip}$ was calibrated by measuring the CPD difference between the probe and the freshly cleaved HOPG, where $\Phi_{tip} = e\text{CPD}_{HOPG} + \Phi_{HOPG}$. After an initial measurement of 25 mins, a cleaning procedure involving chemical and ozone cleaning was performed prior to subsequent KPFM measurements. The chemical cleaning involved gently rinsing the tips in ethanol, isopropanol and milli-Q water, then drying under nitrogen flow. This was followed by treatment in an UV ozone treated (BioForce UV/Ozone ProCleaner) for 30 min. After each cleaning treatment the tip was loaded into the AFM and engaged onto the surface of a freshly cleaved HOPG. The CPD between the Pt/Ir-coated tip and HOPG was recorded at a single location (i.e., without scanning) for 25 mins. A large jump in the probe work function (> 700 meV) is observed after the first cleaning procedure, likely a direct result of removal of any organic layers from the tip surface. Indeed, AFM probes can be contaminated by thin layers of silicone oil transferred from the storage gel pack, commonly made from poly(dimethylsiloxane) (PDMS).[309] After cleaning, a gradual change in the probe work function could be observed in most instances, likely a result of a slow and gradual stabilization of the tip surface chemistry. After allowing the tip to stabilize (in the AFM) for 24 hours (Figure 15(h)) the probe work function had decreased by greater than 150 meV. Figure 15(g) and (h) demonstrate a major problem with performing absolute work function measurements in ambient conditions, as well as the importance of tip cleanliness and equilibration of tip chemistry before performing KPFM measurements in ambient.

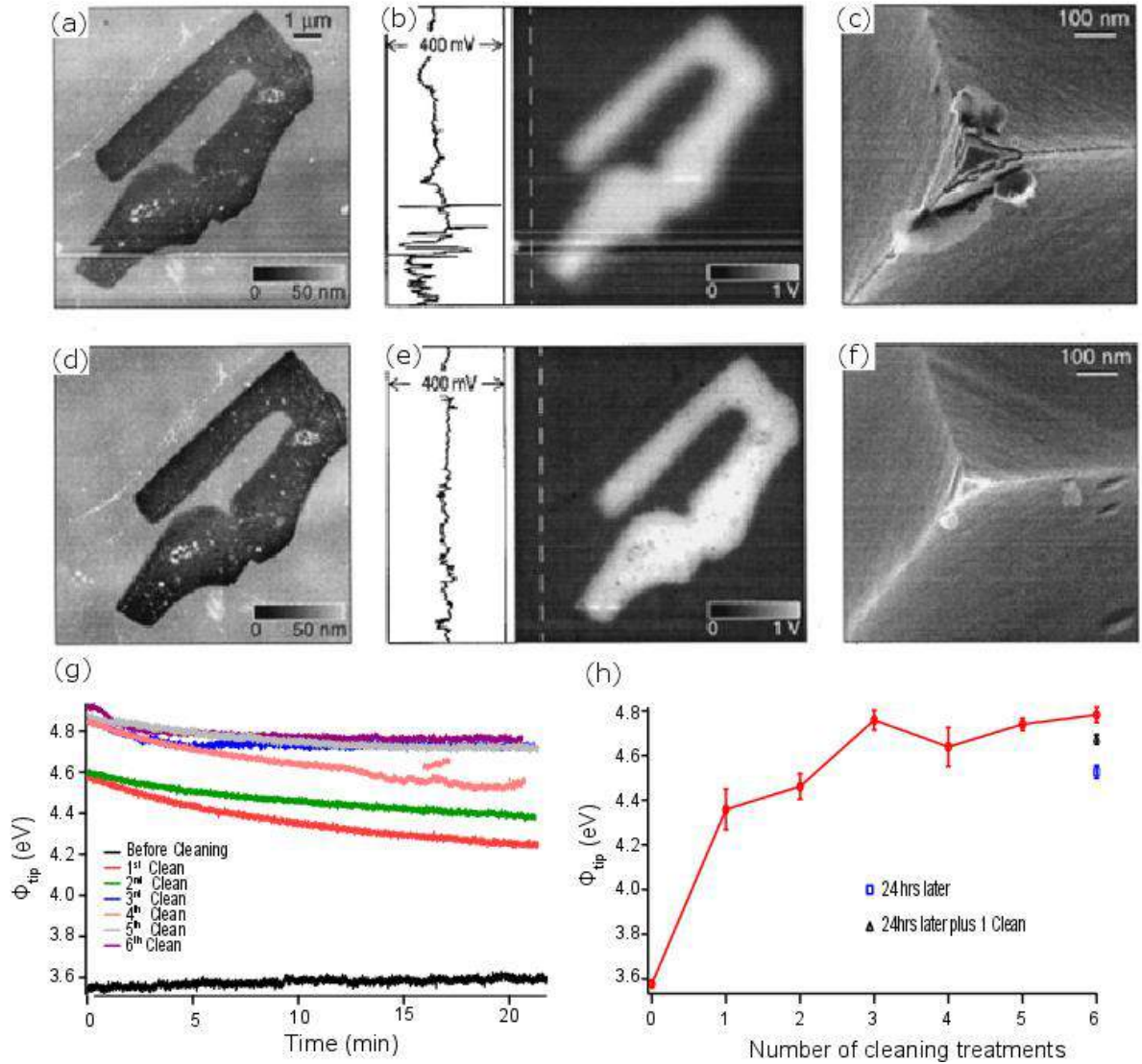

Figure 15. Influence on tip condition and cleanliness of measured work function. (a) The temporal dependence of the work function after consecutive tip cleaning procedures. Reprinted with permission from [291, 297]. Copyright 1999, AIP Publishing LLC. (b) Average and standard deviation (error bars) of the measured tip work function with each tip cleaning procedure.

### 3.3.2 Instrumentation effects

The macroscopic KP community has known as early 1991 that the traditional implementation of the KP method, using a LIA and bias feedback, suffers from two significant sources of systematic error. First the spacing dependence of the feedback gain involved in a LIA technique,[310] and secondly the spacing dependence introduced by stray capacitances, or "micro phonics".[311] Any KP measurements which do not account for and eliminate these effects, may contain essentially unpredictable systematic errors of several hundreds of millivolts and was lead to the development of off-null KP approaches. Almost a decade later similar realizations we made in the field of KPFM and have been comprehensively studied in several papers[118, 143, 286, 312, 313] and book chapters.[314] Hence, the influence of the LIA and bias feedback, as well as their operating parameters merits careful consideration here.

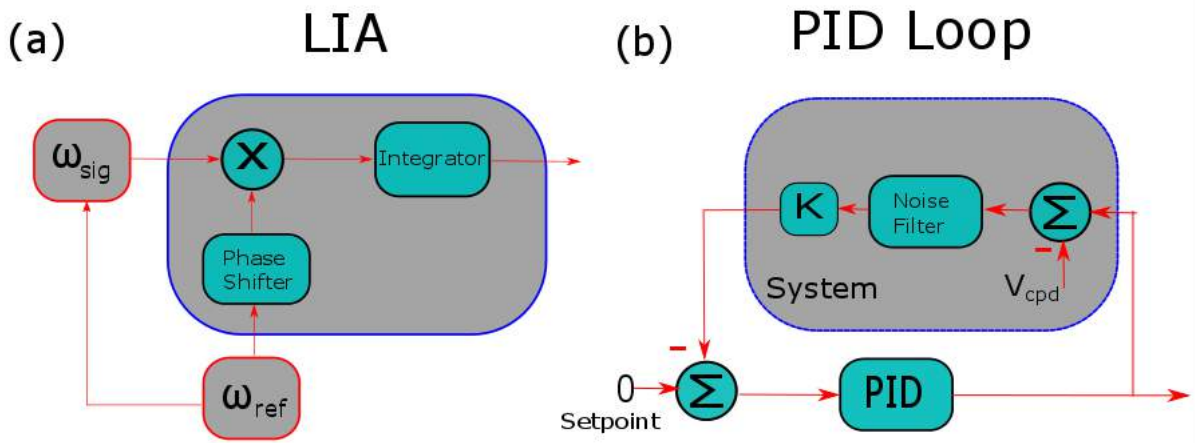

Figure 16. Block Diagram of a (a) LIA and (b) Kelvin controller based on a proportional-integral-differential (PID) controller.

### 3.3.2.1 Lock-in effect

In the SPM field, demodulation techniques have been the excepted approaches to date for extraction of response and attenuation of noise. It can be postulated that this is a remnant of the fact the in the early days of SPM signal to noise was a much bigger problem than today (i.e., low noise AFM and AFM cantilevers are commercially available). This is an important point, as the introduction of heterodyne detection in the KPFM setup has some important benefits, as well as imposing significant limitations. It is useful then to consider for a moment the working principles, as well as imposed limitations, involving the LIA. A LIA is a phase-sensitive detector, which can efficiently extract a signal, with known frequency, from an extremely noisy background. A simple schematic of a LIA working principle is shown in Figure 16(a). The LIA works by mixing the input signal ($\omega_{sig}$) (e.g., photodetector deflection signal) with a reference signal ($\omega_{ref}$) matching that supplied to the tip as an electrical excitation ($V_{ac} \sin(wt)$). For a sine reference signal and an input waveform, $U_{in}(s)$, the dc output signal, $U_{out}(t)$, can be calculated using Equation (22).

$$U_{out}(t) = \frac{1}{T} \int_{t-T}^{t} \sin(2\pi f_{ref} + \varphi) U_{in}(s) ds \qquad (22)$$

Here, $\varphi$ is the phase which can be set on the LIA. If $\omega_{ref} = \omega_{sig}$, the output of the mixer is a dc voltage with a magnitude proportional to the amplitudes of $\omega_{ref}$ and $\omega_{sig}$. If $\omega_{ref} \neq \omega_{sig}$, the output signals of the mixer have frequencies $\omega_{ref} - \omega_{sig}$ and $\omega_{ref} + \omega_{sig}$. This integration procedure suppresses the new frequencies and the output is zero, and hence any and all other information contained in these frequency bins or bands are essentially attenuated and unrecoverable. Most modern LIA are two-phase detectors, where calculation shown by Equation (22) is performed a second time but with an additional 90° phase offset to capture both the in-phase ($X$) and out-of-phase ($Y$) response and can be further converted to the amplitude/phase channels.

In a KPFM measurement, the user defines the amplitude, frequency and phase of the demodulator reference signal which is supplied to the tip as a voltage. In KPFM, to allow both positive and negative CPD to be measured, the input (error signal) to the Kelvin feedback loop needs to be the in-phase amplitude, $X_\omega$, (or quadrature, $Y_\omega$) component (i.e., having a bipolar bias response) of the first harmonic response, with the reference phase adjusted to maximize $X_\omega$ (or $Y_\omega$). This consideration is important as $A_\omega = \sqrt{X_\omega^{\,2} + Y_\omega^{\,2}}$, and thus it can never have a negative response. Correspondingly, the phase is given by $\theta_\omega = -\tan^{-1}(Y_\omega / X_\omega)$). Jacobs *et al.*[297] clearly demonstrated the critical importance of optimization of the LIA phase to achieve proper KPFM sensitivity as demonstrated in Figure 17. In this procedure effect of phase on both open and CL-KPFM operation as the surface potential of an electrode is modulated using a square wave of 50 mV. By monitoring the open loop (OL) output (e.g., EFM output from the LIA) the effect of phase on the magnitude of the amplitude signal is clear. The phase should be optimized in this case to ensure maximum sensitivity to the applied surface potential. The effect of phase on the measured CPD in KPFM is more obvious when the loop is closed. Incorrect phase settings have the effect of reducing the sensitivity of the KPFM measurement, or precluding KPFM operation entirely causing the feedback loop to fail.[297]

Another important consideration in KPFM is the chosen amplitude of reference signal, and hence the applied ac voltage. During an ideal KPFM measurement the amplitude should be chosen to provide a measurable amplitude response, without result in tip-induced modification of the sample electronic properties, particularly important for semiconducting samples where large voltages can induce band bending.[294] However, more worryingly, Kalinin *et al.*,[118] and others,[312, 315, 316] have demonstrated that the measured CPD in AM-KPFM is extremely sensitive to the amplitude of the applied ac voltage. The ac dependence could be explained in the context of non-ideal bias feedback compensation of the electrostatic force.[315] This point is discussed in greater detail in the following section.

Clearly, having correct settings of the LIA reference signal is important for sensitive and stable potential measurements in KPFM. In a more subtle way, noise filter settings of the LIA are of critical importance for defining the sensitivity and noise in the KPFM measurement. The LIA integration time (normally $100~\mu s - 4$ ms) is primarily controlled by the time constant of the LIA low pass filter, where a long time constant will remove more noise from the signal; at the price of reducing the response time of the overall measurement. At the same time, the time constant defines the output (i.e., sampling) rate of the amplitude and phase of the response at the reference frequency and hence is deterministic of the measurement speed.

In summary, although heterodyne detection allows detection of weak electrostatic actuation of the cantilever from the noisy photodetector signal, it does so at the cost of; (a) attenuating all fluctuations in the cantilever dynamics at frequencies other that the fundamental driving frequency, (b) ignores non-sinusoidal responses, and (c) requires long integration times $(0.1 - 4$ ms). Collectively heterodyne detection in KPFM limits the possibility of detecting other channels of information (e.g., harmonics) without additional LIAs, restricts the bandwidth of the KPFM measurement, and renders the approach unsuitable for measuring fast electronic, or electrochemical processes. (e.g., $< 0.1$ ms).

### 3.3.2.2   The feedback effect

As is the case for the LIA, correctly optimized settings of the bias feedback loop is critically important for true measurements using KPFM.[297] In an ideal PID controller, the system minimizes the difference between an input signal (in this case the cantilever amplitude) and a setpoint (in this case 0 V as CL-KPFM is a nulling technique), by varying an output parameter (the dc bias applied to the tip). In particular, for KPFM, the dc bias is equal to the CPD between the tip and the sample when the feedback loop is performing optimally (input signal = setpoint), and where the input signal arises purely due to electrostatic contributions (i.e., does not contain non-electrostatic or parasitic signals). Although it is rarely discussed in experimental methods section of papers, incorrect gains can easily result in deviations of surface potentials corresponding to 100s mV meriting consideration of using automated gain optimization[317] or adoption of more appropriate feedback methods   e.g. based on using the Kalman filtering [156]). Alternatively, by following the procedure by Jacobs *et al.*, KPFM phase offset and feedback loop gains can be optimized by modulating the surface potential of a standard electrode (e.g., HOPG) using a slow varying square wave (~0.1-10 Hz) supplied by a function while performing KPFM (see Figure 17(a)). Optimizing LIA time constant, phase, and feedback gain parameters (e.g., integral and proportional gains in a PIDS loop) can be performed until the correct modulation voltage is measured using KPFM, without excessive noise introduction.

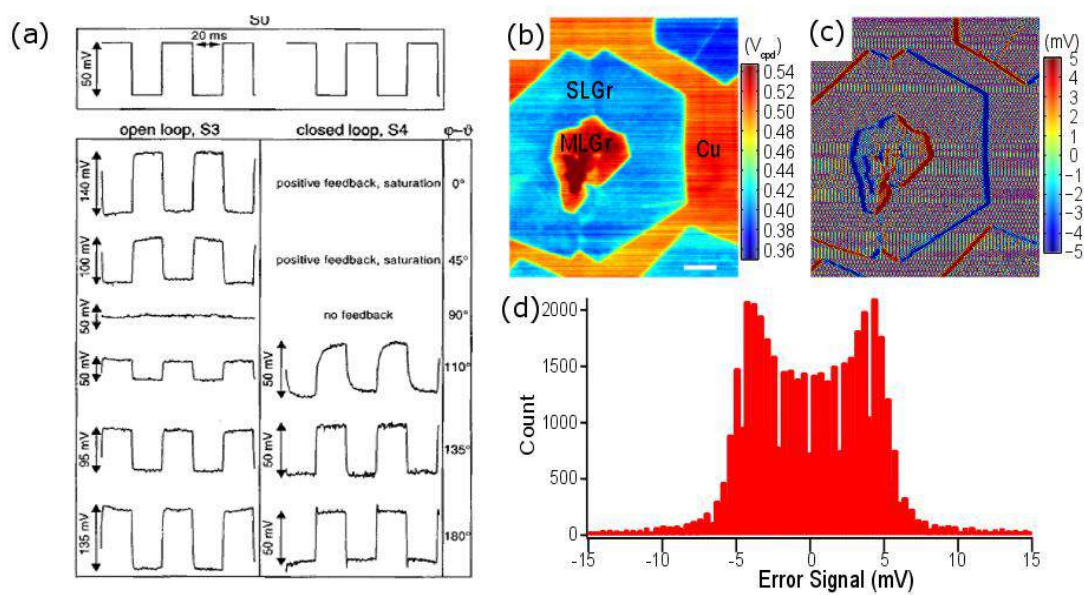

Figure 17. (a) Measured influence of the reference phase θ on the open-loop lock-in output signal *S3* and the closed-loop dc tip potential *S4*. Reprinted with permission from *[297]*. Copyright 1999, AIP Publishing LLC. (b) KPFM image of single and multilayer graphene on $CuO_2$ indicating copper substrate (Cu), single layer (SLGr) and multilayer graphene sheets (MLGr) (scale bar = 5 μm) (c) Image of the remnant amplitude response (i.e., error signal) from the KPFM bias feedback loop. (d) Histogram plot of the error signal in (c). Measurements were performed with $V_{ac}$ = 1 V [66 kHz] applied to the probe at a height of 50 nm above the surface.

However, it is always difficult to have precise feedback operation; in fact, the best a nulling technique can do is minimize the response to the noise in the system. Hence, under certain circumstances there is benefit for performing EFS measurements in determining the CPD. As an example, Figure 17 shows a CL-KPFM image of graphene grown on a Cu substrate. The CPD image shows clear contrast between the single and multilayer graphene, however the error signal shows a finite magnitude in the amplitude, particularly at boundaries of interfaces which can be due to sudden changes in the electrostatic force, or even topographical influences.

Apart from incorrect gain settings, several other parasitic influences can result in incorrect KPFM operation. Figure 18 shows a schematic of an ideal EFS measurement along with an experimental data for an EFS measurement collected 50 nm above a freshly cleaved HOPG surface, using $V_{ac} = 1$ V at a frequency of 12.5 kHz (far from the resonance frequency, $\omega_0 = 65$ kHz). It is clear that the minimum of $A_\omega$ is non-zero and has some finite magnitude, $\delta$. The graph shows that the bias at which the minimum of $A_\omega$ occurs differs from the zero crossing of $X_\omega$ (indicated by a red dashed line). Under CL operation, the feedback loop adjusts the tip bias in order to nullify $X_\omega$, thus defining the CPD as the $X_\omega$ zero crossing rather than the minimum of $A_\omega$ (where $V_{dc} \neq V_{CPD}$). Under conditions where there is an offset between the minimum of $A_\omega$ and the zero crossing of $X_\omega$, this leads to a non-quantitative absolute CPD (an example of the feedback effect).[286] In practice, electronic offsets of the instrumentation used, parasitic signals such as capacitive coupling between ac excitation voltage and deflection output signal,[314] along with experimental limitations, such as thermo-mechanical and electrical noise, all prohibit the recorded amplitude from converging to zero (when $V_{dc} = V_{CPD}$). The result of which is that absolute (i.e., the real values of the system being measured) CPD measurements in CL-KPFM are subject to the feedback effect and can within an instrument-dependent range up to a few volt. This deviation of $V_{dc}$ from the actual $V_{CPD}$ must be taken into consideration for CL-KPFM measurements, as given by equation 3.3[118, 286]:

$$V_{dc} = V_{CPD} + \frac{\delta}{A_{max}(V_{ac}C_z^{'})} \tag{23}$$

where $A_{max} = G(\omega, \omega_0)$ is the transfer function of the cantilever. The second term in equation (23) introduces a dependence on $C_z^{'}$ and $V_{ac,}$ when $\delta$ has some finite magnitude (i.e., in the presence of a feedback effect).

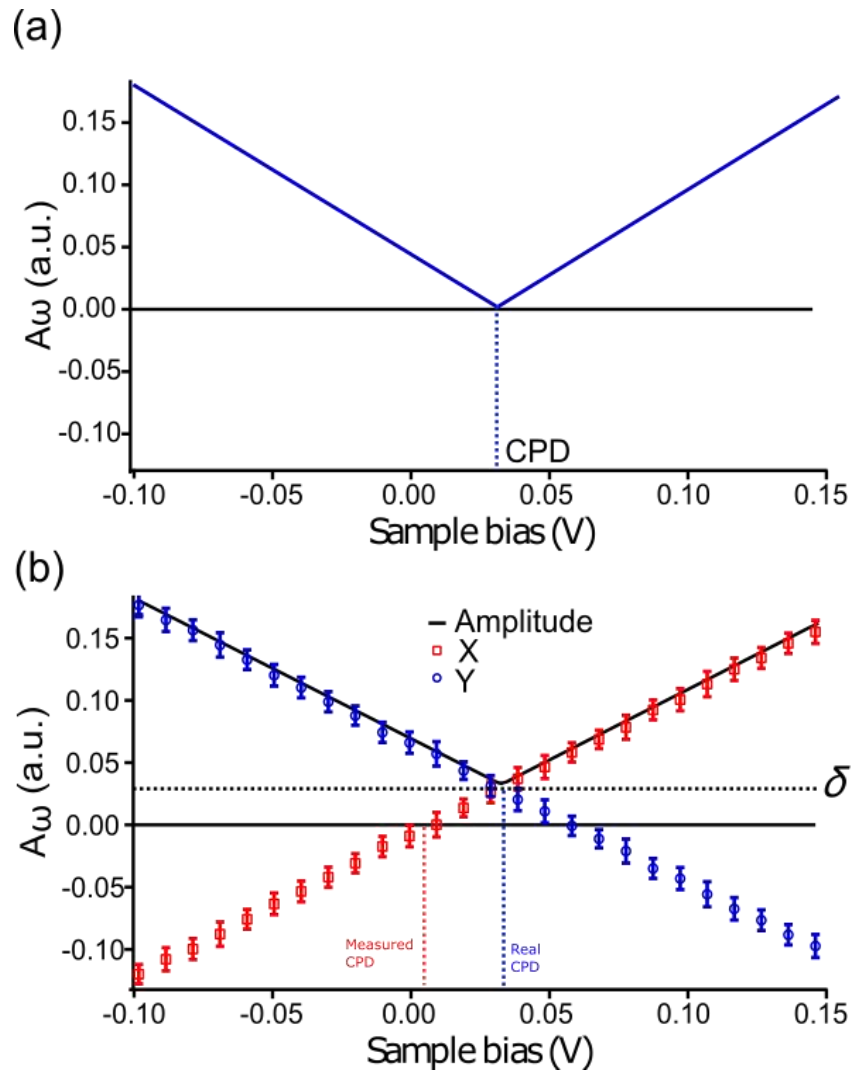

Figure 18. Open-loop bias spectroscopy and the feedback effect. An OLBS measurement showing the electrostatically excited cantilever oscillation amplitude as a function of dc bias applied to the sample. (a) Schematic of the ideal behavior, i.e., zero oscillation when the dc bias corresponds to the CPD. (b) Experimentally measured behavior. (note the phase was adjusted by 90° between collection of $X_\omega$ and $Y_\omega$).

Indeed, it is known that CPD values measured by KPFM are greatly affected by LIA parameters[297] including the frequency,[318, 319] amplitude[118] and phase[297] reference of the excitation voltage.[312, 315] Kalinin *et al.* demonstrated that the feedback effect must be taken into account in describing the distance dependence, and ac amplitude dependence of the KPFM measurement.[118] We demonstrate this by performing EFS measurements for different values of $V_{ac}$, as shown in Figure 19. For visualization purposes we are showing the linear fits of the sample bias dependence of $X_\omega$, as a function of varying ac tip voltage. Again, the zero amplitude crossing of $X_\omega$ is used to determine the CPD recorded in a CL-KPFM measurement and the slope of the bias dependence provides a measure of $C_z'$. Figure 19(b) demonstrates that the feedback effect manifests itself in the form of a $1/V_{ac}$ dependence of the measured CPD. In our setup it was possible to minimize the feedback effect by performing a stringent electronic calibration of each signal channel, as well as adjusting the setpoint of the feedback loop to adjust for parasitic signals on the input channel of the feedback loop. Figure 19(c) demonstrates that, following the removal of all parasitic signals, it was possible to minimize the $V_{ac}$ dependence on the CPD becoming independent for voltages between 500 mV and 8 V. To highlight the range of CPD offsets which could be present in a typical CL-KPFM system, we artificially induce offsets of ±100 mV, not uncommon when multiple digital-to-analogue (DAC) and analogue-to-digital convertors (ADC) are connected in series. In this case, the feedback effect results in CPD variations of approximately 1V in magnitude, highlighting the challenges in obtaining absolute measurements of work function or surface potential using CL-KPFM, and making comparison between experimental results on different instruments or with theory, particularly problematic. Figure 19(d) shows the distance dependence of the CPD for an uncorrected CL-KPFM measurement on a HOPG sample, having homogenous CPD ((190 ± 25) mV over a 10 µm region). The CPD at a single point can be seen to vary significantly with distance, closely following the distance dependence in the second term of equation (23) due to changes in the $C'_z$. The observed distance dependence (on this homogenous sample) is supressed after the corrections have been applied to the CL-KPFM, demonstrating that the feedback effect introduces a tip-sample distance dependence on the recorded CPD. Furthermore, this distance dependence in CL-KPFM prevents the implementation of 3D-KPFM approaches,[313] at least using traditional KPFM methods.[320]

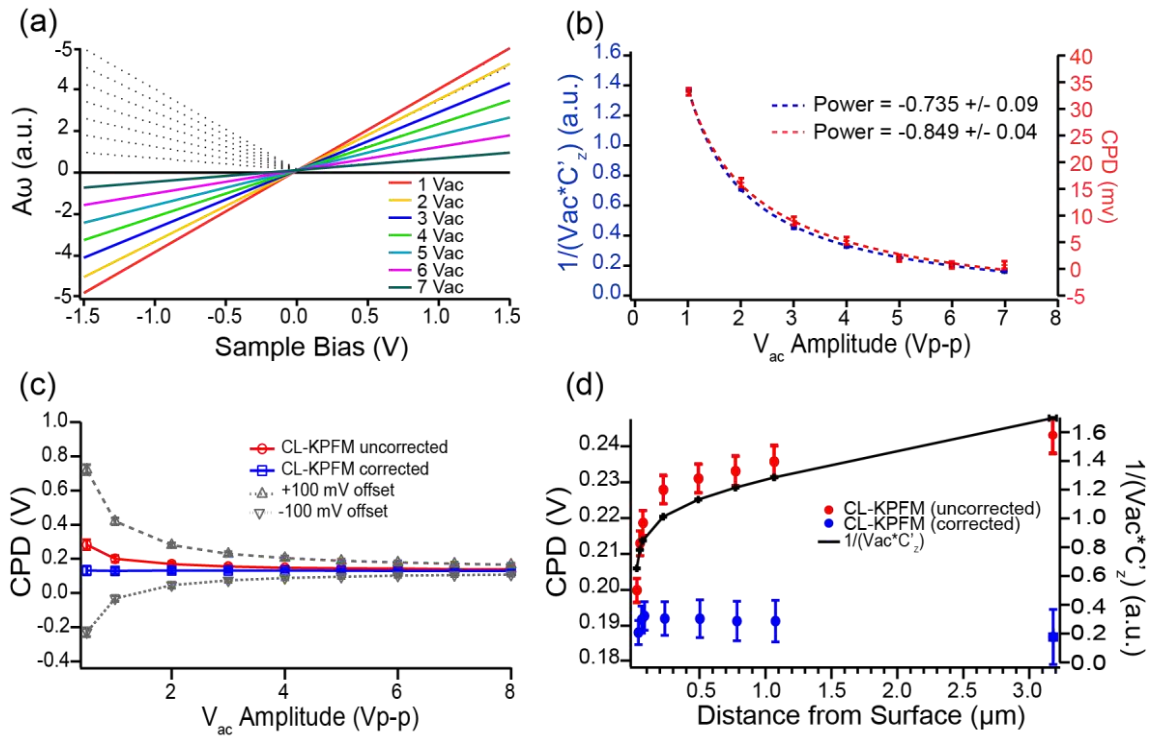

Figure 19. Identifying and correcting the feedback effect in KPFM. (a) Linear fits to $X_\omega$ collected as a function of sample bias for different $V_{ac}$. (b) Corresponding zero point crossing determined from (a) as a function of $V_{ac}$ and remnant capacitance term measured simultaneously. (c) Elimination of the feedback effect using electronic calibrations and setpoint correction procedures (appendix A.3 and A.4) (d) CPD measurements as a function of distance for uncorrected (red circles) and fully corrected (blue circles) KPFM. Black line shows the remnant capacitance term plotted as a function of distance collected using a $V_{ac}$ of 2 V at 12.5 kHz.

Figure 19 demonstrates that the feedback effect can emancipate itself in a very instrument specific way. The group of Melin *et al.*,[312, 314, 321-323] reported ac crosstalk between the electrostatic excitation signal and the photodetector deflection signal, influencing the output of the KPFM feedback loop. They demonstrated that this parasitic contribution can result in CPD measurements, which vary as a function of excitation frequency and phase φ used in the KPFM feedback loop as well as tip-substrate distance. They overcame this effect by using an active crosstalk compensation method to 'eliminate' the feedback effect in their UHV system.[312, 314, 321, 322] Other studies concluded that with sufficient shielding of cables and small modifications to commercial setups, ac crosstalk could be sufficiently reduced.[324, 325]

Furthermore, non-ideal feedback, and hence remnant capacitive forces represent the most obvious source of crosstalk in KPFM, meaning that CPD contrast can be due to variations in surface topography through the tip-sample capacitance gradient, as analyzed by several authors.[118, 326-328] At the same time as Kalinin *et al.*[118] outlined the influence of the feedback effect, Efimov and Cohen[326] investigated artefacts present in KPFM images, which appear as features in the KPFM image due only to tip/surface geometry and not to true surface potential variations which are correlated to with the capacitance gradient, and a result of non-ideal feedback. Shortly afterwards, Okamoto *et al.* [327] demonstrated, experimentally and theoretically, that the feedback effect can result in atomic scale crosstalk between van der Waals forces and electrostatic forces (i.e., topographical crosstalk). Figure 20 shows the topography, KPFM and capacitance gradient images of a Si(111)-7×7 surface. The 7×7 structure was successfully imaged with atomic resolution in all images. Worryingly for quantitative KPFM measurements of Si-(7×7), the topography and capacitance gradient channels show reverse contrast to the CPD channel.[329] This reverse contrast had previously been demonstrated [288] as a successful atomic resolution KPFM image of Si-(7×7), however Okatomo *et al.[329]* realized that this was an artifact which needed to be taken into account. Later, Lee *et al.*[328] modified the KPFM controller to feedback on the ratio of the first and second harmonic force (as opposed to the first harmonic alone) to suppress this geometrical artefact.

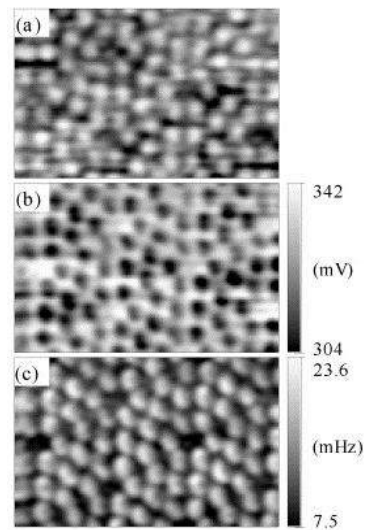

Figure 20. Simultaneously obtained images of (a) topography, (b) CPD and (c) 2$\omega$-signal for Si(1 1 1) 7×7 surface. The adatoms in the arrangement of 7×7 reconstructed structure can be seen. (a) and (c) have the similar contrast, while (b) shows the reverse contrast. The scan area is 102×66 Å. $V_{ac} = 1\ V_{rms}$ and $\omega/2\pi$=1 kHz. Reprinted with permission from [329]. Copyright 1999, AIP Publishing LLC.

### 3.3.3   Influence of environment

Operation across all environments (e.g., UHV, ambient and liquid) is one of the major attributes of AFM. Environments can be chosen in such a way to eliminate undesirable force interactions (e.g., absence of capillary forces in liquid) or even tune interaction distances (e.g., EDL screening). For example, several authors have shown the effect of water shielding on KPFM, [330, 331] Furthermore, they can be chosen to take advantage of particular cantilever dynamics (e.g., high $Q$ environment of UHV) or to precisely control surface chemistry (e.g., humidity control).

#### 3.3.3.1   KPFM in UHV

Kitamura and Iwatsuki [288] were the first to implement KPFM in UHV, which was later improved on by several groups.[251, 332, 333] In UHV, KPFM allows absolute work function with a high degree of accuracy, providing the work function of the metal tip is known. If the work function of the probe is not known, it can be determined from a calibration sample (e.g., HOPG [308]), then the CPD can be used to provide a measure of the work function of the sample under test. In UHV, the sensitivity is increased by the $Q$ of the resonance, which under UHV conditions can reach values up to $10^5$. This makes it possible to excite the oscillation with a small voltage and produce appreciable larger responses than would be achieved in air. However, detection of the mechanical oscillation at the resonance frequency suffers from a long settling time, $\tau = Q/2\omega$, which drastically reduces the permissible scan speed as well as the CPD output rate.

In terms of spatial resolution, UHV-KPFM has been demonstrated to regularly provide measurements down to the nanometer scale.[295] In fact several authors have even reported sub-molecular or atomic resolution CPD, referred to as local CPD based on Wandelt's concept of short-ranged atomic-scale variation of work function on surfaces.[334] Atomically resolved CPD measurements have been realized on a variety of semiconductors, insulators, and ionic substrates including Si,[255, 288, 327, 335] TiO2 [336], and InSb.[337] However that the origin of the contrast on local CPD still attracts significant controversy, in particular the underlying physical mechanism that connects the magnitude of the Local CPD (LCPD) and the surface potential still remains under debate.

Kitamura *et al.*[288] were the first to demonstrate atomic scale variations in CPD on Si (7x7) surface using FM-KPFM which was later reproduced by several groups. [338-340] However, a clear understanding of the derived values or the imaging mechanism is not clearly understood, and several inconsistencies between the measured CPD and the values derived from other experimental techniques or from theoretical calculations have been identified.[157] Okamoto *et al.* [327] demonstrated the possibility that the atomic scale contrast could be caused by artifacts as a result of influences from parasitic crosstalk with the capacitance gradient, whereas Arai *et al.*[341] demonstrated that contrast in NC-AFM on Si (7x7) was influenced by short range attractive force. Sadewasser *et al.* [340] combined KPFM imaging, bias- and force-distance spectroscopy imaging with first principles simulations to investigate the LCPD contrast on semiconducting surfaces. Using this combination they demonstrated that atomic scale variations in LCPD on semiconducting surfaces can arise due to variations of the surface local electronic structure due to a charge polarization induced by the tip-sample interatomic interaction.

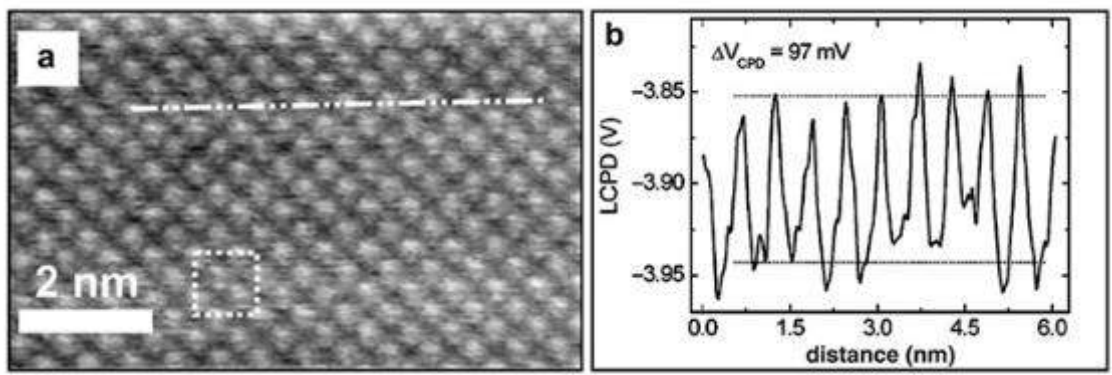

Figure 21. (a) CPD of KBr(001) surface by KPFM. (b) CPD profile along the dashed line (a). Reprinted with permission from [342]. Copyright 2008, Applied Physics Society.

Boucquet *et al.*[342] demonstrated that it was possible to obtain atomically resolved KPFM on the KBr(001) surface in UHV, as shown in Figure 21. They were able to measure variation in LCPD (~100 mV) having atomic features (0.63 nm) in very good agreement with the lattice constant of KBr, 0.66 nm. Using an analytic model they concluded the that measured LCPD on (001) ionic crystals reflected an effective surface potential whose contrast matched the atomic corrugation of the Madelung surface potential of the ionic crystal but with significant variations due to strong influences from the geometry of the tip. Later, Nony *et al.*[343] would build on this work to combine measurements with atomistic simulations the force field occurring between a realistic tip consisting of a metallic body carrying a NaCl cluster and a defect-free (001) NaCl surface. They found that the interpretation of the LCPD on the basis of the Madelung surface potential alone is insufficient and that bias dependent short-range tip bias induced polarization of the ions at the tip-sample interface can influence the measured KPFM contrast.

In recent years, there has been a focus on applying KPFM to the study of molecular systems and devices.[344] In a seminal work, Gross and co-workers[345] demonstrated using a combination of NC-AFM and bias spectroscopy (e.g., Electrostatic force gradient spectroscopy (EFGS)) that it was possible to detect the charge state of adatoms on thin insulating films. Mohn *et al.* [346] used KPFM methods to successfully map the charge distribution within a naphthalocyanine molecules on a thin insulating layer of NaCl on Cu(111), was studied using KPFM. Figure 22(b) and (c) show the LCPD images recorded before and after switching the automerization state of a single naphthalocyanine molecule, whereas d shows the difference between the two states. A clear asymmetry is visible between the H-lobs and N-lobes matching that expected from DFT calculations, Figure 22(d). The asymmetry is even more clearly visible in the difference image (Figure 22(d)) obtained by subtracting the LCPD images of the initial and switched configurations. The DFT-calculated results (Figure 22(e)) reveal that the submolecular resolution in the LCPD images reflects the total charge distribution within the molecule. [346]

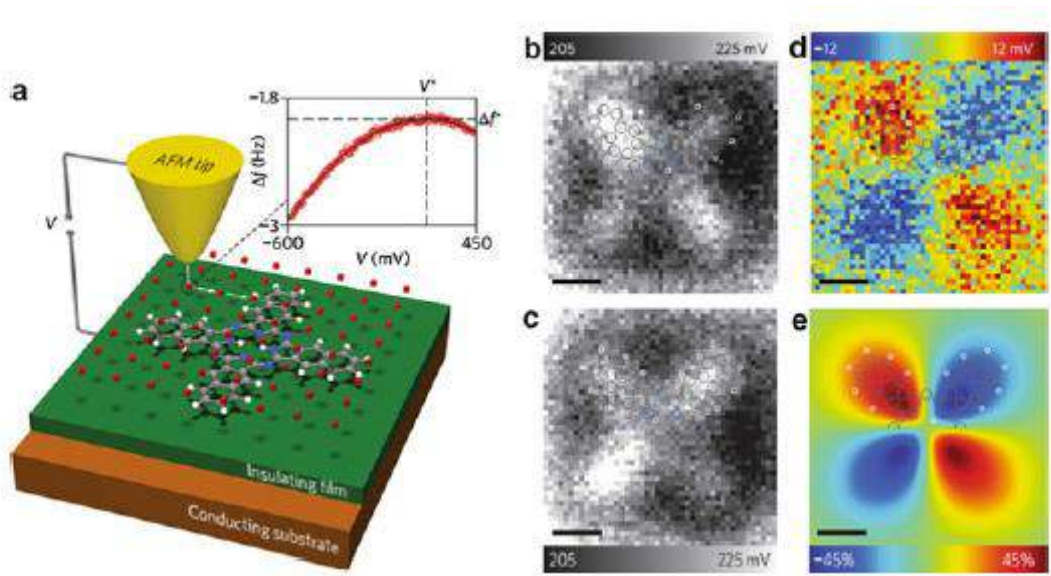

Figure 22. LCPD images of the tautomerization switching or naphthalocyanine. (a) Schematic of the measurement principle. (b) and (c), LCPD images of naphthalocyanine before (b) and after (c) switching the tautomerization state of the molecule. (d) Difference in LCPD between states determined by subtraction of (b) and (c). DFT-calculated asymmetry of the z-component of electric field above a free naphthalocyanine. Adapted with permission from [346]. Copyright 2008, Nature publishing group.

### 3.3.3.2   KPFM in ambient

For many areas of research, or for absolute measurements free environmental influence, operation in vacuum is always particularly desirable. This, along with the relative cost and abundance of ambient AFM compared to vacuum systems, has resulted in the extraordinary popularity of KPFM in ambient and making it accessible to an extraordinary number of scientific fields of research. It might be surprising then to consider that under ambient conditions absolute electronic properties of a material can be difficult, if not impossible, to determine using KPFM. The reason is that both the sample surface and tip surface condition is extremely sensitive to adsorbates and oxidation which strongly influences KPFM measurements.[330, 347, 348] In addition, interpretation of the measured CPD is further complicated by the presence of a thin water layer at the interface, leading to an unknown background potential.[330]

### 3.3.4   Stray capacitance and crosstalk in KPFM

It is well known that the spatial resolution of KPFM is intrinsically limited by the long-range effect of electrostatic interactions (section 2.3.2), which includes contributions from the macroscopic cantilever and the conical part of the tip. A significant increase in the measured CPD with increasing lift height was observed has been observed by numerous authors, recall Figure 13. As has been demonstrated in Section 3.3.2.2, a nonzero $\delta$ leads to a distance dependence in the KPFM measurement through contributions of $C_z'$, as outlined in equation (23) and demonstrated in Figure 18(d). However, the observed distance dependence cannot be attributed entirely to the feedback effect, as a result of the samples inhomogeneous surface potential. Even if the tip is located directly above an area of interest, the response is not obtained solely from the area beneath the tip, but also from the interaction of the cantilever with the adjacent sample area. Following the analysis by Kalinin *et al.*,[118] the backing potential can be described by:

$$V_{dc} = \frac{V_{CPD}^{loc} C_{loc}' + V_{CPD}^{avg} C_{cant}'}{C_{loc}' + C_{cant}'} + \frac{\delta}{A_{max}(C_{loc}' + C_{cant}')V_{ac}}. \tag{24}$$

In this case, the effect of local ($V_{CPD}^{loc}$, $C_{loc}'$) and non-local ($V_{CPD}^{avg}$, $C_{avg}'$) contributions to the CPD and the capacitance gradient are taken into account. The first two terms in equation (24) arise due to the long-range nature of the electrostatic force and the interaction of the non-local (cantilever beam and tip cone) and local (tip apex) cantilever contributions. Consequently, the measured CPD is a weighted average of the local (tip apex) and non-local (cantilever beam) contributions of the tip-sample interactions. Fortunately for KPFM, the relative CPD or the potential difference between two points in an image ($V_1$ and $V_2$) is independent of the feedback effect, as shown by equation (25):

$$(V_1 - V_2)_{measured} = (V_1 - V_2) \frac{C'_{loc}}{C'_{loc} + C'_{cant}} \tag{25}$$

Therefore, the change in the relative CPD between two dissimilar areas on the same image, is anticipated to be due to non-local tip-sample interactions (stray capacitance) and not due to the feedback effect. This stray capacitance effect is a feature of all VM AFM techniques and is not restricted to CL-KPFM.

Clearly, a major obstacle for achieving truly quantitative high resolution KPFM is eliminating the influence from long-range capacitive coupling between the complex tip architecture (i.e., tip apex, tip cone and cantilever) and the sample under investigation. There are three basic routes which can be navigated towards reducing the influence of stray capacitance and topographical contrast in KPFM.

First, the influence of stray capacitance on KPFM measurements can be minimized by adopting force gradient detection scheme, which are less sensitive to the long-range electrostatic force gradients. Alternatively, techniques which are developed to eliminate the influence of the capacitance gradient entirely (e.g., heterodyne-KPFM[349]) which is discussed in the next section.

A second approach is the implementation of deconvolution algorithms which take into account the tip shape to effectively correct the KPFM data. Several groups attempt to develop algorithms for reconstructing a sample surface potential from its KPFM image. Efimov and Cohen[326] were one of the first to attempt to correct the measured CPD in KPFM images on samples which were not completely flat using a fairly straightforward sim erosion routine.[350] Strassburg *et al.* [166, 351] developed an algorithm for reconstructing KPFM images of semiconductors using linear shift invariant procedure assuming a point spread function (PSF). By calculating the PSF of the KPFM probe (tip and cantilever) they could deconvolve measured KPFM images to obtain the surface potential of the sample. Later, Cohen *et al.*[326] would use a similar method to calculate the PSF for both AM- and FM-KPFM modes. Machleidt *et al.*[351] improved this approach by developing a new method to determine the point spread function involving direct integration of the tip shape found using a certified reference structure. Elias *et al.*[155, 352] were the first to include the cantilever in the convolution of KPFM measurements. They concluded that the cantilever has no effect on the measurement resolution, but has a profound influence on the absolute CPD value.

In a different approach to the stray capacitance problem, specialized probes have been developed to improve the resolution of KPFM operating in ambient conditions. While a sharp tip is obviously required to achieve high resolution AFM images, however, to improve EFM contrast and sensitivity high aspect ratio probes are preferred. Several groups have focused on designing such as probes using carbon nanotubes adhered to their tip[353] [354] or direct growth of nanoneedles on the apex of an AFM tip.[269] Brown *et al.*,[355] have developed co-axial and even tri-axial probes which are shown to reduce the influence of stray capacitance effects by confining the electric field to the very tip apex. Coaxial shielded probes have been shown to provide a several factors of improvements in spatial resolution in comparison to unshielded probes. Finally, novel probe designs, such as the conductive encapsulated insulated probes, [305, 306, 356, 357] are expected to have a symbiotic effect towards the development of novel KPFM techniques in liquid; allowing existing force based functional imaging techniques to be extended to the solid-liquid interface.

## 4    Advanced KPFM

The fact that many important nanoscale electrochemical and electrostatic processes can be studied on the relevant time- and length scales using KPFM, have stimulated extensive research towards improving the veracity of the KPFM technique and overcome its inherent shortcomings. In this section, we review these newly developed modes of operation which aim to: (i) eliminate measurement artifacts, (ii) improve sensitivity and resolution, (iii) capture more channels of information (e.g., polarization forces, dielectric properties) and/or (iv) capture system dynamics or kinetics. An exhaustive table summarizing the various KPFM techniques now available is provided in Table 1. Here, we distinguish techniques by their excitation waveform, classifying them as being single frequency (SF) or multi-frequency (MF) approaches.  Noteworthy, in SF methods, the system is perturbed using a single frequency sinusoidal waveform and amplitude and phase response of the fundamental response (and/or harmonic response). SF detection methods, however, do not capture all information about tip-sample interactions. This led to the development of MF SPMs in which the system is excited at two or more frequencies. The MF methods were recently reviewed by Garcia.[358] We further distinguish techniques in terms of detection methods used, be that heterodyne detection (e.g., LIA or PLL) or other using other forms of data acquisition, as well as the drive scheme chosen (i.e., purely electrostatic excitation as in lift mode AM-KPFM, or mechanical and electrical excitation as in FM-KPFM). A description of the sensitivity to either the electrostatic force, force gradient or both is also provided in Table 1. Finally, these advanced modes of KPFM can be further distinguished by the requirement of dc bias regulation for the determination of the CPD. In regards the latter, we refer to techniques which require bias feedback as CL approaches, whereas modes which negate the requirement for bias feedback are regarded as OL approaches.

Table 1 Comparison of advanced modes of KPFM.

| Mode | Excitation | Drive Scheme | Detection | Interaction | Feedback | Output | Ref. |
|---|---|---|---|---|---|---|---|
| AM-KPFM | SF | E | LIA | $F_{el}$ | √ | CPD | [74] |
| FM-KPFM | SF | E + P | PLL | $F'_{el}$ | √ | CPD | [288] |
| HAM-KPFM | MF | | LIA | $F_{el}$ or $F'_{el}$ | √ | CPD | [349, 359-362] |
| SP-KPFM | MF | E + P | LIA or PLL | $F_{el}$ or $F'_{el}$ | √ | CPD | [287, 363] |
| DH-KPFM | SF | E | LIA or PLL | $F_{el}$ or $F'_{el}$ | X | CPD + C' | [143, 364, 365] |
| BE-KPFM | MF | E | FDA | $F_{el}$ and $F'_{el}$ | X | CPD + C' | [286] |
| PthBE-KPFM | MF | E + P | FDA | $F'_{el}$ | X | CPD + C' | [125] |
| IM-KPFM | MF | E + P | FDA | $F'_{el}$ | X | CPD + C' | [366, 367] |
| tr-KPFM | SF | E | LIA | $F'_{el}$ | X | CPD + TR-Aw | [368-370] |
| tr-EFM | SF | E + P | LIA | $F_{el}$ | X | Charge dynamics | [371-374] |
| pp-KFPM | SF | E + P | LIA | $F_{el}$ or $F'_{el}$ | √ | CPD + Charge dynamics | [375, 376] |
| G-mode-KPFM | SF | E | FDA | $F'_{el}$ | X | Fast CPD + C' | [98, 143, 377] |

Acronyms used: Amplitude modulation (AM); frequency modulation (FM); heterodyne amplitude modulation (HAM); single pass (SP); dual harmonic (DH); band excitation (BE); photothermal excitation band excitation (PthBE); intermodulation (IM); time resolved (tr); pump-probe (pp), general acquisition (G); single frequency (SF); multifrequency (MF); electrical (E); mechanical actuation (P); lockin amplifier (LIA); phase locked loop (PLL); fast data acquisition (FDA).

## 4.1 Advances in closed loop-KPFM

As outlined previously, traditional implementations of KPFM utilized heterodyne detection to detect cantilever oscillations resulting from the electrostatic force combined with CL bias feedback to compensate this force. We refer to such techniques as CL-KPFM approaches. In the previous sections, we outlined traditional forms of CL operation, namely

AM- and FM-KPFM. The inherent shortcomings (e.g., stray capacitance) and inherent artifacts (e.g., feedback effect and topographical crosstalk), as described in section 3.3, have resulted in many incremental improvements in the measurement and detection approach.

In 2006, Lee *et al.*,[328] developed an approach to KPFM which was shown to eliminate geometric artefacts in traditional KPFM measurements.[327] In their method, topographical crosstalk was suppressed by dividing the first and second harmonic components of the cantilever response to the electrostatic interaction in both amplitude and frequency modulation modes. The modified KPFM feedback controller nullifies the ratio of the first- and second-order harmonic components instead of the first harmonic component itself, as is common in conventional KPFM. Suppression of topographic features in the CPD map on an equipotential gold surface compared to conventional KPFM was demonstrated using this approach. [328] Furthermore, they demonstrated that the tip-sample separation dependence of CPD was also significantly reduced using this new technique.

More recently, in order to combine the bias sensitivity and lateral resolution advantages of both AM- and FM-KPFM, respectively, Sugawara *et al.*[359] developed Heterodyne (H)-KPFM. This method is based on the heterodyne (i.e., frequency conversion) and AM detection of the electrostatic force, where mixing products are generated by excitation at the first cantilever resonance and some other driving frequency far from resonance. H-KPFM retains many similarities with FM-KPFM but the topographic and electrostatic interactions are separated by hundreds of kHz. This approach retains the potential sensitivity of AM-KPFM but having the improved spatial resolution afforded by FM-KPFM.[349, 359]

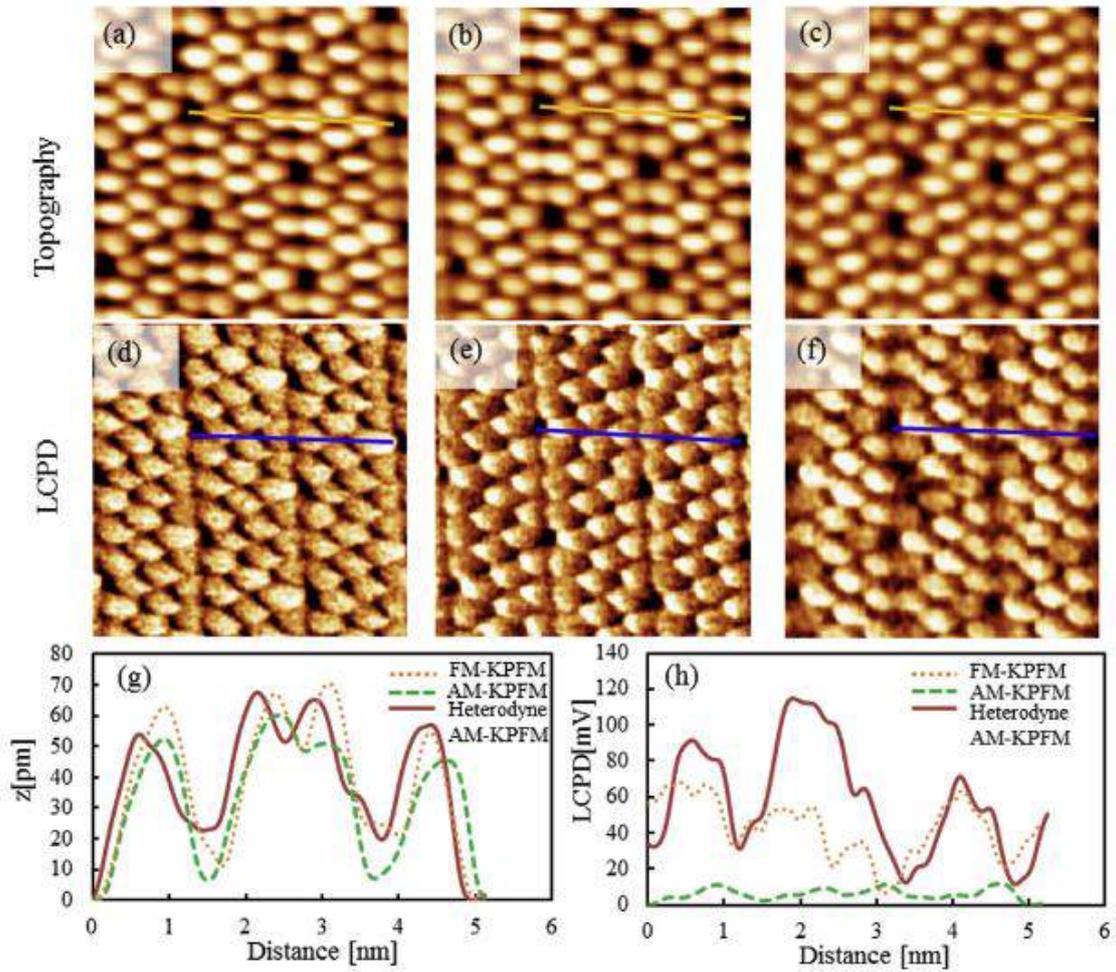

Figure 23. (a), (d) AFM topography and corresponding surface potential images on a Si(111)-7×7 obtained simultaneously using FM-KPFM, (b,c) AM-KPFM and (g,h) heterodyne AM-KPFM. (g,h) The topographic and potential line profiles from the areas indicated in (d,e,f). *Seeking permissions.*

Ma *et al.*,[349] compared the influence of the stray capacitance effect on surface potential on three different modes (AM, FM, and heterodyne AM; Figure 23). They confirmed that the measured CPD was almost the same in FM-KPFM and heterodyne AM-KPFM, though it was quite different in AM-KPFM. Noteworthy, the differences in measured CPD, can also have implications for cross-talk between the electrostatic force and the topography signals.[57, 283] In this work, they found local CPD corrugations in AM-KPFM were significantly weaker than in FM- and heterodyne AM-KPFMs, which they proscribe to a parasitic tip-sample distance variation resulting from topography feedback. Consequently, the atomic resolution in topographic and potential images in FM- and heterodyne AM-KPFMs is expected to be closer to the actual potential distribution on the surface.[349] Gareth *et al.*[360] demonstrated that heterodyne AM-KPFM also had significant speed advantages over FM-KPFM as it is not subject to bandwidth limitations to avoid ac coupling if topography and KPFM bandwidths overlap. In agreement with previous studies,[349] they also demonstrated improved resolution over AM-KPFM and concluded that this approach will facilitate faster and more accurate nanoscale potential measurements than is possible with traditional KPFM approaches and went on in a separate paper to apply heterodyne AM-KPFM to the investigation of patch potentials in Casimir force measurements.[361]

### 4.2 Open loop-KPFM

Open loop (OL)-KPFM has been demonstrated in a variety of ways so far. The first approach is determination of the CPD from the harmonic response of the cantilever to the application of a AC modulation voltage. The second consists of capturing a multidimensional data set, such as electrostatic force (gradient) bias spectroscopy for each pixel on the surface, where the electronic parameters are found from the functional form of the bias dependence of the tip-sample electrostatic interaction. Techniques which can accurately measure the CPD while obviating the requirement for bias feedback present significant advantages for many application of KPFM. First, bias free KPFM approaches are more suitable for non-destructive characterization of voltage sensitive materials such as semiconductors, insulators, or electroactive devices. In such measurements it is imperative that charge transfer between tip and sample is suppressed. Secondly, the lack of a bias feedback loop in the set-up presents an opportunity for the development of fast modes of KPFM,[143, 371, 373-376] reduces artifacts related to non-ideal feedback,[118, 286] and avoids the need for cumbersome tuning of the KPFM bias feedback loop.[297] In addition, open loop approaches can often access additional information relating to the tip-sample capacitance which is nullified in CL-KPM. Lastly, and most importantly, bias feedback free approaches are extremely promising for operation in liquids where the finite conductivity of the solution produces complex electrochemical effects at relatively low dc bias voltages, violating the principles of closed loop KPFM.[46] Importantly, techniques which do not utilize active bias feedback cannot compensate for the electrostatic force between tip-sample; however, in many applications such as the ones described, their advantages outweigh this shortcoming.

#### 4.2.1   Dual harmonic-KPFM

Takeuchi *et al.*[364] proposed a KPFM approach in vacuum which could be operated in a similar fashion to FM-KPFM (e.g., operated during standard FM-AFM imaging while monitoring the frequency shift), but which obviated the requirement for bias voltage feedback. This approach, referred to here as Dual harmonic (DH)-KPFM, uses only ac voltage (i.e., $V_{dc} = 0$) with a low frequency excitation (e.g., 1-3 kHz below bandwidth of PLL) far from the resonance frequency. In their setup, they introduced a second LIA allowing both first and second order derivative of the frequency shift to be measured simultaneously. Monitoring the ratio of the first two derivatives of the electrostatic force allowed mapping CPD variation across semiconducting materials. [364] This approach was later compared with conventional CL-KPFM in ambient for measurements on voltage sensitive ferroelectric materials.[365] In this study, both KPFM techniques were implemented using AM detection during dual pass imaging, where the actuation of the cantilever, far $(10 - 100$ nm) from the surface was purely electrostatic in nature (no mechanical excitation of the cantilever). DH and conventional KPFM on charge-patterned bismuth ferrite found agreement between recorded relative surface potential values between domains, demonstrating that DH-KPFM can be used for quantitative mapping of relative surface potentials. [365] A constant offset between techniques was observed raising questions over the agreement of absolute CPD measurements in OL and CL approaches. In this study, the DH-KPFM technique was also used to simultaneously map changes in surface potential and dielectric properties resulting from surface modification of a strontium barium niobate film after application of a dc bias between tip and sample.[365]

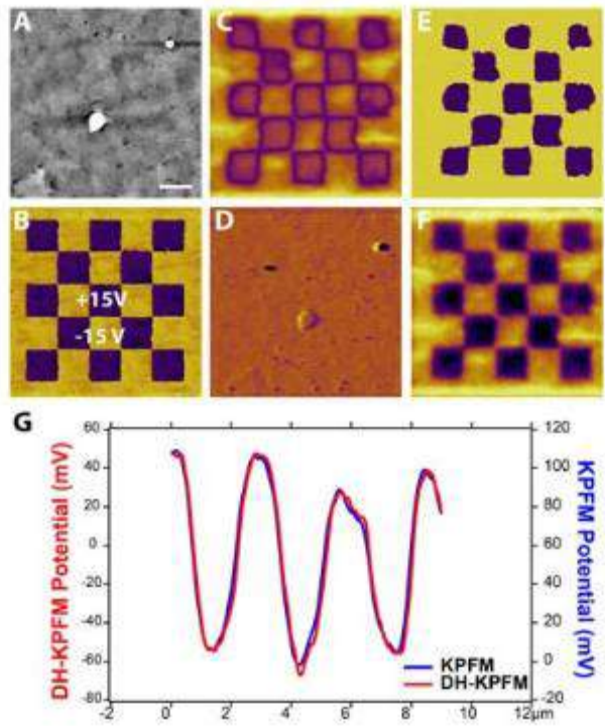

Figure 24. (a) Topography (z-range = 8.8 nm) and (b) PFM phase images of a patterned BFO thin film where bright color represents domains with out-of-plane polarization pointing towards the sample surface. Images of (c) first and (d) second harmonic amplitudes, (e) first harmonic phase, and (f) recorded DH surface potential. (g) Line profile cross section of the same area, recorded using conventional KPFM and DH-KPFM and plotted on different axis scales. Scale bar represents 2 μm. *Seeking permissions.*

In initial implementations of DH-KPFM, the influence of cantilever transfer function gain was neglected,[48, 364, 365] which will contribute to the absolute CPD measured by a magnitude related to the modulation frequency.[49] However, with knowledge of cantilever transfer function gain, $X_{gain}$, and the known $V_{ac}$, the ratio of the harmonic responses to the applied voltage can be used to determine the $V_{CPD}$, as described by equation (26): [49]

$$V_{CPD} = \frac{A_\omega \cos(\varphi_\omega)}{A_{2\omega}} \frac{V_{ac}}{4X_{gain}} \qquad (26)$$

The polarity of $V_{CPD}$ is given by the phase ($\varphi_\omega$) of the cantilever response at $\omega$. Recently, Polak *et al.*,[325] demonstrated that, after careful calibration of parasitic signals for both cases, CPD measured by CL-KPFM and DH-KPFM can be made independent of frequency and amplitude of the excitation signal. Finally, recently Kuo *et al.*,[378] achieved atomic resolution of surface potential using DH-KPFM in UHV in dual pass mode. This study was the first to realize the atomic resolution imaging of Si(111) 7x7 by dual pass mode.

Interestingly they concluded that dual harmonic-KPFM is stable and suitable for high speed KPFM, as it avoids instability related to the bias feedback loop. [378]

To further press this last point we compare the measurement speed of open and closed loop KPFM. First, we note optimize the feedback loop parameters following the loop tuning procedure outlined in the seminal work by Jacobs et al. Briefly, the tip was engaged onto a freshly cleaved HOPG surface which was connected to a function generator. A square wave bias waveform having amplitude of between 100 mV -1 V amplitude and frequency between 1 Hz – 10 Hz was typically used to modulate the sample surface potential (see Figure 25(a)). Before the feedback loop was closed (operating in open loop) the phase of the LIA was adjusted to maximize $X\omega$. Furthermore, the time constant of LIA was chosen smaller enough such that $X\omega$ tracked the applied bias waveform exactly, but large enough so that the signal was not excessively noisy. The next step involved closing the KPFM bias feedback loop and carefully adjusting the gains to minimize the $X\omega$ (i.e. the error signal) and maximize the CPD signal as best as possible (see Figure 25 (a) and (b)). Cross sections of a finely tuned feedback loop are shown in Figure 25 (a). In the results presented in Figure 25b) the transfer functions of the open and closed loop system were measured by applying a chirp function to the electrical excitation applied to the tip and the detected response is evaluated by the calculation of a cross correlation function. The -3dB point is used to determine the effective bandwidth of open loop transfer function was found to be 53 Hz which was significantly diminished when the loop was closed (5.6 Hz). Note, in this case the closed loop the bandwidth is determined by the time constant of the LIA and the optimized gain settings, so it should be measured each time CL-KPFM is operated to determine the appropriate scan speed. For DH-KPFM on the other hand, the bandwidth is effectively governed by the time constant of the LIA (user defined) and limited by the cantilever bandwidth. The trade-off between bandwidth and error in the measurement is on which must be considered however. This is demonstrated in Figure 25 (c), showing the error in recorded CPD as a function of measurement bandwidth in a typical DH-KPFM measurement. Here a cantilever with a resonance frequency of 70.59 kHz and Q factor of 150.3 was used, leading to an effective cantilever bandwidth of 469 Hz. It is clear from Figure 25(c) that as the DH-KPFM bandwidth approaches the cantilever bandwidth the error signal rises steeply. However, this does allow the bandwidth of the DH-KPFM measurement to be chosen to achieve the desired error in the CPD. Furthermore, adopting ultra-small cantilevers with high bandwidth (>100 kHz) could allow KPFM imaging for fast scanning applications. Additionally, the measurement speed could be increased even further by overcoming the requirement for heterodyne detection having long integration times (e.g. G-Mode KPFM, see Section 4.3.4).

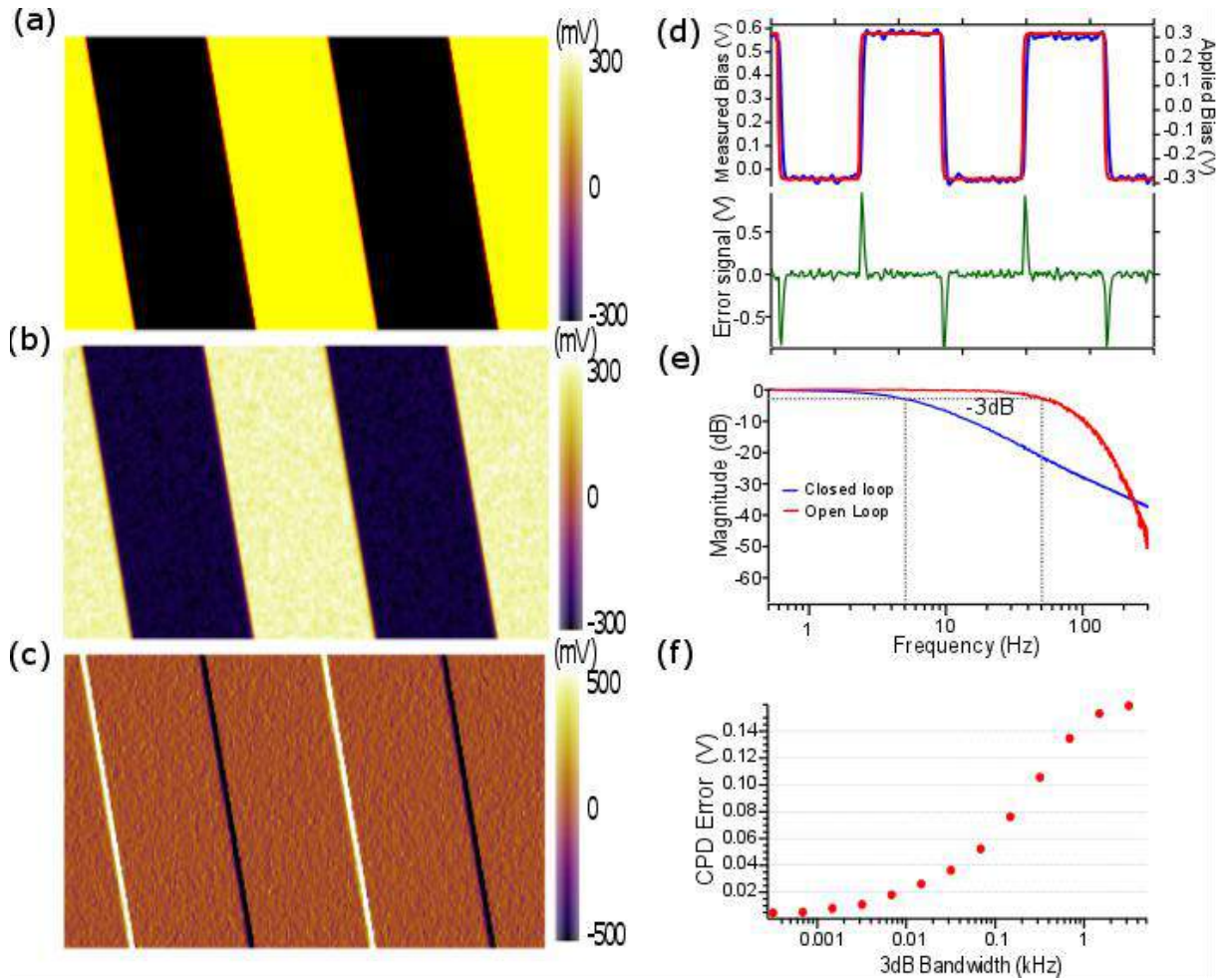

Figure 25. Loop tuning procedure in CL-KFPM. (a) Applied bias (z scale 600 mV) used to modulate the HOPG surface potential, (b)recorded CPD (z scale 600 mV) and (c) error signal (z scale 1V). (right) Optimized KPFM bias feedback loop (b) open and closed loop transfer function of the KPFM system. (c) Error in CPD determined by DH-KPFM as a function of measurement bandwidth.

### 4.2.2 Intermodulation-EFM

One disadvantage of DH-KPFM compared to conventional CL-KPFM, is its limitation to off resonance frequencies, arising from complications in determining $X_{gain}$ when the harmonics are positioned on the cantilever resonance frequency itself (i.e., indirect cross talk due to changes in the resonance peak[285, 286]). To overcome this, MF techniques such as BE [87-89] and IM [90] have been configured to perform OL-KPFM. Intermodulation IM-EFM[366] is operated on-resonance during single pass imaging, and consists of driving the cantilever with two pure drive tones close to the resonance, one used to mechanically actuate the cantilever and another used to modulate the electrostatic force. This approach has been applied to spatially resolve photogeneration of charge in a photovoltaic thin film, as shown in Figure 26, realized by measuring the CPD during both the absence and presence of illumination in consecutive trance and retrace measurements. The surface photo-voltage ($V_{SPV}$) was determined from the difference between trace and retrace measurements, allowing visualization of domains having different $V_{SPV}$ related to variation in concentration of donor sites.[366] In a different study, IM-EFM has been applied to investigate local charge injection and extraction on surface-modified $Al_2O_3$ nanoparticles in a low-density polyethylene matrix.[367] Since the responses of interest all lie close to the resonance frequency, this approach leads to increased signal to noise over off resonance DH-KPFM and is expected to lead to higher spatial resolution as the tip is closer to the sample when operated in single pass.

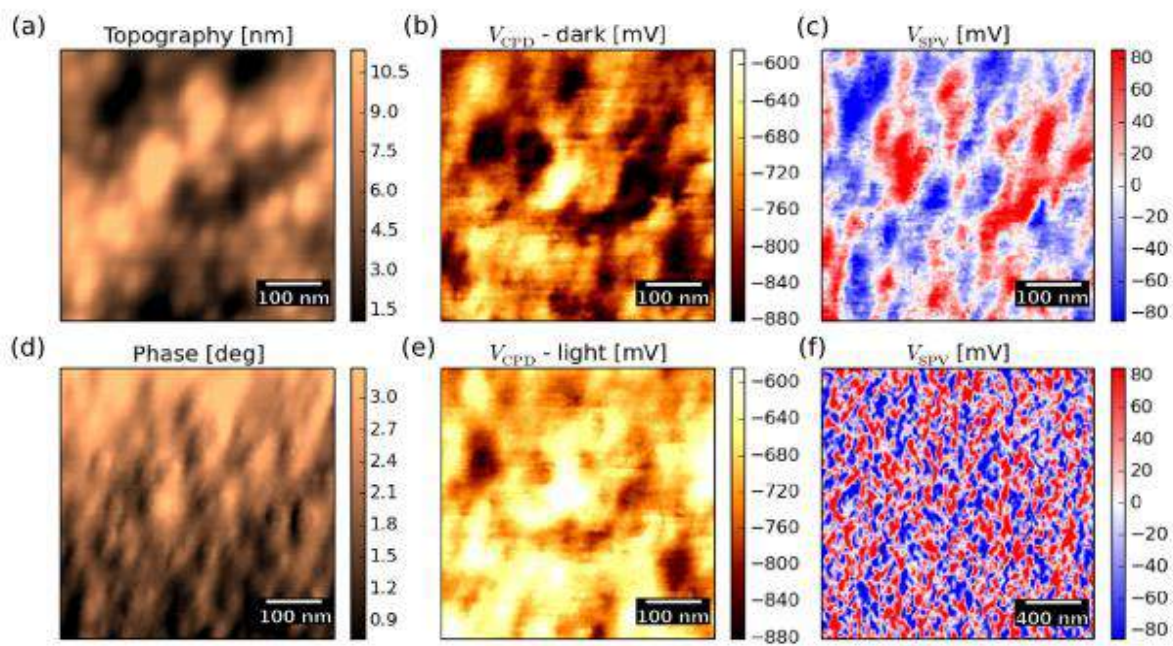

Figure 26. IM-EFM on a TQ1:PCBM:C60 sample demonstrating only small variation in (a) topography and (d) phase imaging. CPD measurements performed in the (b) absence and (d) presence of sample illumination were used to determine (c) surface photo-voltage. (f) Regular domain contrast was observed across the sample surface associated with variation in photo-generated charge carriers. *Seeking permissions.*

### 4.2.3 Band excitation-KPFM

The BE method is an alternative to traditional single frequency LIA detection by exciting and detecting response at all frequencies within a specified frequency range simultaneously.[87] A typical example of an excitation and response signal in Fourier and time domains in single frequency and BE methods are shown in Figure 27. In BE a voltage waveform with predefined amplitude and phase content across a continuous range of frequencies is generated by an arbitrary waveform generator and used to drive the tip either electrically (as in PFM and KPFM), mechanically (as in tapping mode AFM, magnetic force microscopy (MFM), and EFM), or to drive an external oscillator below the sample (Atomic force acoustic microscopy)[109]). The center frequency of the detection (and excitation) band is normally centered on the cantilever resonance (or higher eigenmode) to allow capture of the entire resonance response to the applied stimulus. As is the case in IM-EFM,[366] operation on resonance has the added benefit of amplification and much higher signal to noise in comparison to off resonance approaches. When operated in non-contact (or lift mode) the behavior of the cantilever can be well described by a SHO model and thereby described by three parameters: $\omega_0$, amplitude at the resonance, $A(\omega_0) = A_0)$, and $Q$, as given by equation (7)and (8). Maps of $\omega_0$, $A_0$, and $Q$ are then deconvoluted from the measurement and stored as images.

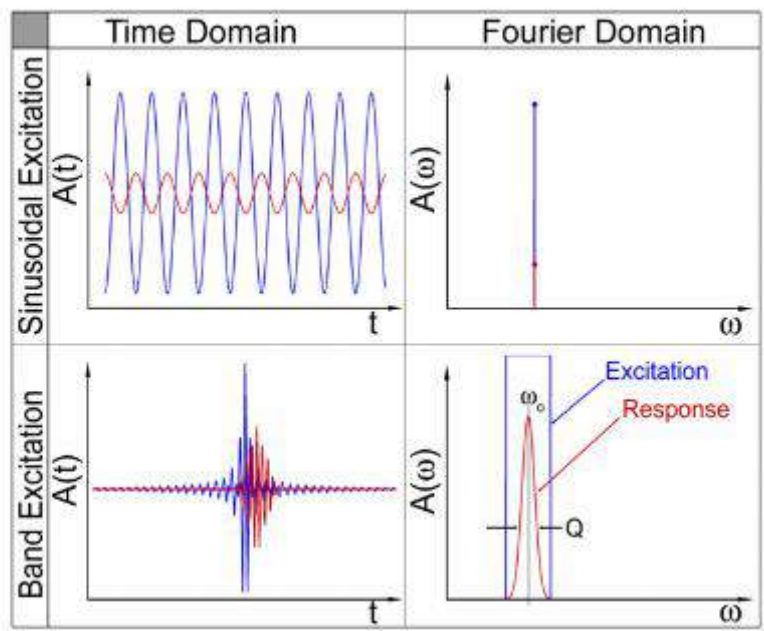

Figure 27. Excitation (blue) and response (red) signals in standard SPM techniques in (a) time and (b) Fourier domains. Excitation (blue) and response (red) signals in BE SPM in (c) time and (d) Fourier domain. In BE, the system response is probed in the specified frequency range (e.g., encompassing a resonance), as opposed to a single frequency in conventional SPM. *Seeking permissions.*

In the past couple of years, BE-KPFM has been realized in many different forms including the half harmonic (HH)BE approach,[379] a voltage spectroscopy mode referred to as open loop band excitation (OLBE),[286] force volume (FV)BE-KPFM,[320] as well as being BE-KPFM with photothermal excitation.[125] The basic idea for these modes is that electrostatic interactions are captured through monitoring the response of the cantilever resonance peak to an applied VM (either pure ac (e.g., HHBE-KPFM, or a combination of ac and dc voltage (e.g., OLBE-KPFM)).

### 4.2.3.1 Half harmonic BE-KPFM

In half harmonic BE (HHBE)-KPFM the BE method is applied in a similar manner as DH-KFPM but in a MF embodiment.[379] as illustrated in Figure 28. The cantilever is electrically excited in a frequency band with a central frequency positioned at $\omega_0$, and the response is recorded for the same frequency band. The second harmonic response of this excitation is located in a band centered at twice the excitation frequencies ($2\omega_0$), falling outside of the cantilever resonance and detection band. To detect the second harmonic response, a subsequent BE electrical drive centered at half the resonance frequency ($\omega_0/2$) is applied to the cantilever. Thus, a second harmonic response is generated in the frequency band around the resonance peak (referred to as Half BE (HBE)). The BE and HBE excitations are applied sequentially for each image pixel. In this manner, the first and the second harmonic components of the response can be compared directly since they fall within the same frequency band of the cantilever transfer function and $X_{gain} = G(\omega)/G(2\omega) = 1$. In essence, HHBE-KPFM is similar to its single frequency analogue (DH-KPFM) with the added benefit of resonant amplification and built-in cantilever transfer function gain correction.

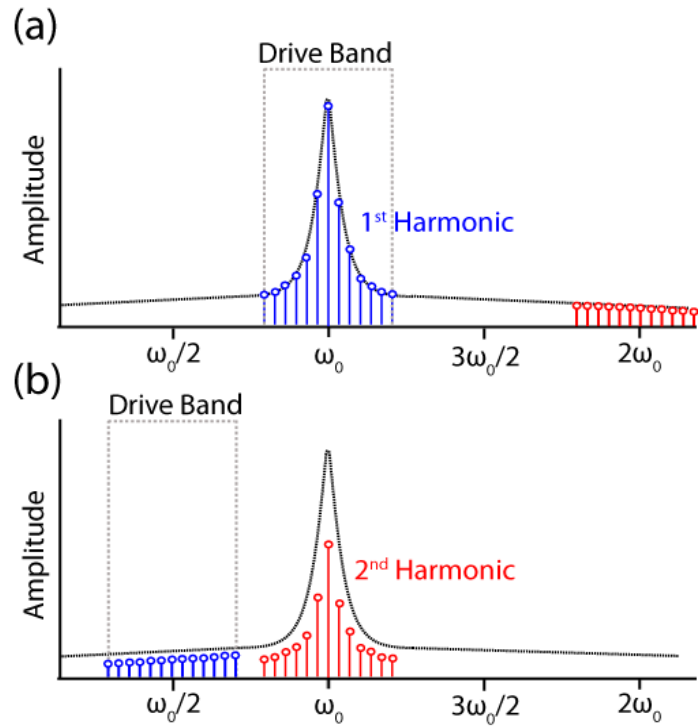

Figure 28. Schematic of HHBE-KPFM operation. (a) BE is performed with electrical excitation in a band having a central frequency at $\omega_0$, the first harmonic response is recorded within the same band but the second harmonic lies outside the detection bandwidth at $2\omega_0$. (b) HBE ($\omega_0/2$) is performed sequentially and the second harmonic response is recorded in the frequency (doubled) band around $\omega$. In this way, the first and second harmonic response lies in the same frequency band. *Seeking permissions.*

Work function measurements recorded using OL (DH-KPFM and HHBE-KPFM) and CL (AM-KPFM) KPFM techniques between an $Al_2O_3$-coated single layer graphene on a copper electrode have been demonstrated.[316] All techniques were used to measure the same region of a single layer graphene which exhibited a large defect across the width of the hexagon structure. The CPD values recorded by DH-KPFM and HHBE-KPFM were shown to agree, whereas an offset of 55 mV was recorded between OL and CL techniques (Figure 29. (d)). All techniques, however, recorded a CPD difference of ~120 mV between the $Al_2O_3$-coated graphene and the $Al_2O_3$-coated Cu substrate (see Figure 29. (d)). The offset between techniques was attributed to the previously described feedback effect (see Section 3.3.2.2). [316] Agreement of the relative values is also expected while considering the influence of the feedback effect, since it does not influence the ability of CL-KPFM to make relative measurements.[118] It is expected that offset between CL- and OL-KPFM could be reduced if the setpoint correction was taken into consideration by the manufacturers (via introduction of suitable correction protocols), and highlights the difficultly in performing quantitative CL-KPFM without taking these effects into account.[325]

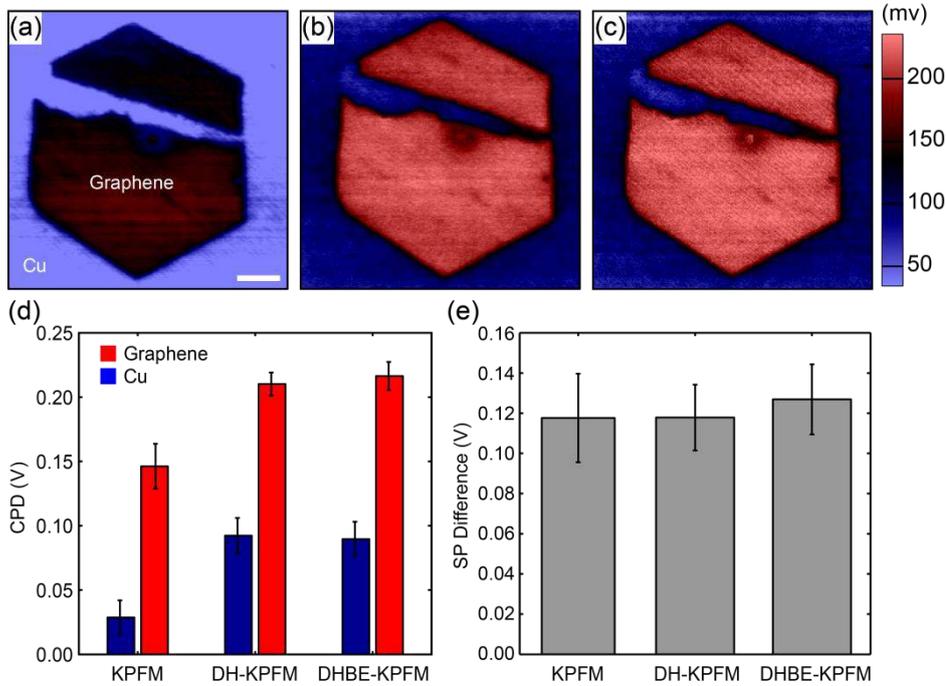

Figure 29. Comparison of CL-KPFM with single and multifrequency DH-KPFM. CPD maps of a single hexagonal graphene layer collected with (a) KPFM, (b) DH-KPFM, and (c) HHBE-KPFM (vertical scale = 200 mV, offset = +135 mV for all images). Bar chart showing (d) CPD (mean ± std. dev.) for graphene and Cu and (e) the SP difference (mean ± std. dev.) measured using CL-KFPM, DH-KPFM, and DHBE-KPFM *Seeking permissions.*

### 4.2.4 Open loop-BE-KPFM

In a different approach, Guo *et al.*[286] were the first to combine OL bias spectroscopy with BE detection which was later improved upon by Collins *et al.*.[320] In OLBE-KPFM, the $V_{ac}$, is applied in a frequency band centered on the cantilever resonant frequency, $\omega_0$. The bias dependence is probed using a train of BE pulses with varying dc voltage offset, as shown in Figure 30(a). The full amplitude and phase response as a function of frequency and bias is acquired at each spatial location. Figure 30(b) shows the amplitude response vs. frequency and bias for a tip located 10 nm above a HOPG surface.[320] This 2D response contains information both on the electrostatic tip-sample interactions and cantilever transfer function in the vicinity of the resonance. Figure 30(c) shows the recorded cantilever response amplitude vs. frequency for negative bias values. The response amplitude demonstrates a strong bias dependence. Figure 30(d) shows the results of fitting the BE data to the SHO model and plotting $A_0 \cos(\varphi)$ and $\omega_0$ as a function of bias. Note that since OLBE-KPFM detects both the conservative $(\Delta\omega_0)$ and dissipative $(\Delta A_0)$ interactions it is sensitive to both the electrostatic force and force gradient detection simultaneously, hence combines the advantages inherent in both AM- and FM-KPFM.[125, 286, 320] This is confirmed by the bias dependence of the responses in Figure 30(d), where the $A_0 \cos(\varphi)$ demonstrates a linear bias dependence, while $\omega_0$ demonstrates a parabolic bias dependence, as described by equation (12) and equation (3), respectively. In essence, adoption of MF approaches has allowed capturing both electrostatic force (nonlocal) and force gradient (local) response.

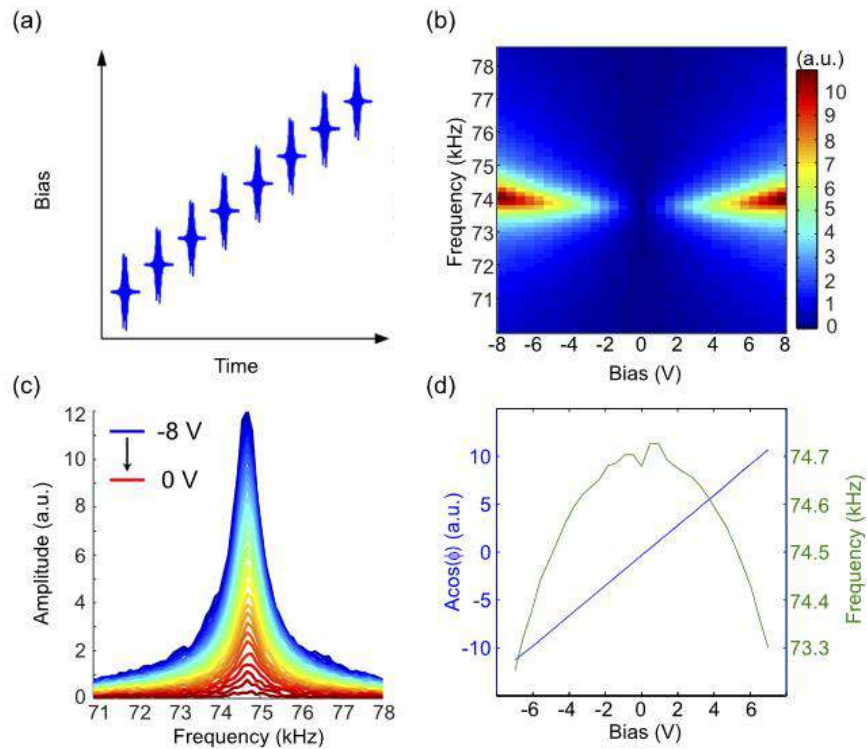

Figure 30 (a) Operating principle of OLBE-KPFM. (a) A train of BE waveforms superimposed on a linear voltage ramp is applied to the tip. (b) Cantilever response amplitude versus bias and frequency is acquired at each location. (c) Cantilever response amplitude versus frequency for negative bias values. (d) Bias dependence of cantilever mixed amplitude response and resonant frequency axis. *Seeking permissions.*

OLBE-KPFM imaging was first demonstrated on a model sample Au/Si surface, with a charge-written pattern on the Si surface containing topographic (step edge), work function (Au vs. Si) and charge contrast, as shown in Figure 31. The topography and corresponding classical AM-KPFM images are shown in Figure 31(a,b). The capacitance gradient image, which is shown in Figure 31(c) clearly illustrates changes in tip-sample capacitance gradient at the step edge due to the interaction of the tip with side wall. The nulling potential of the amplitude response is illustrated in Figure 31(d) demonstrating similar contrast and CPD values as AM-KPFM but having an absolute offset between both approaches.

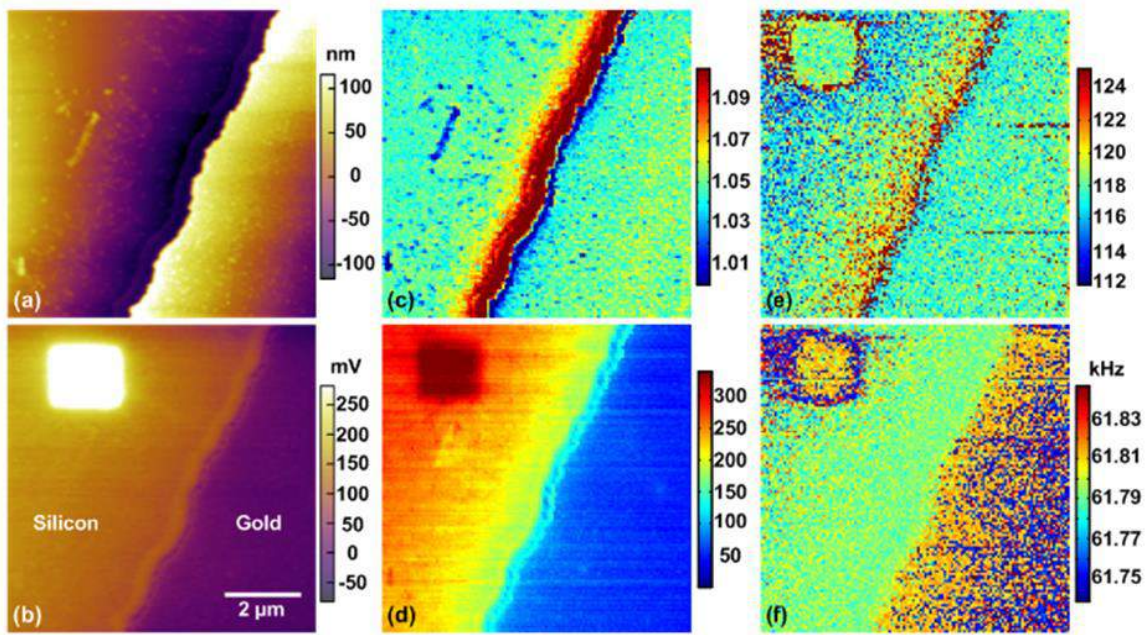

Figure 31(a) Surface topography and (b) KPFM images of charge-written Au/Si sample with an interleave scan of 50 nm. KPFM image is zero-order flattened. (c) The OL-BE-KPFM slope image and (d) nulling potential with an interleave scan of 100 nm. (e) $Q$ factor and (f) resonant frequency maps averaged for all biases. Seeking permissions.

### 4.2.5   Photothermal BE-KPFM

In an attempt to improve the bias sensitivity to the electrostatic force gradient, Collins *et al.* later developed photothermal BE (PthBE)-KPFM.[125] Figure 32 shows a schematic outlining the differences between OLBE and PthBE-KPFM setups. Common to both is dc bias pulses applied directly to the cantilever, used to induce changes in the electrostatic interactions between probe and sample. BE waveforms are used to detect the resultant changes in the dynamic cantilever response by recording the resonance peak of the oscillating cantilever. The difference between OLBE-KPFM and PthBE-KPFM is simply where the BE signal is supplied for excitation. In OLBE-KPFM, the BE excitation waveform is applied as a voltage directly to the conductive probe, as shown in Figure 32(a). Conversely, in PthBE-KPFM, the BE waveform is used to modulate the photothermal laser module directly. In photothermal excitation, the laser modulates the temperature of the cantilever base, resulting in cantilever oscillation based on thermal expansion.

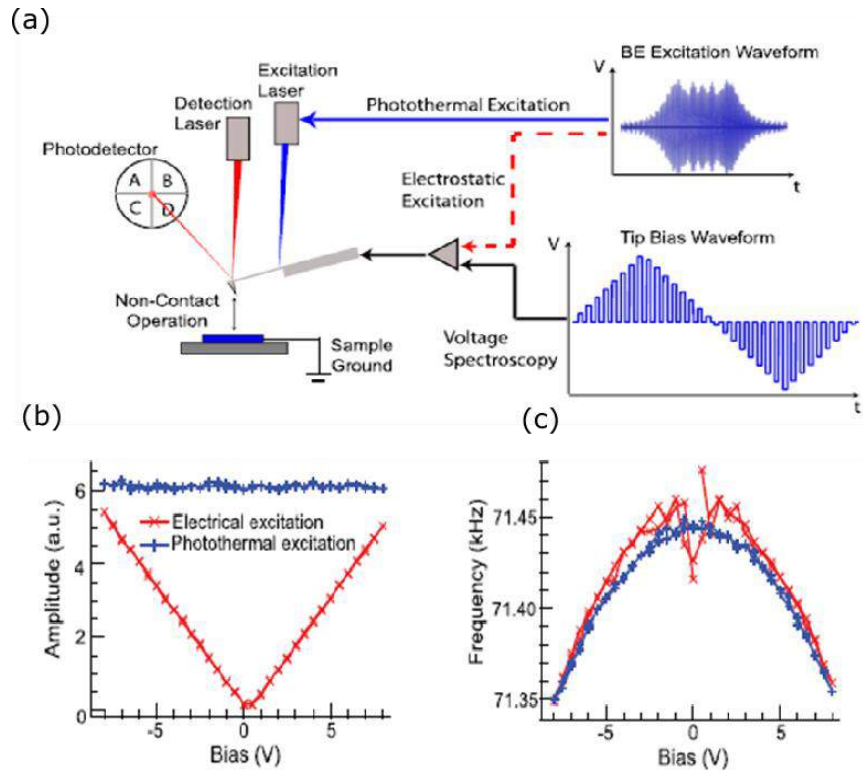

Figure 32. (a) Schematic of the working principle behind both OLBE (red dashed line) and PthBE (blue line)—KPFM utilizing electrostatic and photothermal excitation of the cantilever, respectively. Voltage spectroscopy is performed in this case using a dc bipolar square wave applied between a conductive probe and sample. Bias dependence of (b) cantilever response amplitude and (c) resonant frequency for OLBE (red line) and PthBE (blue line) -KPFM, respectively. *Seeking permissions.*

When relying on capacitive actuation of the cantilever, the response becomes increasingly smaller approaching the CPD, eventually vanishing at $V_{CPD}$ as shown in Figure 32(b). The small amplitude at bias values close to zero can decrease the accuracy of the SHO fitting procedure used to determine the resonance frequency which is why large deviations in frequency can be observed. On the contrary, photothermal actuation was shown to be highly sensitive to the electrostatic force gradient, and capable of detection small changes in resonance frequency at bias values close to the CPD in comparison to OLBE-KPFM. PthBE-KPFM was shown to provide a more localized measurement of true CPD in comparison to AM-KPFM as well as having the added benefit of containing information relating to local dielectric properties and electronic dissipation between tip and sample available through subtle changes in the amplitude and $Q$ maps.[125]

### 4.2.6 3D BE-KPFM

A major consideration when studying electrostatic forces at either the gas or liquid solid interface is the long-range nature of the interaction. This poses a major complication in interpretation of KPFM measurements, and ultimately limits the maximum achievable resolution of the technique. This distance dependence of the tip-sample interactions has motivated continued research focused on the development of 3D mapping techniques to capture not only the spatial variation in the X-Y plane, but also as a function of tip-sample distance. Wang *et al.* first developed 3D-KPFM for characterizing of in-plane piezoelectric potential of laterally deflected ZnO nanowires.[380] This technique was developed to eliminate artifacts induced by high topographical variations along the edges of micro/nanowires which made characterization by conventional SKPM inappropriate or impossible. In addition, it allowed the electrical potential to be mapped in a 3D spatial volume above the sample surface. However, since their setup relied on bias feedback, parasitic contributions to the measured Z profile of the surface potential can be expected as a result of the feedback effect (see section 3.3.2.2).[286]

This inherent distance dependence was overcome through the development of force volume (FV)BE-KPFM,[320] where rather than raster scanning in the plane of the surface, as with conventional scanning probe techniques. An example of a FVBE-KPFM measurement on a model Si/Au sample is demonstrated in Figure 33. FVBE-KPFM profiles the electrostatic interactions as function of tip-sample distance across a fine grid of points on the surface. In this manner, a 5D data set comprising response vs. frequency, bias, and position (X,Y,Z) is acquired. Bias vs distance spectroscopy for both amplitude and frequency are shown in Figure 33(a) and (b) respectively for a single location on the Si substrate. Notable, each plot provides two independent methods of determination of capacitance through evaluation of either the distance or bias dependence. The resulting multidimensional data can be compressed by fitting the bias dependence of the electrostatic force (linear bias dependence) and electrostatic force gradient (parabolic bias dependence) respectively. The fitting parameters set can be used to determine the 3D CPD and capacitance gradient determined using the electrostatic force and force gradient, respectively. Figure 33(c) and (d) depict maps of the CPD for a single XY plane as well as single XZ plane from a 45x45x50 grid, providing a wealth of information on the overall tip-sample interactions present in KPFM measurements. In this regard, FVBE-KPFM is expected to offer a useful insight between KPFM experiment and modelling which can ultimately be used to completely deconvolute the effect of the probe geometry.

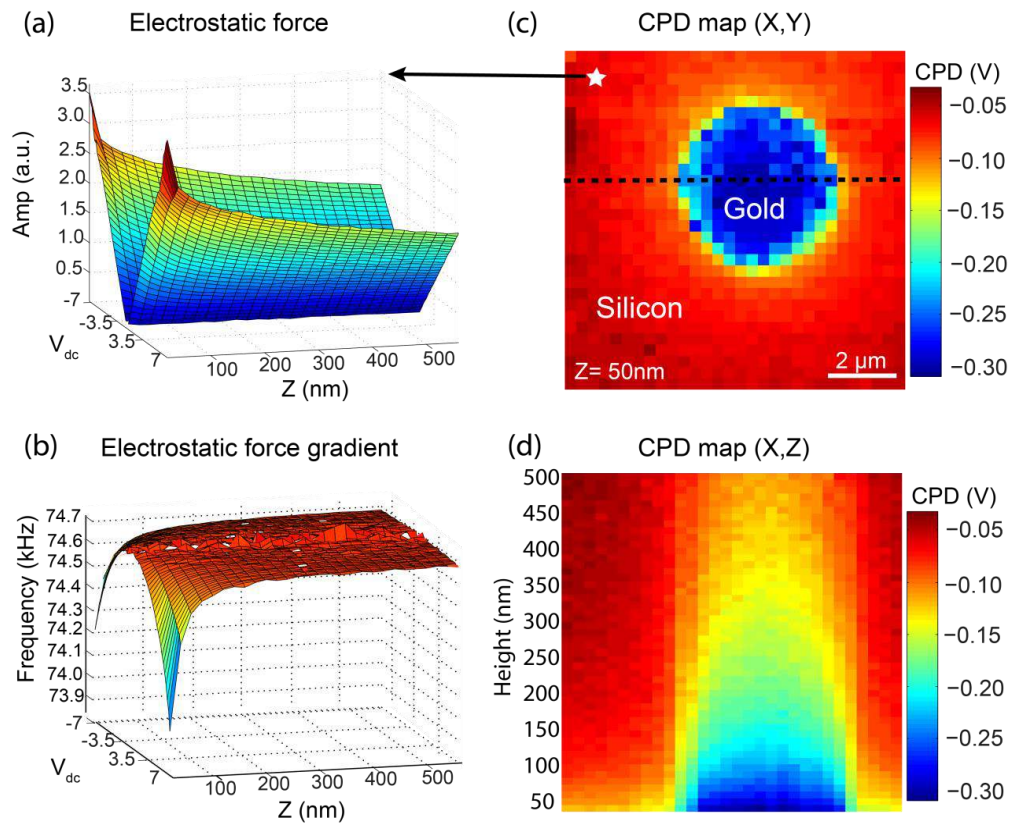

Figure 33. 3D CPD mapping using force volume BE-KPFM. Single point (a) electrostatic force and (b) electrostatic force gradient measurement as a function of tip sample distance (Z). Spatial mapping of the CPD in the (c) X-Y plane (Z=50nm) and the (d) X-Z plane from (X plane indicated by a dashed black line in image (c)) determined from fitting to the electrostatic force of a 45x45x50 (X,Y,Z) grid recorded on a gold/silicon test structure. *Seeking permissions.*

## 4.3 Time resolved measurements

While classical EFM and KPFM methods have seen extraordinary success in the characterization of static or quasi-static processes, in a whole range of materials and devices, the surface potential and capacitive information are averaged over the measurement time, and hence mask much of the relevant information on fast processes which take place at timescales below the bandwidth of the measurement itself. Surprising then that only recently has the importance of capturing the local electro-dynamic processes been considered by the KPFM community.[368] This has however lead to the development of novel modes of KPFM, or related techniques, to achieve a locally and time-resolved analysis of the surface potential, charge or ionic transport phenomena. Such approaches provide information beyond the time averaged CPD, on fast local charging, or ion dynamics and have proven useful in probing the time dependent ionic transport in lateral devices,[381] surface photo-voltage and charge carrier generation,[371, 372] as well as charge screening and ion dynamics at the solid-liquid interface.[46, 47]

### 4.3.1 Time resolved-EFM

In the recent years, the group of Ginger has made great strides in the development of time resolved SPM measurements processing microsecond and even sub-microsecond temporal resolution. Their work has been pivotal towards improving the time resolution of mechanical detection methods to probe transient processes such as carrier dynamics with sub-50 nm spatial resolution, and microsecond-scale temporal resolution. In particular, they developed time resolved (tr)-EFM to help understand correlations between nanoscale structure, charge generation, recombination, and transport in thin film solar cells.[371]

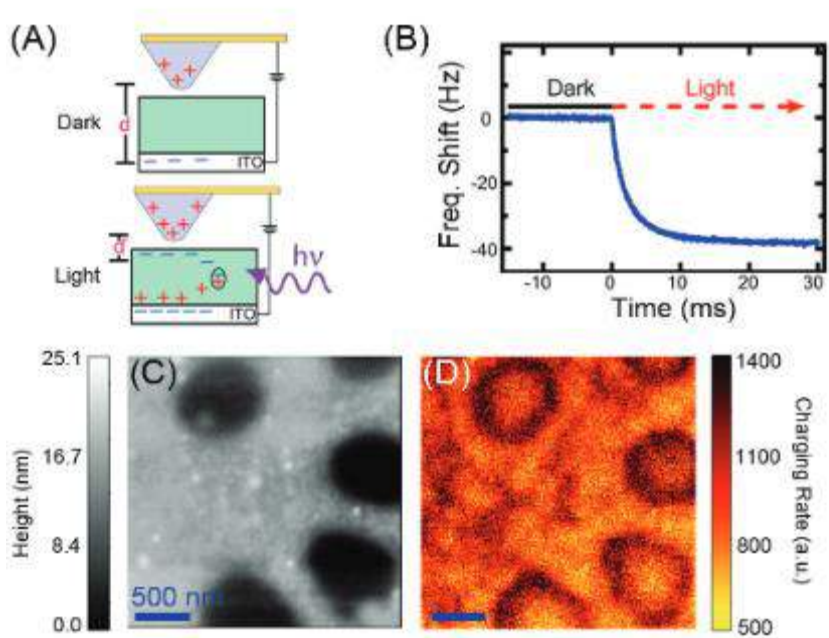

Figure 34 (a) Schematic depiction of the operation of tr-EFM. tr-EFM captures the frequency shifts corresponding to changes in capacitance gradient and surface potentials resulting from photo generated charge carriers. (b) Representative plot of the resonance frequency shift vs. time following photoexcitation. At time t=0 ms, the LED is turned on, causing an exponential decay in the frequency shift. tr-EFM provides a measurement of the relative charging rate determined from the time constant of this decay. (c) Topography and (d) charging rate image for the same area of a PFB:F8BT sample, dissolved in xylenes with 1:1 composition. *Seeking permissions.*

Figure 34(a) depicts the operation of a tr-EFM experiment to measure photo generated charge.[371] Upon illumination, charge carriers in the photovoltaic material are generated in the sample by the photoelectric effect. The VM-AFM tip causes the photo generated charge carriers to migrate to opposite sides of the active layer. The resulting accumulation of charge changes the capacitance and electrostatic force gradient, in turn causing a resonance frequency shift according to equation 15. By detection and continuous capture of the frequency shift using a PLL and home built detection platform. At each location on the sample a charging curve can be measured and the local charging rate in the material can be determined (Figure 34(b)). In this implementation, the authors claim that tr-EFM has a spatial/time resolution of ~100 nm/~100 μs.[371, 372] The limitation in maximum achievable temporal resolution being the inherent signal/noise stability limit of frequency shift feedback loop. In more recent work by the same group,[373, 374] this bottleneck has been overcome by adopting an OL (i.e. feedback-free) approach to obtaining similar information about fast local force transients. This new method was demonstrated to be capable of discerning local dynamics with transient rise times as short as ~100−200 ns and in certain implementations allowing sub cycle detection of dynamic events by analysis of the raw photodetector stream without demodulation. In fast free (FF) tr-EFM[374] the cantilever oscillation is digitized, while a light pulse is triggered to initiate the local dynamics of interest. A triggering circuit is employed to phase lock the trigger event to the cantilever motion so that the trigger always occurs at the same point in the cantilever oscillation, thereby improving the efficacy of signal averaging and thus the time resolution. In a post processing step, the cantilever motion is analyzed using numerical demodulation of the digitized cantilever signal. The instantaneous phase (and instantaneous frequency) is then extracted from the demodulated signal via a Hilbert transform of the cantilever position vs. time data.

FFtr-EFM has been used to measure photoexcited charge in polymer films with a lateral resolution of 100 nm and temporal resolution on the order of 100 µs. tr-EFM has proven particular useful for making local quantum efficiency maps as a function of material properties and preparation,[371] degradation,[372, 382] and excitation wavelength.[382] Since this technique utilizes the derivative of the instantaneous phase with respect to time to yield the instantaneous frequency of the AFM cantilever quantitative extraction of capacitance gradient or surface potential is lost in favor of information on recombination rates. [374]

### 4.3.2  Time resolved-KPFM

While conventional KPFM has been useful in characterizing the surface potential of a variety of static or quasi-static processes in organic electronic devices, parameters such as the time averaged surface potential and capacitive gradient fail to provide direct information about the local electronic and ionic transport and their dynamic behavior. To address this limitation, Strelcov *et al.*[383] developed time-resolved (tr)-KPFM for mapping surface potential in the both space and time domains. Schematics of the tr-KPFM, is shown in Figure 35(a). Measurements are taken in grid mode at a specified distance above the device surface. At each point of the grid, a two-step low-frequency probing voltage waveform is applied to the lateral electrodes, whereas the tip bias is modulated by a high-frequency ac signal (ω), as shown. In this way, the dynamic response of the cantilever is perturbed by bias induced activation (step P) and relaxation (step R) of electrochemical processes. As the surface potential at each grid point between the electrodes changes over time, so does the tip surface force, which is reflected in the deflection signal and is detected by the lock-in. This measurement is repeated on a dense grid across the device surface and calibration of the surface potential can be performed over the flat grounded electrode. An example of the resultant CPD map reflecting surface ionic motion is shown in Figure 35(b).

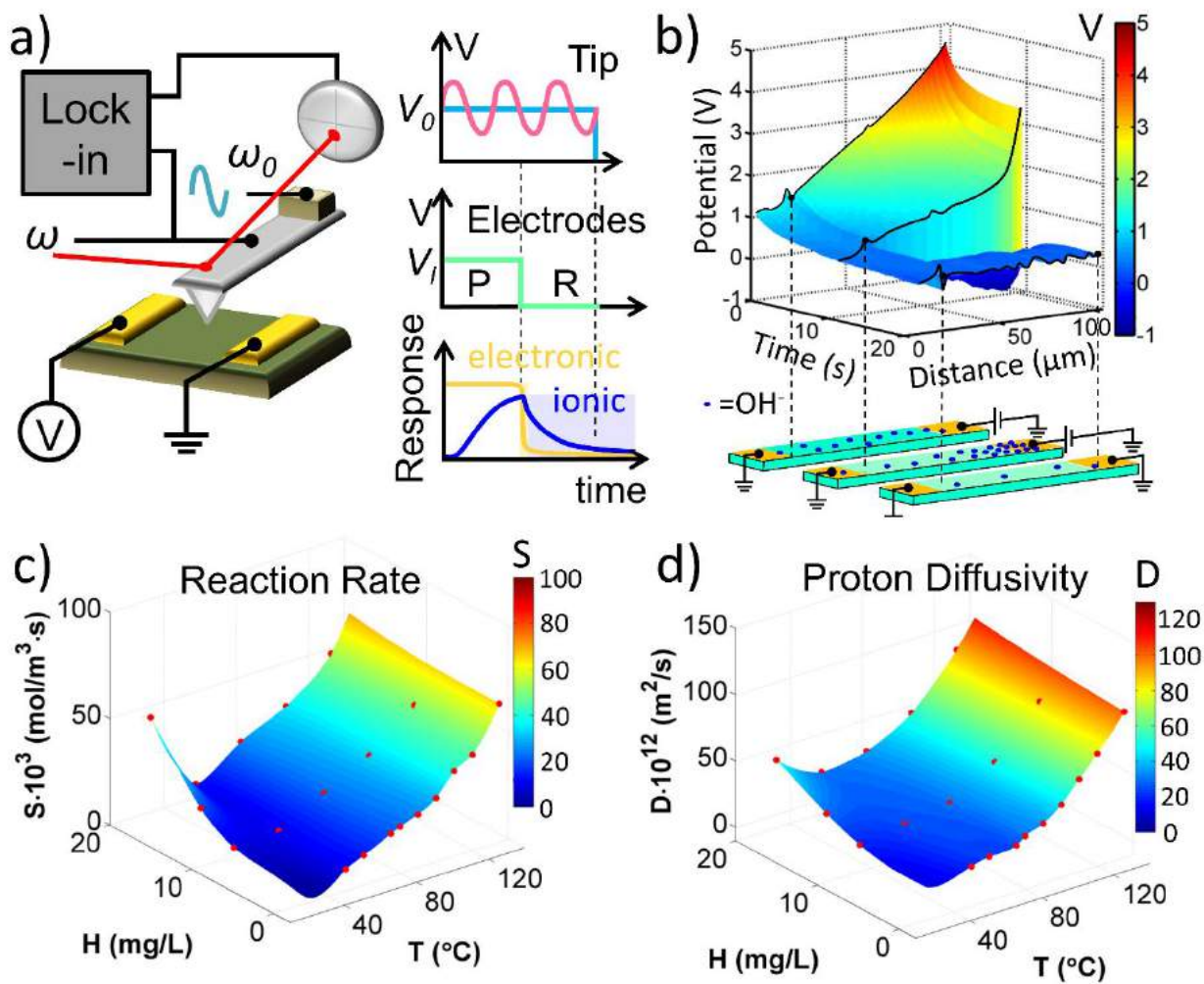

Figure 35. Time-resolved KPFM: a) A schematic of tr-KPFM method showing a sample with two lateral electrodes. The AFM tip oscillations are influenced by local CPD and are monitored via a lock-in amplifier. Graphs on the right display voltages applied to the tip and electrodes during the polarization (P) and relaxation (R) periods. Separation of electronic and ionic responses is possible due to the difference in their time constants. b) an example of tr-KPFM response in a Ca-doped BiFeO3 sample. CPD (averaged over dimension parallel to electrodes) as a function of interelectrode distance and time changes as negative surface ions (presumably, adsorbed OH- groups) are re-distributed in response to applied lateral electric field. c) and d) modeling the measured tr-KPFM response allows separation of different contributing mechanisms. Here, phase diagrams show how water splitting reaction rate and proton diffusivity in nanostructured ceria thin film change as a function of temperature and air humidity. *Seeking permissions.*

This implementation has been shown to successfully separate surface vs. bulk ionic activity in the time domain on insulating surfaces, where flowing currents are well below the detection limit of modern current amplifiers. This noncontact, scanning probe microscopy technique enables detection of surface potential with submicron spatial resolution, on the ten milliseconds to tens of seconds time scale.[368] tr-KPFM has been successfully used on memristive and ferroelectric samples revealing the details of polarization switching.[369, 383] More recently, this approach has proven useful in probing electrochemical reactions on platforms consisting of nanostructured ceria where the surface potential change over time could be attributed to the transport of protons and hydroxyl groups.[370] Modeling of tr-KPFM responses in this system allowed determination[384] of different contributing mechanisms and extraction of relevant electrochemical parameters (such as water splitting reaction rate and proton diffusivity on ceria surfaces) from the tr-KPFM data (Figure 35(c) and (d)).

### 4.3.3 Pump-probe-KPFM

Recently developed Pump-probe (pp)-KPFM [375] has been demonstrated to simultaneously detect the time averaged CPD and nanosecond changes in surface charges. In this approach, the enhanced temporal resolution is achieved by operating FM-KPFM in an electrical pump-probe scheme. As shown in Figure 36, short probing pulses are modulated by a slower sinusoidal envelope, and synchronized to electric pump pulses applied to the device under investigation. This scheme allows tracking charge carriers injected into the transport channel where the achievable time resolution is determined by the width of the probe pulses. A second KPFM loop is used to minimize the average electrostatic force, at the same time, providing a measure of the time-averaged potential distribution as in classical KPFM. This approach has been used to spatially map "speed bumps" in organic field-effect transition devices.[376] pp-KPFM relies on single frequency heterodyne detection and bias feedback, and hence is subject to the standard assumptions required for KPFM operation, albeit adding the benefit of temporal dynamic information.

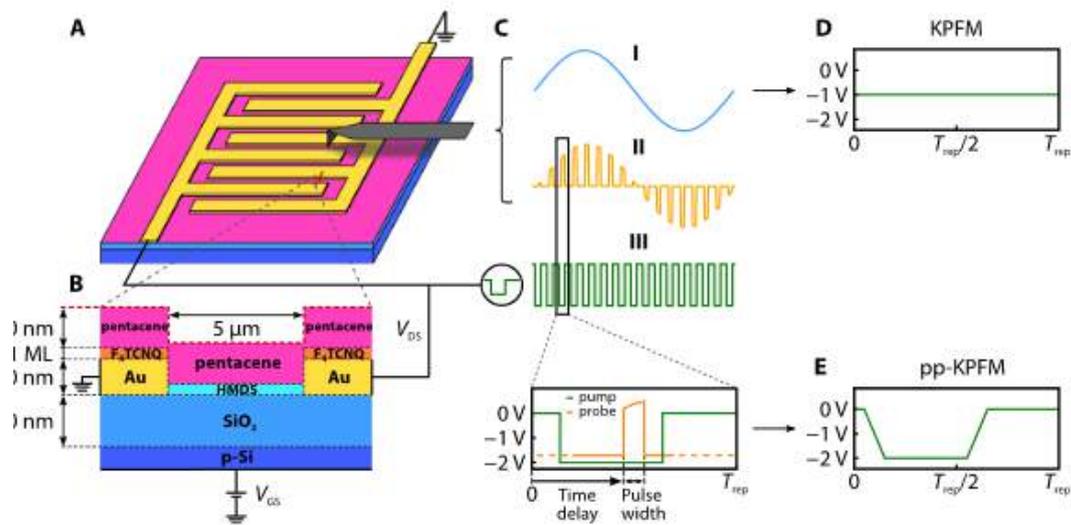

Figure 36. Schematic describing the operation of PP-KPFM on a pentacene-based bottom-gate organic field effect transistor (OFET). (b) Cross-sectional view across the OFET transport channel displaying the different fabrication layers. (c) In pp-KPFM a fast electrical pulses with a slower sinusoidal envelope (signal II) is used. The electrical pump pulses (signal III) are synced with the pp-KPFM tip excitation and applied to the device in order to measure surface potential dynamics. (d) Standard KPFM does not resolve the dynamics providing only the time averaged surface potential, whereas (e) the dynamics can be resolved by pp-KPFM. *Seeking permissions.*

### 4.3.4   G-Mode KPFM

At their core, traditional KPFM techniques rely on heterodyne detection, which has the effect of limiting both the information captured (i.e. attenuation of all response outside detection frequency) as well as imposing bandwidth limitations (i.e. LIA time constant). In 2015, general acquisition mode (G-Mode) [143, 385, 386] SPM has been developed, which allows full exploration of the cantilever deflection with extremely high temporal resolution.[385-387] In the short time since its inception, the G-Mode approach has been shown to have advantages for tapping mode AFM,[385] PFM,[386, 388] MFM [98] and DH-KPFM.[143] The approach overcomes the drawbacks of existing SPM techniques by capturing and storing cantilever dynamics at high speed and over a broad range of frequencies trading heterodyne detection and closed loop feedback for large multidimensional data sets.

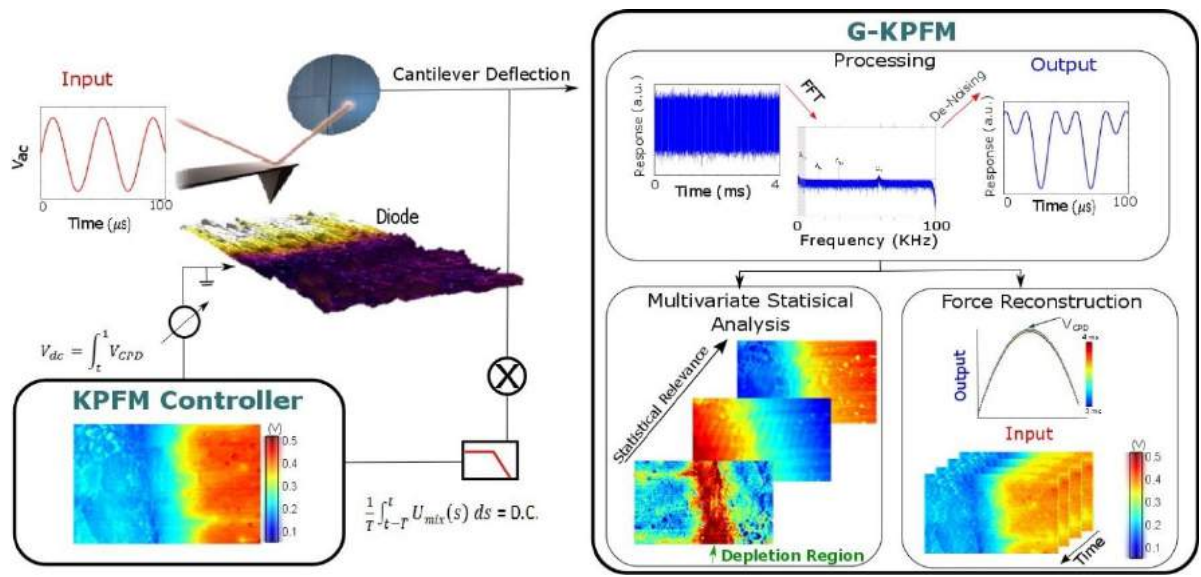

Figure 37. Principles of G-Mode SPM. G-mode captures the complete raw signal from the photodetector thereby adding the time dimension for each spatial pixel. In contrast, the traditional LIA paradigm integrates the product of the excitation and response signals over the time constant and produces a single pair (amplitude and phase) of values at each spatial pixel. *Seeking permissions.*

G-Mode KPFM, as depicted in Figure 37, works by capturing, storing and compressing the AFM photodetector signal at the sampling rate limit (~4-10 MHz). Combining full data acquisition with adaptive filtering and statistical cleaning methods (e.g. Principle component analysis) has been shown to be an effect method for recovering a densely sampled permanent record of the dynamic cantilever trajectory. A range of analysis methods including emulation of conventional heterodyne methods,[143] as well as physics-[377] and information-based[98] approaches have been demonstrated to extract quantitative measurements of electronic properties (i.e. surface potential and capacitance gradient) as well as data mining the cantilever dynamics for additional hidden information.

As a proof of principle, G-Mode KPFM was first shown to be able to emulate KPFM, or more specifically DH-KPFM, using a purely digital approach.[143] This classical analysis approach of G-Mode KPFM data was shown to have significant advantages in terms of increased flexibility in data exploration across frequency, time, space, and noise domains.[143] An obvious advantage to collecting and storing an entire spectrogram, such as in G-Mode, is the ability to unravel the cantilever response at each frequency, without repeating the measurement. This has recently leveraged to simultaneously probe magnetic and electrostatic properties of the high entropy (HEA) alloy CoFeMnNiSn.[98] In traditional approaches subsequent MFM and KPFM measurements would have to be performed to assess local electric and magnetic domains separately, as show in Figure 38(c,d). However, the availability of the full cantilever spectrum in G-Mode KPFM can be leveraged to map multiple material properties simultaneously, as shown in Figure 38.[98]

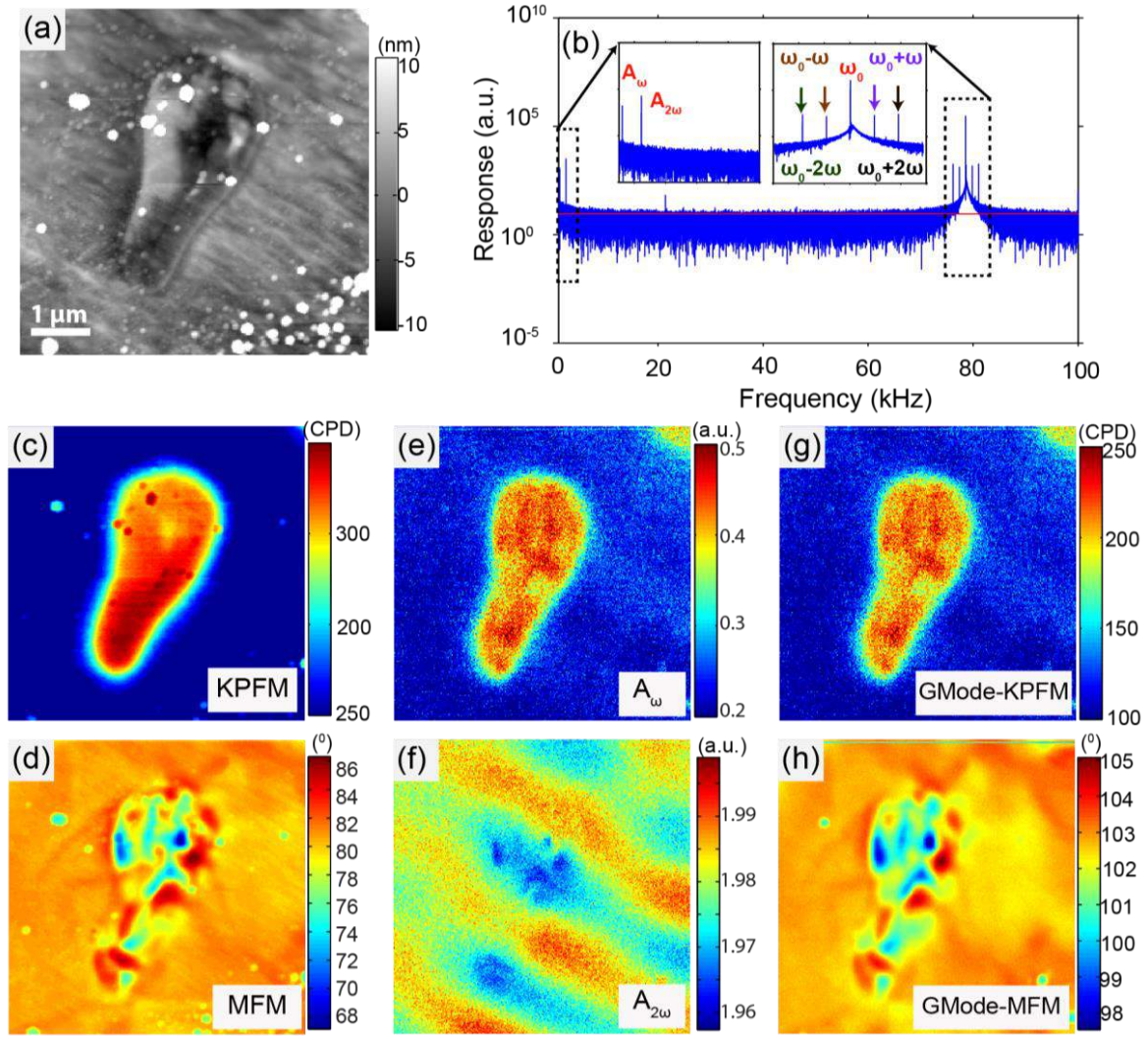

Figure 38. (a) AFM topography image of a CoFeMnNiSn high entropy alloy (HEA). (b) Single pixel photodetector response, as translating into frequency domain using Fourier transform, demonstrating complex tip-sample interaction with multiple harmonics and sidebands evident. Variation in (c) CPD and (d) magnetic domains determined using conventional KPFM and MFM consecutively. Digital LIA analysis of G-Mode data showing (e) $A_\omega$ and (f) $A_{2\omega}$ used to determine the open loop (g) calculated CPD. (h) The simultaneously recorded magnetic domain response from LIA analysis of the mechanical excitation at fundamental resonance ($\omega_0$).

In a different and more powerful approach, the fast temporal component of G-Mode data was harnessed to allow direct recovery of the parabolic bias dependence of the electrostatic force (Figure 39). Fitting the bias dependence of the response then leads to readout of the CPD and capacitance gradient at speeds corresponding to the period (or half period) of the AC oscillation. In this way, G-KPFM is much faster than conventional methods, where demodulation of the response is typically performed over hundreds of periods of oscillation. As such, G-KPFM has the ability to quantify important information beyond a temporally averaged CPD (as in classical KPFM) relating to fast charge dynamics.

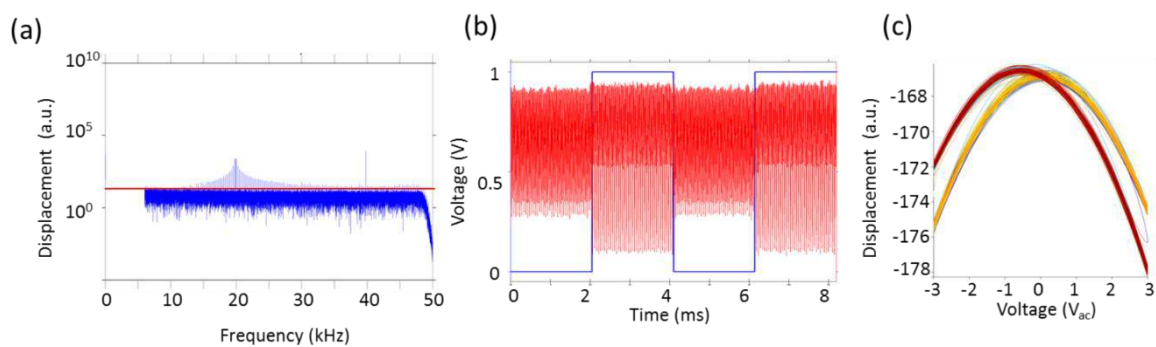

Figure 39. (a) G-Mode KPFM data in the (a) frequency domain and (b) time domain after noise thresholding (red line in (a)) and low pass (50 kHz) and band pass filters (0.1–6 kHz) were applied to the raw data. The excitation voltage consisted of a single frequency $V_{ac}$ superimposed on a square wave modulation (DC bias applied to the tip is shown in blue) resulting in the presence of many harmonics across a broad frequency range. (c) Recovered parabolas and inset of shows the measured CPD determined from parabolic fitting of the parabolas correspond to a time resolved measurement of the applied DC bias.

# 5    The solid-liquid interface

## 5.1    Electric double layer theory

As an introduction, we concentrate our theoretical description of an idealized 'flat' blocking electrode in which no penetration of mobile ions across the solid-liquid interface can occur (i.e., no Faradaic current). This assumption generally works well at high-frequencies, where ions do not participate in conduction phenomena.[389] In essence this simplifies to purely EDL capacitive interactions, neglecting pseudocapactive effects, reactions and the influence of any topographical microstructure.[390] A more comprehensive discussion on the theory behind EDL charging[9-11] as well as electrochemical processes at non-blocking electrode interfaces[391-393] and in the presence of porous electrodes can be found elsewhere.[390, 394, 395]

### 5.1.1   Electric double layer under equilibrium

When an interface is exposed to a polar liquid, such as water, the interface will become charged due to a combination of ionization of surface groups, ion adsorption, and/or incomplete dissolution of surface ions.[396] This charged interface will in turn affect the distribution of ions in the solution which are in proximity to the interface, such that there will be a greater concentration of counterions than co-ions.[397] The earliest theoretical study of the formation of ion structure at the solid liquid interface, is generally attributed to Herman von Helmholtz,[40] who in 1853 first coined the term "electric double layer". Helmholtz postulated that the entire countercharge in the EDL was confined to a single compact, rigid layer, immobilized on the electrode surface by electrostatic attraction such that the surface charge of the electrode was completely neutralized.[40] A visual representation of the EDL as described by Helmholtz can be seen in Figure 40(a). Almost two decades later, Louis Georges Gouy[41] and David Leonard Chapman[398] independently extended the Helmholtz EDL model by taking into consideration that mobile ions near a charged surface are driven by the coupled influence of ionic diffusion and electrostatic forces[9, 399] They suggested that the capacitor plate arrangement be replaced by a diffuse cloud of counterions adjacent to electrode surface and extending into the bulk solution, as shown in Figure 40(b). Soon after the introduction of the GC model, Stern[42] combined the Helmholtz model with the GC model into a single theory, referred to as the Gouy Chapman Stern (GCS) model. He proposed that at high voltages ($\psi > \psi_T$), it must be possible to form a compact layer of adsorbed ions at the surface, as first depicted by Helmholtz. Stern's model, depicted in Figure 40(c), introduced a finite molecular size into the model by requiring that there was some distance of closest approach to the surface, commonly called the compact Stern layer.[9]

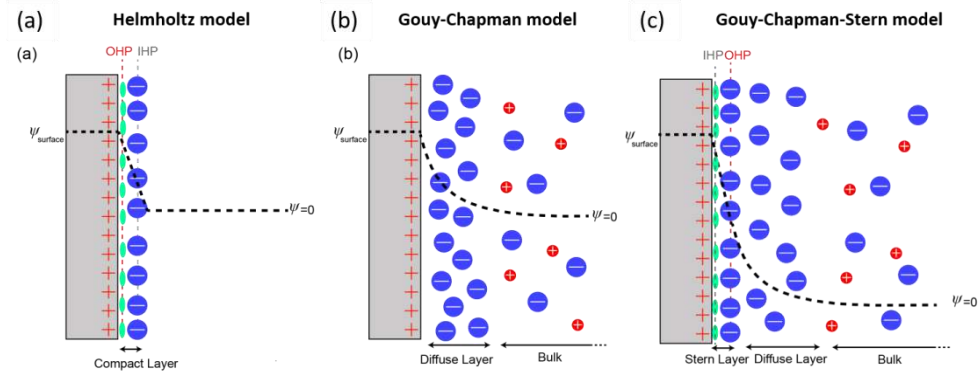

Figure 40. Schematic representations of the (a) Helmholtz, (b) Guoy-Chapman and (c) Guoy-Chapman-Stern models of the electric double layer.[400] *Seeking permissions.*

In the mathematical description of the GC model, the mean-field Poisson-Boltzmann (PB) theory is used to calculate the potential, electric field and counterion density as a function of separation from the surface.[401] In the PB theory, ions are treated as point charges, where at equilibrium the ion concentration is described by a Boltzmann distribution:

$$c(x_2) = c(x_1)\exp\left\{\frac{-z_i e[\psi(x_2) - \psi(x_1)]}{k_B T}\right\} \quad (27)$$

where $\psi$ is the local electrostatic potential, $z_i$ is the valency, $e$ is the elementary charge, $c_i$ is the molar concentration, and $k_B$ is the Boltzmann constant. Poisson's equation, which is derived from Gauss' law, is used to relate charge ($\rho_e$) and electrostatic potential. The mean field approximation of the Poisson equation can be written as:

$$\nabla^2 \psi = -\frac{\rho_e}{\varepsilon_0 \varepsilon_r} = \sum_i \frac{z_i e c_i}{\varepsilon_0 \varepsilon_r} \quad (28)$$

Summing the Boltzmann distribution (equation (27)) from the surface of the electrode to the bulk of the solution ($x=0$ to $x\rightarrow\infty$), and substituting with equation (28), we arrive at the Poisson-Boltzmann (PB) equation, describing the equilibrium distribution of a thin diffuse layer of charge near a solid surface.

$$\nabla^2 \psi = \sum_i \frac{z_i e c_i}{\varepsilon_0 \varepsilon_r}\exp\left\{\frac{-z_i e \psi}{k_B T}\right\} \quad (29)$$

The nonlinear PB equation is difficult to solve analytically owing to the summation of the charge density terms as well as their strongly nonlinear behavior; however for surfaces of low surface potential ($\psi_0 < 25$ mV) the potential, $\psi$, as a function of x, can be estimated by the Debye-Huckel approximation[400]

$$\psi \approx \psi_0 e^{(-\kappa x)} \tag{30}$$

where the decay of the potential is controlled by the Debye length, $\kappa^{-1}$, which is a measure of the thickness of the EDL and is defined for a symmetrical electrolyte by

$$\lambda_D = \sqrt{\frac{\varepsilon_0 \varepsilon_r k_B T}{2 z_i^2 e^2 C_b}} \tag{31}$$

where $\varepsilon$ is the permittivity, $k_B$ is Boltzmann's constant, T is temperature, e is an electron charge, $z_i$ is the valency and $C_b$ is the concentration of counterions. This low potential limit is defined as the regime within which the electrostatic potential is far smaller than the thermal voltage (i.e., $\psi << \psi T = k_B T/e \approx 25$ mV).

### 5.1.2 Breakdown of the Poisson-Boltzmann model

The PB model remains the most popular model for describing equilibrium EDL potentials to date. However, application of the PB model is only valid under assumptions of; (i) ions act like point like charges, interacting through their mean field approximation, (ii) the permittivity is constant throughout the bulk and EDL regions and (iii) the solvent is treated as a continuum.[9] All three of these conditions are likely to be violated under conditions of high surface potential, high ion concentrations and when using VM techniques to probe the solid-liquid surface in confined geometries (e.g., VM-AFM based measurements) where (i) the nonlinear PB model results in unphysically high concentrations of counterions at the interface [10, 11, 402] requiring additional consideration of non-linear steric effects [402, 403] and ion depletion at the solid-liquid interface,[10] (ii) confined solvents and hydrated ions have been shown to exhibit non-continuum behavior [404-406] and (iii) polar solvents (e.g., water) have been shown to have varying permittivity adjacent to interfaces due to orientation [407-412] and depletion effects.[410, 413] Numerous modifications have been made to the PB equation to account for steric constraints at high surface potentials[414, 415] (see Grochowski [416], Kilic [10, 11] and Bazant [9, 403, 417] for reviews).

### 5.1.3 Electric double layer dynamics

Ultimately, the establishment of the EDL is not an instantaneous process. EDL dynamics are governed by charge driven diffusion and electro-migration of mobile charge species from the bulk electrolyte[9] as well as a diverse set of electrochemical reactions that may take place at the solid-liquid interface in the case of non-blocking electrodes.[418] Migration (or electro-migration) is relevant only to charged particles and involves the transport of ions driven by an electric field or more specifically by a violation of local charge neutrality. Diffusion, on the other hand affects both charged and neutral species, where the net diffusion of ions is governed by the chemical concentration gradient. Notice that both processes are intrinsically related, but controlled by fundamentally different parameters. The charge relaxation of the EDL is generally described by the Debye time, $\tau_D = \lambda^2/D$, where $\lambda$ is the Debye screening length and $D$ is the ion diffusivity across $\lambda$, [9] whereas, the bulk diffusion takes place on much longer timescales, $\tau_L = (L/2)^2/D$ where L is surface separation. Here, diffusivity is used as a matter of convenience. Charging of the EDL capacitor happens through the electrical resistance of the bulk electrolyte, and can be shown to be governed by; $\tau_C = \sqrt{\tau_D \tau_L} = \lambda(L/2)/D$ .[9]

In most biological or energy applications it is crucially important to have a strong understanding of the EDL charging processes and their individual timescales. This necessitates a clear description of the bias and time dependent migration and diffusion of ions, particularly at larger voltages ($\psi \gg \psi_T$) or at relatively high ion concentrations (e.g., 100 mM $-$ 1 M). To date, the classical PB theory combined with the Nernst Plank (NP) equations has been the most widely used model to describe electrolyte behavior and ion transport dynamics to a time-dependent applied voltage. This combination is referred to as the Poisson-Nernst-Planck (PNP) equations.[246, 254, 256].

$$\frac{\partial c_i}{\partial t} = -\nabla . F_i \quad = D_i[\nabla^2 c_i + \frac{z_i e}{kT} \nabla . (c_i \nabla \psi)] \tag{32}$$

where the Einstein relation, $D_i = b_i/k_B T$, relates ion mobility to its diffusivity. The PNP equations are commonly used to model ion transport in biological systems[249] or in micro- or nano-fluidic devices.[419] General solutions of the PNP equation, however, are only possible upon linearization under conditions where the potential satisfies the Debye-Falkenhagen equation.[10, 420]

$$\frac{1}{D} \frac{\partial \rho}{\partial t} = (\nabla^2 - \kappa^2)\rho \tag{33}$$

where $\rho$ is net charge density.[421]

The validity of the low potential assumption for measurements where the applied voltage is typically $\ll \psi_T$ remains unknown; however this simplification continues to be used for lack of a simple mathematical alternative.[422, 423] Understanding the electric field at low frequencies requires numerical treatment of the PNP equation. Again, the fundamental assumption which must be made here is that the electrolyte acts like a dilute solution of point like charges interacting through mean-field electrostatic force.[10, 417] Under these assumptions the electrolyte dynamics can be described by the Nernst-Plank (NP) equation; In the high potential limit ($\psi >> \psi_T$), the assumption of a dilute solution of point like charges, breaks down near the electrode surface. This has motivated researchers to modify the standard equations to overcome significant limitations in concentrated solutions and at large surface potentials.[11, 424-426] Only recently, has the nonlinear transient electrochemical response to large dc [10, 11] and ac voltages [424, 427] been analyzed. In addition, whilst modern numerical and statistical methods can be used to develop more complex theories [422] experimental verification remains challenging. Traditional electrochemical observables such as impedance, differential capacitance, zeta-potential, surface charge and dielectric constants for nanoscale objects are experimentally accessible but the need to quantify the influence of timescales, local topography, solvation/hydration forces and non-linear changes in permittivity adjacent to interfaces are persistent challenges to a fully verified system of equations.

Here, we have only considered idealized flat blocking electrodes. However, porous electrode materials that facilitate Faradaic reactions at their interface, are attractive materials for a large range of industrial applications [428] including for battery devices [429, 430], desalination processes,[431] and electrochemical capacitors [432, 433]. It is likely that VM-SPM will play a key role into probing these processes *in situ* at the nanoscale to increase our understanding of these technologically important electrochemical processes.[434]

## 5.2 Electric double layer forces: DVLO theory

It has been recognized since the early 20th century that colloids in an aqueous medium coagulate after a threshold concentration of salt is introduced and that the valency of the ions present influences dispersion efficiency. In the 1940s, Derjaguin and Landau [435] and Verwey and Overbeek [436] independently developed a quantitative theory which could be used to explain this phenomena. In their theory, the repulsive electrolytic entropic force of the EDL was summed with the attractive van der Waals interactions (including Keesom, Debye, and London dispersion forces)[401] to describe the stability of, and forces acting between, charged colloidal systems in aqueous environments. This continuum theory, which became known as the DLVO theory, has been used since its discovery as a tool for describing the behavior of a vast array of systems, and remains a valuable predictive theorem in use today.[437, 438]

The forces present can be understood when one considers that the local ionic concentrations due to the EDLs formed between two charged objects, resulting in a build-up of osmotic pressure, in turn generate an electrostatic force between the two objects. This electrostatic force acts over distances that are comparable to the Debye Length and the strength of these forces increases with the magnitude of the surface charge density (or the electrical surface potential). For two similarly charged objects, this force is repulsive and decays exponentially at larger distances. For unequally charged objects and for similarly charged objects at short distances, these forces may also be attractive. At low ion concentration, the double-layer repulsion is strong enough to keep colloidal particles apart, but the electrostatic repulsion becomes screened with increasing ion concentration and ion valency. The coagulation of charged colloids will happen after some threshold concentration has been reached at which point the van der Waals attraction overcomes the repulsive electrostatic barrier.[439]

The EDL force can be calculated using continuum theory, based on the GCS theory for EDLs. In DVLO theory, the potential distribution is determined from the PB equation and generally using one of two boundary conditions. Either it is assumed that the surface charges remain constant (constant charge, cc)[440-442]

$$F_{el}^{cc} = \frac{2\pi R \lambda_D}{\varepsilon \varepsilon_0} \left[ 2\sigma_S \sigma_T e^{-D/\lambda_D} + (\sigma_S^2 + \sigma_T^2) e^{-2D/\lambda_D} \right] \qquad (34)$$

where $\sigma_S$ and $\sigma_T$ are the surface charge densities of a flat plate and a spherical particle of radius $R$, or that the surface potentials potentials of the flat plate $\psi_S$ and the particle $\psi_T$ remain constant (constant potential, cp)[443]

$$F_{el}^{cp} = \frac{2\pi R \varepsilon \varepsilon_0}{\lambda_D} \left[ 2\psi_S \psi_T e^{-D/\lambda_D} - (\psi_S^2 + \psi_T^2) e^{-2D/\lambda_D} \right] \qquad (35)$$

In both cases, it is assumed that the surface potentials are low ($\psi_S, \psi_T < 50$ mV) and the geometry has been adapted for that of an AFM tip interacting with a flat interface.[444] These boundary conditions have a strong influence on the electrostatic force at small distances $z \leq \lambda_D$. For example, two surfaces with constant charge of equal sign always repel each other for $z \to 0$. Two surfaces with constant potential are attracted for $z \to 0$ even when the surface potentials have the same sign (except for the hypothetical case that the potentials are precisely equal in magnitude and sign).[445] Experiments often demonstrate a force-distance profile that occurs between these two boundary conditions.[446, 447] Ninham and Parsegain developed a third boundary condition of charge regulation,[440] where the surface charge would vary as a function of separation, which generally matched experimental data.[446, 447]

Recent papers from the group of Mugele have resolved the internal structure of the Stern layer using AFM [448] and combined charge regulation DLVO theory with the concept of tip-induced adsorption/desorption of ions and protons at small tip-sample separations.[449] These results highlight the need to account for the tip-induced suppression of the dissociation kinetics and the approach developed can be applied to quantify adsorption/desorption equilibrium constants.[449]

While the simplicity and elegance of DLVO theory is appealing, an inherent assumption of DLVO is that the system can be described as a continuum model with point charges in equilibrium, and thus fails to describe any system which exhibits discretized, dynamic or ion specific behavior. Furthermore, DLVO theory relies on being able to accurately describe the behavior of the electric potential of the EDL by the nonlinear PB equation, which is a valid approximation only under conditions of low ($< 50$ mV) potential that limits its applicability to salt concentrations below 10 mM.[439] These known limitations, combined with observable experimental results, have inspired many authors to extend DLVO theory in the decades since its inception by accounting for the effects of solvation/hydration forces,[450-454] surface roughness,[455-458] surface charge heterogeneity,[459, 460] specific ion effects[461-464], and dynamics.[465] For reviews of extensions applied to account for non-DLVO forces see Ninham [466] and Bergendahl [467].

Despite the widespread adoption of DLVO theory and the work of many authors to both extend and verify its application, some argue that the fundamental premise of the theory, that a nonlinear PB description of the EDL can be combined with the linear (Lifschitz) theory for van der Waals forces is inherently flawed.[468] For an interesting historical perspective on the development of DLVO theory and many of the non-DLVO extensions as well as a recent overview of the limits of its application see the work of Ninham.[466, 469]

## 5.3 Experimental investigations of DLVO

Huisman and Mysels were the first to measure double layer forces, and their decay according to debye length, using soap films.[470] Experimental insight into the equilibrium behavior of aqueous systems on the length scales of the EDL was greatly enhanced with the development of the liquid SFA.[471] Based on seminal work of Israelachvili [401, 471-473] it was shown that by fitting the force-distance relationship to the DLVO model,[474] it was possible to quantify EDL interactions [444, 475] with agreement between experimental data and theory for larger separations ( $z > 1$-2 nm) for a range of systems. SFA experiments were also the first to provide direct evidence of deviations from DLVO theory at small separations [406, 476] due to solvation/hydration forces,[453] hydrodynamics,[477] specific ion effects,[451] and steric hindrance.[478] Soon after measurements of DLVO interactions using SFA became dominant, researchers began to adapt these measurements to its successors,[479] such as the atomic force microscope (AFM),[480, 481] optical tweezers [482] and others [483, 484] for experimental verification of models and theories underlying the interaction forces and EDL structures.[444, 445, 485-496]

The development of AFM[497] yielded an important tool for imaging surfaces at the nanoscale [498] due to a unique combination of pN force sensitivity combined with atomic scale lateral resolution.[499, 500] Unlike the SFA, which generally requires optically transparent and molecularly smooth substrates, AFM was the first instrument which could map topography and interaction forces at the nanoscale for virtually any system. The transition to imaging at the solid-liquid interface[481] occurred soon after the development of AFM which then led to authors demonstrating a unique ability to probe the interaction forces [501] for a wide range of samples from inorganic clays and crystals, to biological surfaces such as cells, proteins, DNA and bacteria.[502-505] This ability of AFM to operate with virtually any substrate / liquid combination led to a new and much expanded opportunity to verify the applicability of DLVO theory for a wide range of systems.[506] Since AFM measurements yield interaction forces as a function of tip-sample separation, it became possible to probe a wide range of parameters including surface mechanical properties,[507] adhesion,[501] Debye length,[506] Hamaker constants[501, 507-509] and hydrodynamics.[510]

Initial investigations were limited to the materials suitable for cantilevers to be made from and experiments typically consisted of a sharp tip interacting with a flat surface. Butt[511] measured the interaction of a silicon nitride cantilever with a mica surface as a function of KCl concentration and found good agreement with the predicted Debye length. The interaction of an alumina tip on mica as a function of pH demonstrated a transition from repulsive to attractive interactions as predicted by DLVO theory. Force vs. distance curves obtained at different $MgCl_2$ concentrations with a silicon nitride tip on mica demonstrated a divergence from DLVO theory at high salt concentrations, attributed to hydration forces, in agreement with the SFA observations of Pashley.[512] pH dependence on force-distance curves can be seen in Figure 41.

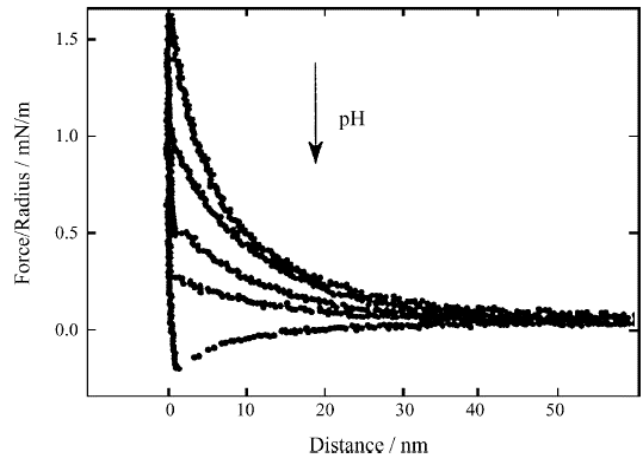

Figure 41. AFM force measurements in an aqueous medium recorded with a colloid probe above a titania surface as a function of pH. Electrostatic repulsing decreases with decreasing pH. When the pH is 3.0 there is an overall attraction.[513, 514] *Seeking permissions.*

Such deviations at small separations were later investigated in detail as AFM instrumentation improved to be able to routinely resolve force distance curves at the 1-2 nm length scale and allow hydration layers to be detected (see Figure 42).[448, 515-517]

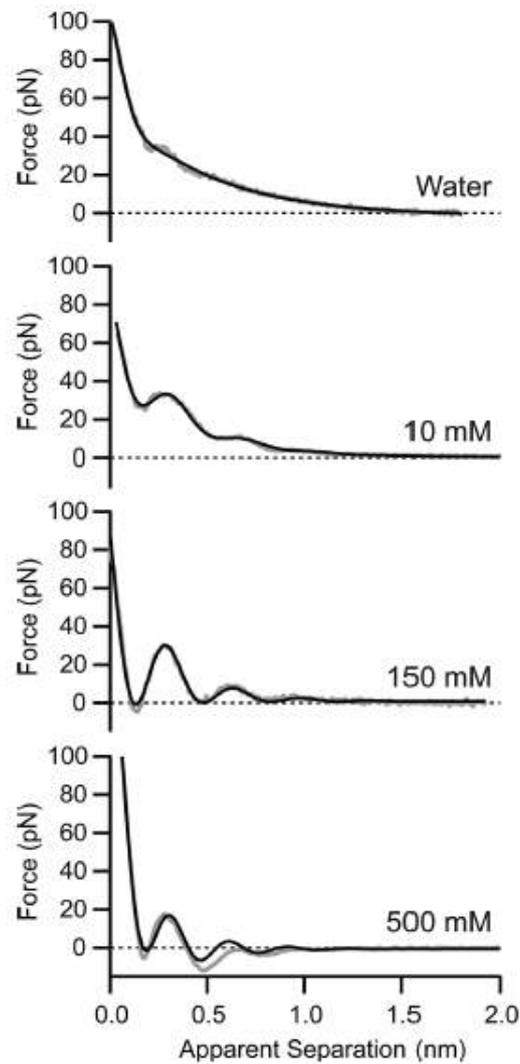

Figure 42. Force-distance curves measured in water and 10, 150, and 500 mM NaCl solutions showing the detection of hydration layers.[515] *Seeking permissions.*

With the development of the colloid probe technique[487, 518] the interactions that could be measured by AFM vastly expanded to any material with a spherical geometry that could be attached to a cantilever, providing significant advances in the understanding of DLVO interactions in fields including surface science, materials engineering and biology.[445] Several groups have since demonstrated measurements of charge density and Debye length using the AFM as a function of pH, electrolyte type, and concentration; finding agreement with the DLVO theory for large separations and in the low potential regime.[445, 485, 488, 492, 495, 518-520]

In addition, the flexibility of AFM further allows the investigation of EDL interactions with solid surfaces with surface potential well above 100 mV using VM-SPM techniques. This can be achieved by applying a voltage to a conducting sample surface or to a conductive cantilever and tip. Ishino *et al.* used this methodology to observe repulsive interactions at negative potentials and attractive interactions at positive potentials between a gold-coated tip and a steric acid monolayer on a gold electrode.[521] More conventionally, the experimental setup consists of an insulating probe and a conducting sample, which serves as a working electrode for the study of the effects of bias on the EDL.[490, 513, 520, 522-525] Prieve *et al.* demonstrated that by applying an ac potential to a conducting sample surface, a colloidal particle would move in a sinusoidal motion due to the electroosmotic flows generated in the EDL.[526]

Beyond measuring the distance dependency using a single point force curves, it is also interesting to investigate how tip-sample forces change from one location on the sample to the next. The AFM is an ideal technique for such investigations at it can acquire isoforce and/or isodistance images across a surface.[527] Sneddon employed an isoforce technique to map the interactions spatially at three forces (EDL, hydration and Born) and found that the resulting data is a combination of topography and charge density effects.[528] To separate topographical from charge contributions, Heinz and Hoh developed a protocol, called D-minus-D mapping, where isoforce images were collected at different salt concentrations.[529] A similar approach was also employed by Hafner *et al.*[530]

In addition to isoforce and isodistance imaging, AFM can also be used to collect force-distance curves at every point corresponding to a pixel of the AFM image (force-volume). [528, 531, 532] Using this technique, isoforce and isodistance images can be reconstructed from FV data sets to reveal physicochemical spatial variations across a sample. Pixel densities close to those of standard imaging modes can be achieved but the acquisition time required generally prohibits higher than 64 x 64 pixel FV images being acquired due to sample lifetime and drift limitations. FV analysis has primarily been used in the study of biological systems [533, 534] where mechanical properties,[535-538] molecular recognition,[539, 540] and DLVO interactions [541] are used in combination to derive biological function. The force distance relationship for a gibbsite crystal on an oxidized silicon surface in 20 mM NaCl is shown in Figure 43.[448]

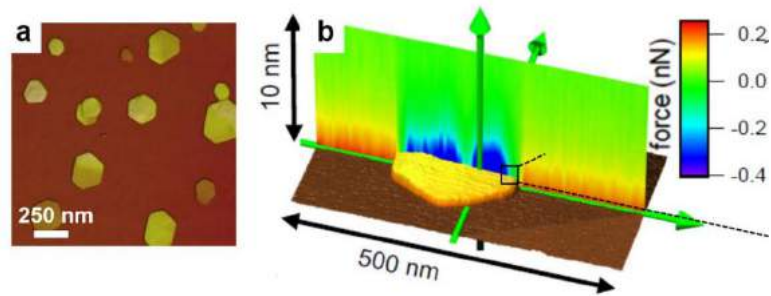

Figure 43. (a) AFM topography image of gibbsite on oxidized Si. (b) Topography and two dimensional force plot obtained from force-distance curves recorded in 20 mM NaCl.[448] *Seeking permissions.*

Spatially resolved AFM measurements have elucidated the adsorption behavior of hemimicelles on hydrophobic substrates [542] as well as charge heterogeneity on colloid supports. Alternatively, the static deflection of the cantilever is monitored as it is scanned near a charged surface.[528, 530, 532] Combined with a simplified model for the tip-sample interaction, this approach has been used to map surface charge densities of a range of biological membranes.[543, 544] Such imaging approaches significantly decrease the acquisition time of the measurement compared to FV approaches, and hence allow increased lateral resolution afforded by the AFM. Beyond equilibrium conditions, probing systems under non-equilibrium conditions has allowed the study of dynamic phenomena such as hydrodynamics [545] and electrokinetics.[546] Recent developments have led to an ability to form 3D force map images of the discretized solvent and ion structure in proximity to the solid-liquid interface at similar scanning speeds to conventional imaging (see Figure 44).[547-549] Note that since the diffusion rate of these species are orders of magnitude faster than that of the measurements, the data generated from such techniques is likely to be more of a probability map than a direct measurement of the position of any individual molecule or ion time.[405] This new ability to locate the position of species at charges interfaces and extending into the EDL has provided significant insight into the solid-liquid interface in recent years with studies of water structure,[550] ion adsorption,[551], EDL structure,[552] protein adsorption,[553] and interactions between water and lipid headgroups.[554] Such studies have so far required hardware and software modifications combined with high performance instrumentation,[555] limiting their widespread adoption. As with the development of all other AFM modes, it is expected that instrumentation enabling 3D force imaging will be commercialized in the near future. One possible candidate for commercialization is PeakForce Tapping™, which already implements many of the required hardware and software modifications.[556]

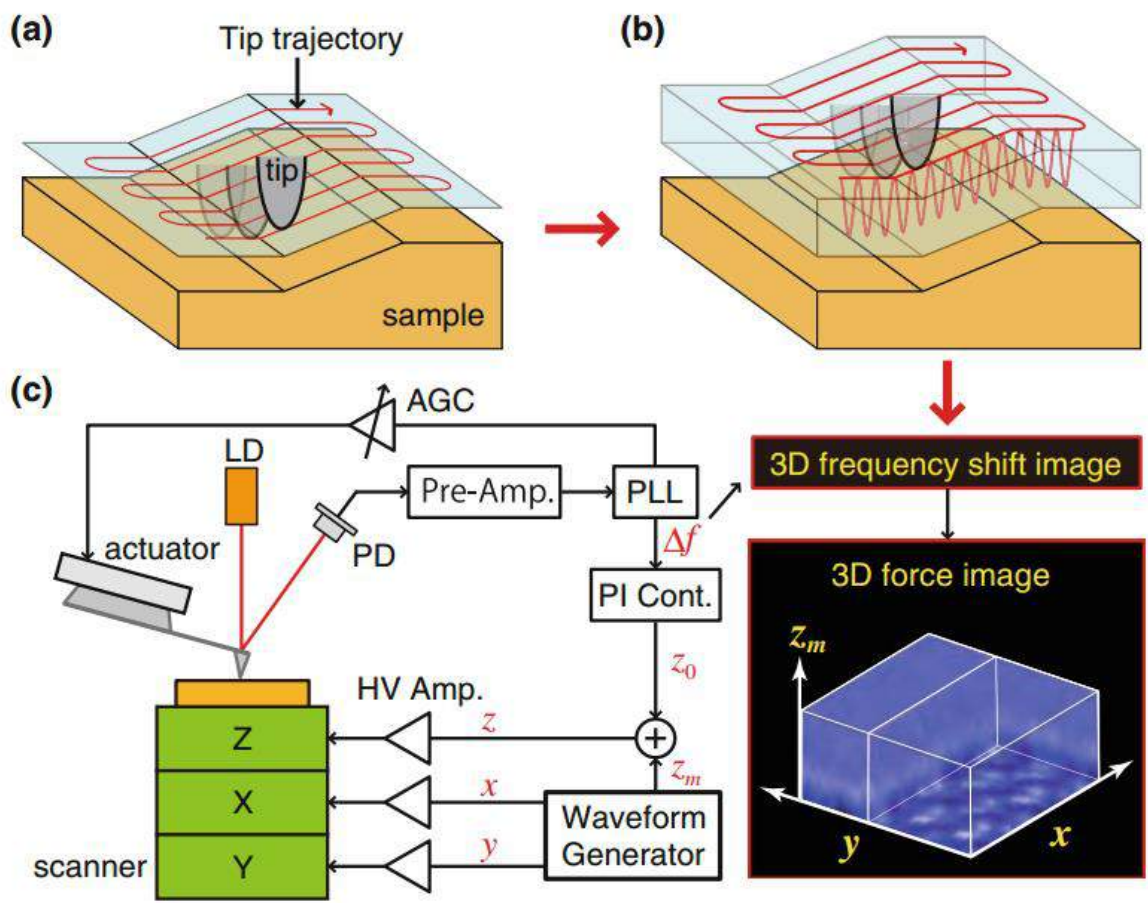

Figure 44. Schematics of (a) conventional scanning and (b) 3D force imaging along with corresponding (c) experimental setup for 3D force imaging and example 3D force image recorded above a 2 nm × 2 nm mica surface in water. [557] *Seeking permissions.*

# 6   Liquid measurements

So far in this review, we have discussed investigation of electronic functionalities at the solid-gas (e.g., ambient and UHV) interfaces using EFM and KPFM. As has been outlined in the previous sections, EFM and KPFM has been widely used to either qualitatively or quantitatively map surface charge density, dielectric properties and surface potentials of nanostructures and molecular systems in ambient and UHV environments. However, in order to study and ultimately understand processes in living systems or to further improve energy storage such as electrochemical batteries, *in-situ* nanoscale electrical measurements are required. Indeed, most fields of science would benefit greatly from a KPFM-like measurement in liquid which could provide electrochemical information equivalent to that available with bulk macroscopic electrochemical measurements but on the nanoscale. The spatial resolution and nature of the information obtained from these approaches place then in a pivotal position to further applications in nanotechnology and bio-nanotechnology, as well opening the opportunity to further our understanding of fundamental structures (e.g., EDL) and processes (e.g., adsorption, corrosion, etc.) pertinent to physics, biology and chemistry.

Despite the importance and broad applicability of liquid EFM/KPFM measurement, limited success has been reported. Clearly, if this was a simple transition from gas to liquid operating conditions such an approach would already be readily available. Interestingly, the macro KP community attempted a similar transition from gaseous to liquid operation as is being attempted today in the KPFM community.[558, 559] Although studies are limited they noted through distance- and time-dependent measurements in a variety of solutions that CPDs could only be reliably made in nonpolar solvents and with deviations observed in polar solvents attributed to their increased conductivity.[559] These findings are indicative of the complex range of electrostatic and electrochemical phenomena that occur at the solid-liquid interface which underpin, e.g., corrosive processes, charge storage in the EDL, interactions between cells and their environment, etc. In the following section, we describe the various implementations of KPFM-based approaches for measuring surface potentials and dielectric properties in liquid environments. We provide practical guidelines for when such can be measured and describe what other information can be obtained.

## 6.1 Capacitive forces in electrolyte

In section 2.2 we outlined the underlying physics pertaining to the electrostatic force and force gradient detection in EFM and KPFM, based on simple description of the expected capacitive forces acting between a VM AFM probe and a metal sample. Noteworthy, equations (11)-(13) describing the electrostatic force implicitly rely on the presence of a linear lossless dielectric between probe and sample. This has several major implications, which will underpin the operation of any EFM/KPFM measurement.[51] First, for quantification of the surface potential using either OL bias spectroscopy methods or CL bias feedback approaches; a linear bias dependence of the first harmonic electrostatic force is a prerequisite. Secondly, for all meaningful measurements, the electrostatic force must be measured under equilibrium conditions, stated differently, the measurement time must be faster than the kinetics of the system (i.e., quasistatic equilibrium). Clearly the introduction of a conductive liquid (i.e., lossy polarizable dielectric) between tip and sample, require careful consideration into the veracity of the fundamental assumption of successful EFM/KPFM operation.

### 6.1.1 Electrostatic actuation in microfluidic MEMS

As a starting point for understanding capacitive forces in liquid EFM or KPFM, it natural to consider established theory governing the more mature field of micro electromechanical systems (MEMS and NEMS). While actuation in MEMS devices can take many forms including; thermoelastic forces, piezoelectric forces, and even magnetic forces, actuation by means of electrostatic forces is the most commonly employed method.[560] In this field, electrostatic actuation under dry conditions offers many advantages including moderate displacements (10-100 μm), fast response times (ns − ms), low power consumption as well as ease of design and fabrication.[561] Significantly, in liquids having high dielectric constants (e.g., water), electrostatic actuation gains of two orders of magnitude can be obtained over operation in dry conditions.[561, 562] This makes electrostatic actuation a promising candidate for MEMS devices in liquid. At the same time, the presence of ions in solutions creates problems of shielding and electrochemical reactions.[563] Notably, many of the challenges facing electrostatic actuation of MEMS devices in liquid, such as electrolysis, anodization,[564] electrode polarization screening, are expected to be pertinent in attempts of liquid EFM/KPFM measurements. In MEMS research, several approaches have been explored to overcome these challenges including; operation in low conductive liquids, use of short voltage pulses (ms − μs) and by using sufficiently low electrode potentials.[564] Sounart et al, [357] first demonstrated electrostatic actuation in polar liquids by using ac drive waveforms, at potentials low enough to avoid electrochemistry. They demonstrated that ac voltages can be used to prevent electrode screening and thus enable electrostatic actuation in many liquids. In this seminal work, the frequency dependence of the actuation of interdigitated MEMS devices in electrolytes of various conductivities were presented, the results of which are shown in

Figure 45. In the absence of charge screening, the applied electric field is expected to be uniform in the gap between interdigitated electrodes, and electrolyte behavior similar to that of a linear lossless dielectric.[561, 562] As can be seen, by using an AC excitation wave of sufficiently high frequency the actuation saturates to a maximum, and the magnitude of this actuation decreases with decreasing frequencies until finally no electromechanical response is observed. Measurements were repeated using different solutions of increasing conductivity, used to demonstrate that a critical excitation frequency can be observed for all solutions studied, above which the device can be successfully actuated using ac voltage.[565, 566] At low frequencies, where the excitation timescale >> response time of the solution (or the ions in solution), the system will be in equilibrium and the EDL screening will be fully established, and hence no electrostatic force or actuation will be detected. In addition, at intermediate timescales, the magnitude of actuation increases over a broad frequency range, suggesting the dynamic processes (ion migration and diffusion) will result in partial screening of the electrode and attenuation of the actuation is observed. At high frequencies above the critical frequency, the frequency is such that the polarity of the electric field changes at a rate faster than that of ionic response and thus avoids electrostatic shielding. In other words, at excitation timescales much shorter than the RC time of the equivalent electrical circuit for passivated electrodes immersed in an electrolyte solution, the electrolyte will behave like a linear dielectric. It is believed that this analysis will be valid in a thin EDL limit.[561]

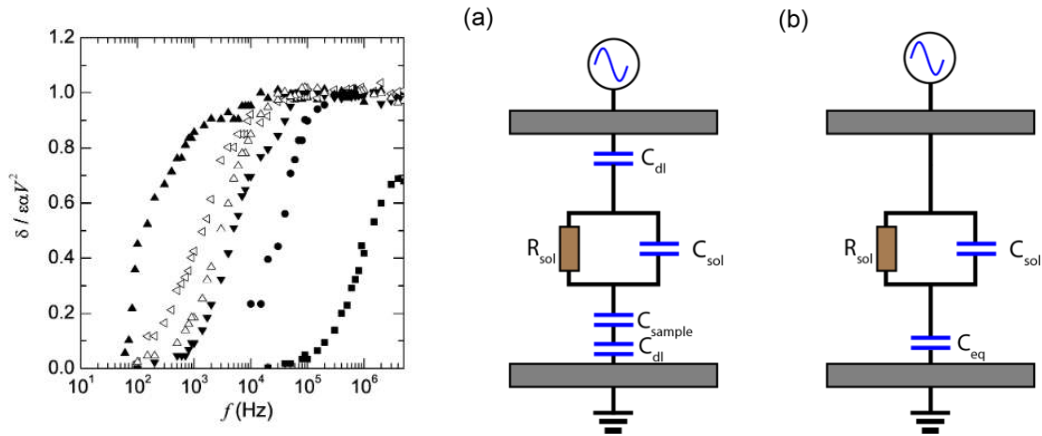

Figure 45. (a) Actuator frequency response in several conductive liquids ($\blacktriangle$ HeOH)($\triangleleft$ IPA)($\blacktriangledown$ EtOH) ($\wedge$ EG)($\bullet$ $H_2O$) ($\blacksquare$MeNO). (b) Equivalent circuit model where $C_{dl}$ is the EDL capacitance, $C_{sample}$ is the sample capacitance, $C_{sol}$ and $R_{sol}$ are the solution capacitance and resistance, respectively. (b) Simplified equivalent circuit model where the EDL and sample capacitance are represented as an equivalent capacitance ($C_{eq}$). *Seeking permissions.*

Sounart *et al.*[561] provided a simple model to describe mechanism of electrode charge screening due to ion migration and formation of an EDL when dc and ac potentials are applied across silicon electrodes. Using equivalent circuit models, see

Figure 45(b) the EDLs formed at the tip and sample ($C_{dl}$) are treated as capacitors in series with the sample capacitance ($C_{sample}$) and solution resistance ($R_{sol}$). For a parallel plate model:[55]

$$R_{sol} = \frac{z}{\sigma_{sol} c} \qquad (36)$$

$$C_{sol} = \frac{\varepsilon_0 \varepsilon_{r,sol}}{z} \qquad (37)$$

$$C_{sample} = \frac{\varepsilon_0 \varepsilon_{r,sample}}{h} \qquad (38)$$

where, $z$ is the electrode separation, $c$ is the ion concentration, $\sigma_{sol}$ is conductivity, $h$ the height of the sample oxide, $\varepsilon_0$ is the permittivity of free space and $\varepsilon_r$ represents the relative solution or resistance permittivity. Using common convention used in *RC* filter circuit analysis,[562] approximate critical frequency, $f_c$, can be calculated as:

$$f_c = \tau_{RC}^{-1} = \frac{1}{2\pi R_{sol} C_{eq}} = \frac{h\sigma_{sol}}{\pi \varepsilon_0 \varepsilon_d z} \qquad (39)$$

where $C_{eq}$ is the equivalent capacitance of the system, as shown in

Figure 45(c). It is argued,[54, 55, 561, 562] that $C_{EDL}$ is typically much higher than the capacitance of the passivation layers $C_{sample}$, and can be neglected since they are in series (i.e., $C_{eq} = C_{sample}$). In part this route is chosen due to difficulties in expressing the dynamics of the EDL using simple circuit model using passive elements.[9, 10]

### 6.1.2  Capacitive forces in liquid EFM/KPFM

Noteworthy, the application of electrostatic actuation in conducting liquids for MEMS devices using ac drive signals suggests that a similar approach can be used in EFM/KPFM measurements to prevent electrode polarization (at low potentials) or undesirable electrochemistry (at high potentials). Gramse *et al.*,[52, 54] extended the simple circuit analysis model outlined for MEMS devices, [561, 562] to EFM in liquid. In their initial analysis, for simplicity, they assumed a parallel plate geometry and an equivalent circuit model for the tip-solution-sample system. The electric force acting on parallel plates in response to an applied ac voltage, $V(t) = V_0\sin(\omega t)$, can be written as:

$$F_{elec}(z,t) = \frac{1}{2}\frac{\partial C_{sol}(z)}{\partial z}V_{sol}{}^2(z,t), \tag{40}$$

where $V_{sol}$ is the voltage drop across $C_{sol}$. Note that as $R_{sol}$ and $C_{sol}$ are distance dependent, a more exact circuit model would include many more coupled circuits (i.e. transmission line type model) to describe the complex probe geometry. The first harmonic of the electrostatic force per unit area can be written as:

$$[F_\omega(z,t)] = \frac{1}{4}\frac{\partial C_{sol}(z)}{\partial z}\frac{f/f_{eq}}{1 + f^2(1/f_{eq} + 1/f_{sol})^2}V_0^2 \tag{41}$$

By taking $\partial C_{sol}(z)/\partial z = -\varepsilon_0\varepsilon_r / z^2$ it is possible to approximate the frequency dependence of the electrostatic force as shown in Figure 46(a) for various electrolyte concentrations. Following the work by Gramse $et\ al$, a native oxide of 3 nm on the $SiO_x$ substrate is assumed. Figure 46 demonstrates a stereotypical sigmoidal [55, 561, 562] behavior in all solutions with the force saturating at high frequencies and going towards zero at low frequencies. The critical excitation frequency is defined as the frequency at which the force has reached half its high frequency saturated value (dashed lines).

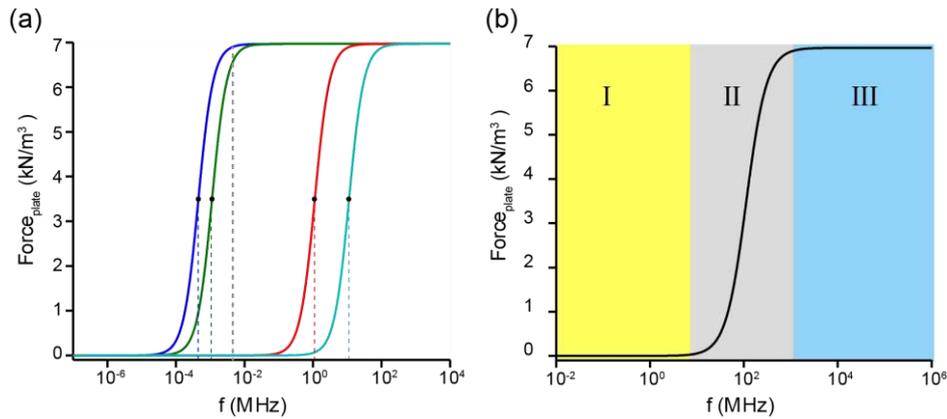

Figure 46. Calculation of the critical actuation frequency. (a) First harmonic electrostatic force per unit area of parallel played predicted by equation (41) where $z$ = 5 nm, $\varepsilon_{r,sol}$=80, $\sigma$ = 13.3 S m$^{-1}$, $\varepsilon_{r,sample}$= 4, h = 3 nm for (blue) molar concentration, $c$ = 4x10$^{-4}$ mM, (green) 1 μM, (red) 1 mM and (cyan) 10 mM. (b) Frequency dependence of the electrostatic force per unit area using the same parameters[530] as where $c$ = 100 mM. The shaded regions indicate the different electrochemical regimes where (I) Faradaic and (II) charge dynamics are expected to dominate the response and (III) where ion dynamics are largely absent (i.e., quasistatic equilibrium). *Seeking permissions.*

Gramse *et al.* experimentally evaluated this behavior by performing a form of EFM in liquid of different concentrations as a function of excitation frequency and distance, on a model sample comprised of striped $SiO_2$ on a $Si^{++}$ substrate as shown in Figure 47(a). In this measurement, an amplitude modulated excitation having a high excitation frequency ($f_{exc}$ is in the MHz range) as well as a low modulation frequency ($f_{mod}$ of a few kHz) component is used to improve signal to noise which suffers from attenuation of high frequency responses above the natural resonance frequency of the cantilever. The electric force in response to the applied voltage is captured using demodulation at the modulation frequency, and in turn the effective capacitance gradient is calculated as:

$$C^{'}(z) = 4 \left| F_{elec}(z,t) \right|_{f_{mod}} / V^2 \qquad (42)$$

Figure 47(a) shows characteristic capacitance gradient approach curves measured on $Si^{++}$ part of the sample at a salt concentration 0.1 mM and for increasing frequencies of the applied voltage from 100 kHz (light purple line) up to 20 MHz (black line). For frequencies below 0.5 MHz, the electrostatic force demonstrates little or no change as a function of distance, only becoming distant dependent at high frequencies. In Figure 47(b), the frequency dependence of the capacitance gradient (at a fixed distance of 50 nm) is plotted for differing salt concentrations (0.1, 1 and 10 mM). The results follow the expected sigmoidal behavior described. Notably, the electric force transitions between low and high frequency limits over a frequency range covering nearly two orders of magnitude, reaching a maximum plateau at very high frequencies (> 1 MHz for 0.1 mM and > 3 MHz for 1 mM). It is not clear if the electrostatic force has reached maximum values for 10 mM, however, the authors estimate it to be about 100 MHz (well above the detection frequencies used in AFM).[54] It was noted that the characteristic frequency at which the force becomes independent from frequency depends roughly linearly on salt concentration. Note that the transition from low to high frequencies is much broader and smoother than that predicted by the parallel plate model in Figure 46(a), which the authors show are in in agreement with numerical simulations employing a more realistic geometry.[55] This last point raises questions about the veracity of a quantitative evaluation of EFM using such simple parallel plate models and suggests more robust numerical methods are required.[55, 567]

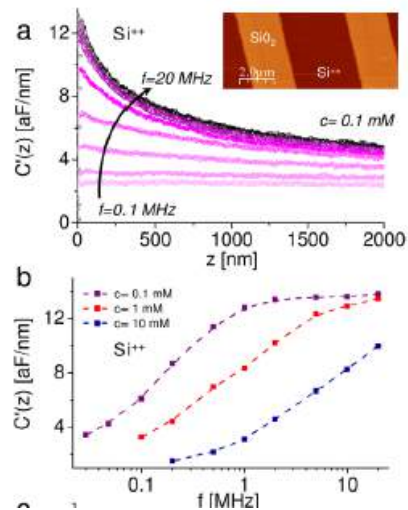

Figure 47. (a) Capacitance gradient vs distance curves measured on positively doped Si substrate for different modulation frequencies (inset: sample topography). Capacitance gradient values measured as a function of frequency (distance = 50nm) for different ion molarities. *Seeking permissions.*

## 6.2    Electrostatic actuation in liquids

The idea of AFM cantilever actuation using both dc and ac voltages has been investigated for several different applications. In the seminal work by Raiteri *et al*.[489] bias induced surface stress was measured using a voltage modulated probe in electrolyte. This work was the first to introduce the concept that imaging mechanism of VM-AFM in liquid consists of competing Maxwell stress tensor and osmotic pressure. Noteworthy, this work demonstrated promise for probing electrochemical kinetics using the probing tip and force based detection. Rodriguez et al[568] used EFM to detect and image periodic electrostatic forces in liquid. In a different approach, electrostatic actuation has been leveraged to drive the AFM cantilever for AFM imaging, which has been implemented in liquid environments for both low (< tens of kHz) and high (> 1 MHz) frequency regimes. It is generally believed that the effects of electrochemical processes in VM-SPM techniques in liquid will be mitigated for ac voltages applied within a certain frequency range. Umeda *et al.* reported that below tens of kHz, the electrochemically induced surface stress dominates the tip-sample interaction, while the electrostatic force dominates the interaction above this value.[50, 94]

In a later work, Umeda *et al.* [567] undertook an in-depth analysis of the capacitive force acting and proposed criteria for local dielectric measurements using EFM in polar liquid media. They noted that the cantilever deflection is not only induced by the electrostatic force, but also by the surface stress, which does not include the local dielectric information.

Umeda proposed that the modulation frequency must be much higher than the characteristic dielectric relaxation frequency as well as the transition frequency at which the electrostatic force contribution originating from the capacitance between the tip and sample becomes predominant. The capacitance per unit area of the tip apex and sample surface must be much lower than the Stern layer capacitance per area. They highlighted that although the capacitive force measurement becomes possible if these criteria are fulfilled, its spatial resolution is much less than that in air or vacuum environment without the modulation frequency being even higher than the relaxation frequency of the polar molecules in the solvent is used.

## 6.3    Applications of classical PFM in liquid

The important aspect of any voltage modulated technique is the spatial distribution of the dc and ac components of electrostatic field produced by the tip, both in the tip-sample gap and within the material. These distributions illustrate the signal generation volume and hence are directly related to the resolution, and are directly connected to the forces experienced by the tip via the Maxwell stress tensor. In most cases, these distributions in the tip-sample junction are not accessible experimentally and can only be deduced based on direct modelling or observed spatial resolution using well defined calibration samples.[569, 570] However, an important insight into the structure of electric fields in liquid can be obtained using liquid PFM.[571-574]

PFM is contact voltage modulated technique that emerged in the last two decades as a primary tool for nanoscale studies of ferroelectric,[575, 576] piezoelectric,[577, 578] and electrochemical [579-581] systems. PFM can be considered a contact mode implementation of EFM and OL-KPFM.[239] Like KPFM, the tip in PFM is modulated using a superposition of dc and ac voltage; however, in the contact regime, the origin of the signal is fundamentally different. First, in contact with the surface, the mechanical stiffness of the system is much higher, and hence tip sensitivity to electrostatic forces is much lower.[81, 582] Secondly, the potential drop now occurs both in the tip surface junction and in the material, with the latter dominating for ideal imaging conditions.[583] Third, the potential drop in the material results in the surface deformation due to piezoelectric [584-590] (or more complex electrochemical [591-594]) effects, and this displacement is translated to the tip and is detected as the PFM signal. Notably, in the non-contact regime, the field drops predominantly in the tip-sample junction due to the dielectric constant mismatch, and the surface is mechanically uncoupled from the cantilever. Thus, analysis of the PFM signal necessitates considering both electrostatic and electromechanical interactions, whereas KPFM is generally sensitive to electrostatic interactions only.

Several attempts [305, 306, 572-574, 595-599] have been made to apply PFM in liquid. As is the case for EFM and KPFM modes,[46, 47, 51] PFM can be affected by the presence of polarizable or conductive media with moveable ions that cause the detected signal to be highly concentration and frequency dependent and be impacted by issues related to EDL formation, charge screening, and Faradaic reactions.[572-574, 597] The tip geometry down to the nanoscale tip-sample contact necessitates considering multiple length and timescales. While PFM and KPFM are generally complementary in terms of imaging mechanisms, it is this complementarity that allows deriving several important conclusions of KPFM mechanisms based on PFM observations. Namely, for PFM imaging in liquid environment, the observability of PFM contrast implies that the ac field is not attenuated by the presence of liquid between the tip and the surface, and the degree of this attenuation can be derived from the observations of PFM signal as a function of frequency and solution concentration. Similarly, PFM domain switching experiments yield information on the dc field distribution in the liquid. Thus, PFM has proved to be a useful technique for probing and understanding the localization of electric fields in liquid environments, and in that sense, can provide some insight to liquid KPFM measurements. Below, we discuss both relevant aspects of PFM imaging and switching in liquids.

As mentioned above, PFM has been widely used to study ferroelectric materials,[100, 600, 601] whereby the highly inhomogeneous electric field resulting from bias applied to the probe can lead to polarization reversal in the ferroelectric material that can be detected either via hysteresis loop measurements [602] or via imaging of pre-existing and final domain structures.[584, 585, 603] Domain formation is a complex process involving nucleation and wall motion stages, and is well explored in the context of PFM.[604, 605] It is no surprise that the first PFM measurements performed in liquid were also conducted using ferroelectric materials.[597, 606]

Early work, motivated by extending PFM of biological systems [577, 607] to physiologically-relevant liquid environments, identified potential benefits of imaging ferroelectric domains in liquid, reporting an improvement in resolution resulting from the minimization of long-range electrostatic interactions.[597] The authors observed that PFM imaging at sufficiently high (>100 kHz) frequencies is possible and leads to images of high quality and high spatial resolution, despite the fact that the applied biases were outside the electrochemical stability window of the aqueous electrolyte. Importantly, imaging at high frequencies suppressed the effect of the liquid environment on the cantilever dynamics.[608, 609] Overall, based on the images it was not possible to distinguish between ambient and liquid scanning. In fact, with long-range electrostatics effectively screened by mobile ions, the electric field resulting from the applied ac voltage was localized to the tip-sample junction, allowing an improved resolution compared to measurements performed on the same sample in air (Figure 48).[597] On transition to lower frequencies, image quality degraded, and in the ~kHz range electrochemical reactions became obvious. Notably, both in-plane and out-of-plane PFM signal has been measured as a function of ion concentration[573] and frequency, in single and multifrequency modes[573]; in-plane PFM amplitude and phase images recorded for bismuth ferrite are shown in Figure 49.

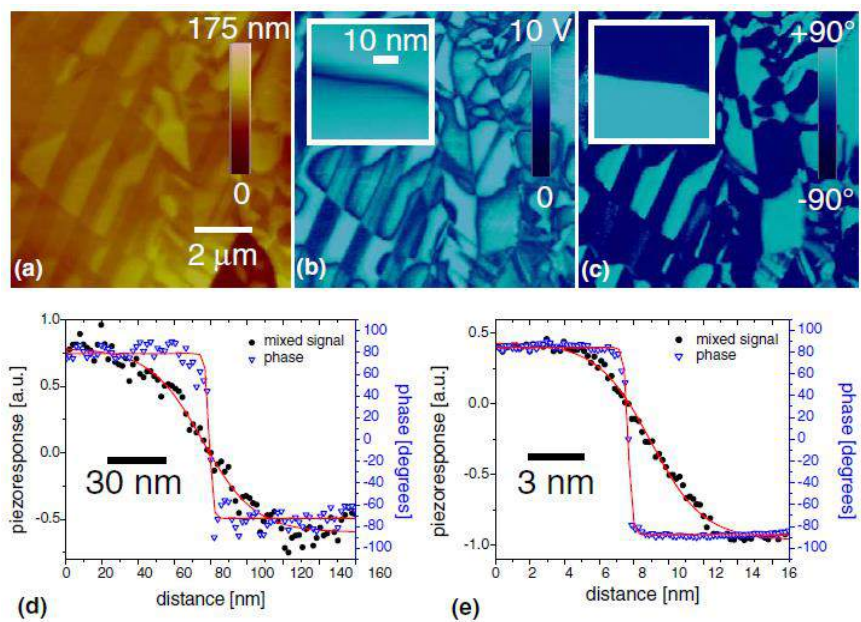

Figure 48. (a) Topography, (b) PFM amplitude, and (c) PFM phase images of ceramic lead zirconate titanate recorded in water. High resolution PFM images are shown as insets. Domain wall cross-sections of mixed PFM and PFM phase in (d) air and (e) water. Reproduced with permission from [597]. Copyright 2006, American Physical Society.

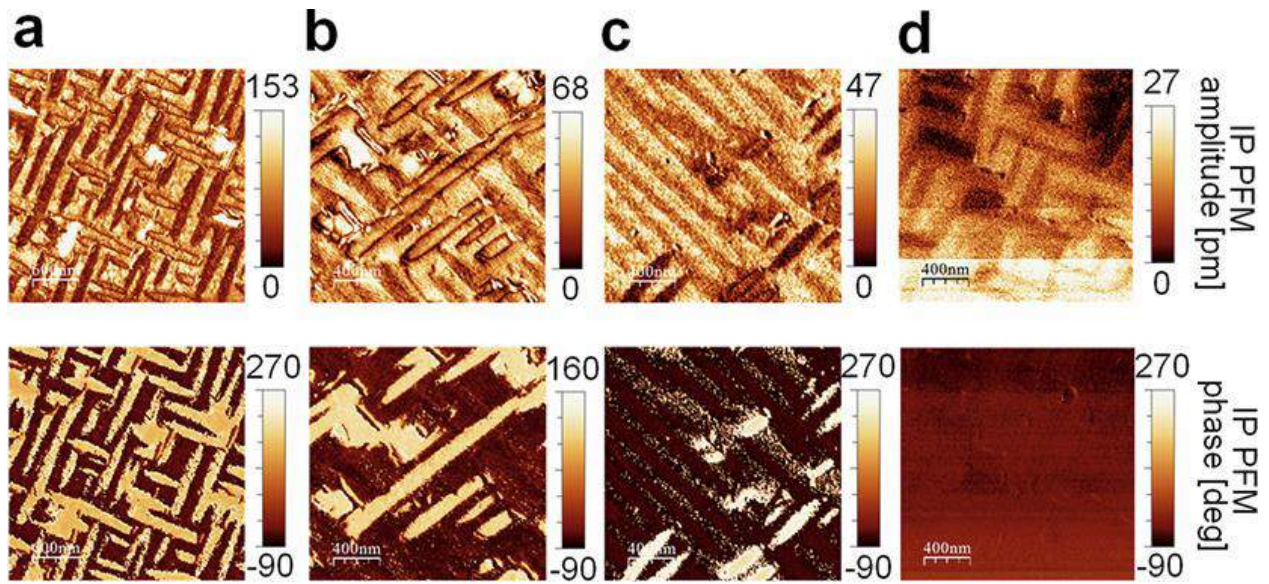

Figure 49. In-plane PFM amplitude and phase images of bismuth ferrite measured as a function of ion concentration; (a) $10^{-6}$ M $Na_2SO_4$, (b) $10^{-4}$ M $Na_2SO_4$, (c) $10^{-2}$ M $Na_2SO_4$, and (d) 1 M $Na_2SO_4$. Reproduced with permission from [573]. Copyright 2012, American Chemical Society.

One motivator for liquid PFM is that it might be applied to biological materials, however, despite the demonstration of liquid PFM on model ferroelectric systems, only a handful of attempts of applying liquid PFM to biological materials have been reported, including on lysozyme, insulin, and amyloid fibrils and adenocarcinoma and bacterial cells.[595, 596, 599] The relatively low piezoelectric constants reported for biological materials, the presence of EDLs in solution, which might prevent the detection of shear piezoelectric deformations and reduce the field applied to the sample, the presence of other electromechanical couplings such as flexoelectricity in membranes, and high ion concentrations of physiological environments are all challenges for implementing liquid PFM. Early reports further indicated the presence of an elastic contribution to the PFM signal of biological materials in a liquid environment and topographic crosstalk, which might be minimized by using BE PFM. Notably, Nikiforov *et al.* used liquid BE-PFM along with principal component analysis and a recognition neural network to distinguish between *Micrococcus lysodeikticus* and *Pseudomonas fluorescens* bacteria based on their broadband electromechanical response (Figure 50).[599]

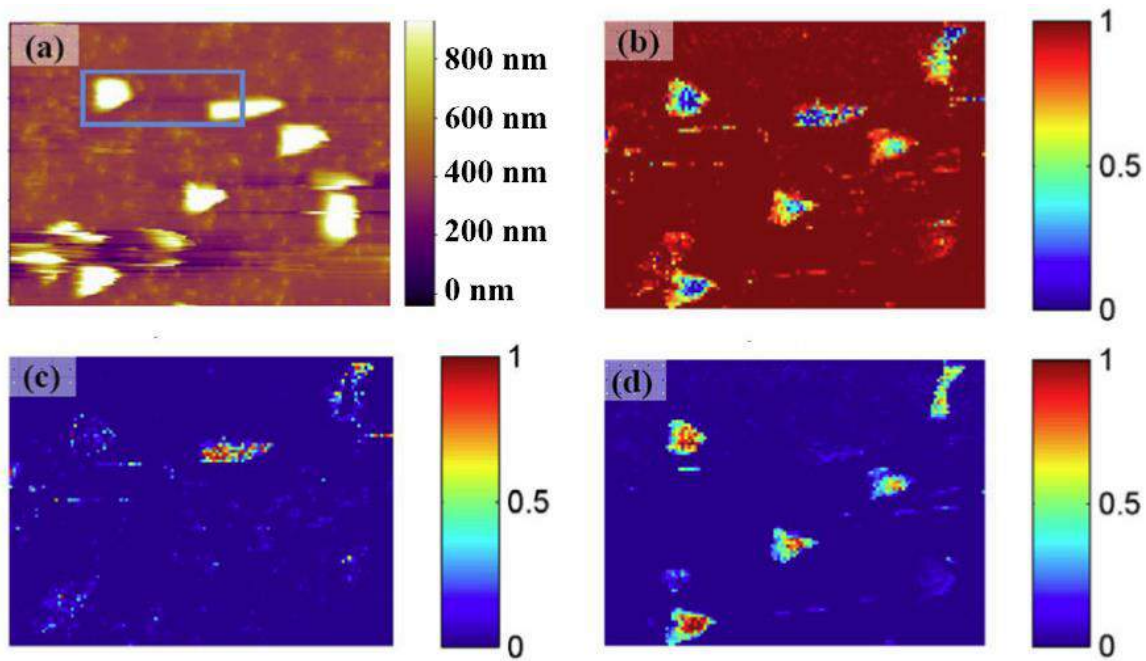

Figure 50. (a) AFM height image of bacteria on poly-L-lysine-coated mica and (b-d) neural network recognition maps for (b) mica, (c) *P. fluorescens*, and (d) *M. lysodeikticus*. The scale in (b-d) refers to the likelihood that the pixel corresponds to the target value based on the neural network training input (rectangle in (a)). Reproduced with permission from [599]. Copyright 2009, Institute of Physics.

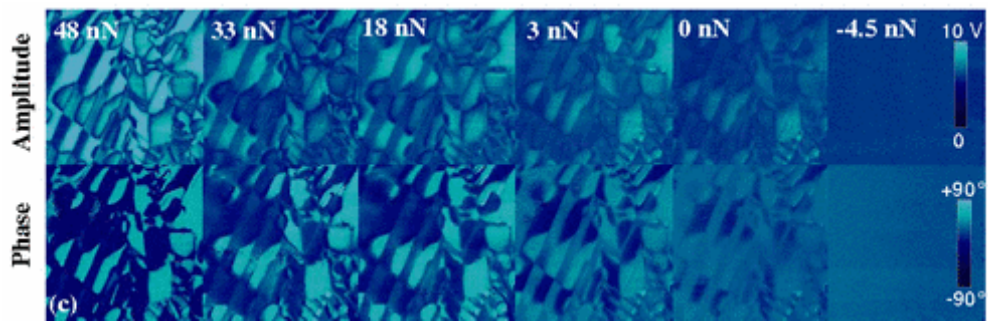

Figure 51. PFM amplitude and phase images of ceramic lead zirconate titanate recorded in water for different loading forces. Reproduced with permission from [597]. Copyright 2006, American Physical Society.

Liquid PFM has also been performed as a function of loading force on model ferroelectric surfaces,[574, 606] whereby contrast is reduced when the tip is no longer in contact with the sample (Figure 51). Note that when the tip is lifted above the surface in liquid PFM, a version of EFM in liquid is being performed.[597] Thus, PFM gives some information on the penetration of the applied ac field in liquid environments (Figure 52) and suggests a route to implementing an intermittent contact mode version of PFM in liquid suitable for providing screening of electrostatics and imaging soft biomaterials.[598] Note that when the Debye length of the solution becomes comparable to the intrinsic tip-sample gap, the PFM contrast is expected to decay. Unexplored at that time [597] was the role of the impedance of the excitation source on the signal strength, potentially accounting for the capacitive losses in the cell. Nevertheless, such studies clearly demonstrated that even in nominally conductive solutions, PFM imaging is possible at high biases and high frequencies, implying that the localization of the ac field is sufficiently close to the ambient case.

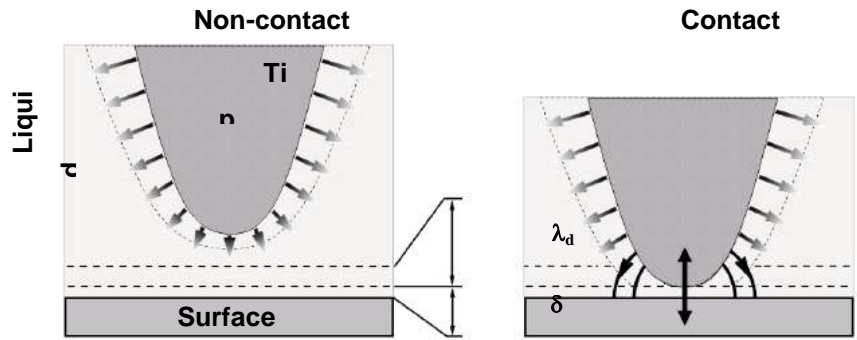

Figure 52. (a) Field penetration into liquid and electrostatic forces acting on a probe are minimized in liquid. $\lambda_D$ is the Debye length and $\delta$ is tip-sample separation, e.g., due to water layer at surface. (b) When the probe is in contact with the surface, electromechanical coupling (double arrow in (b)) may be present. Thus, when considering the full motion of the probe during intermittent contact mode, electrostatic/electromechanical forces are screened when the probe is away from the surface and electromechanical signal may be present during the contact portion of the probe motion. Adapted with permission from [597]. Copyright 2006 by the American Physical Society.

Liquid PFM of ferroelectric surfaces has provided information on the localization of the ac field at the tip surface junction and has revealed that electromechanical imaging can be performed across a range of frequencies and ac voltages.[573, 574] Similarly, ferroelectric switching experiments in liquid provides information about the structure of the dc electric field.[606] In ambient, a dc bias applied to the AFM probe in contact with a ferroelectric can lead to polarization reversal directly under the probe. In liquid, Rodriguez *et al.* reported a transition from local (similar to switching in ambient) to non-local (similar to switching in ferroelectric capacitors) switching depending on the solvent used and the bias applied (Figure 53).[606] Traditionally, in ambient, the tip-induced switching produces well-defined round or convex domains, albeit in some cases back switching and formation of bubble[610] and more complex[611, 612] domain geometries was observed. It is generally assumed that domain nucleation requires a certain critical bias; correspondingly, changes in the nucleation bias and the character of switching between liquid and electrolyte provides information on the field distribution in the solution. In other words, assuming the polarization of the ferroelectric material itself switches at roughly the same bias each instance, the visualization of the switched area under the tip via PFM phase indicates the spatial extent of the dc field in different solutions, which is relevant for the implementation of KPFM modes in liquid environments. For instance, localized switching could only be observed in non-aqueous solvents such as isopropanol or methanol, whereas in aqueous solutions, reactions can occur.[606] Liquid PFM was further explored by Balke,[573, 574], who performed detailed studies of polarization switching of $BiFeO_3$ in liquid environments and elucidated primary quantitative aspects of observed behaviors. These in turn reflect the evolution of tip-induced dc potential in solutions as a function of electrolyte concentration. Notably, the earliest attempt to perform PFM in a liquid environment was reported by Ganpule [613] wherein a highly conductive liquid was used as a top electrode similar to the uniform switching experiments reported in ref. [606].

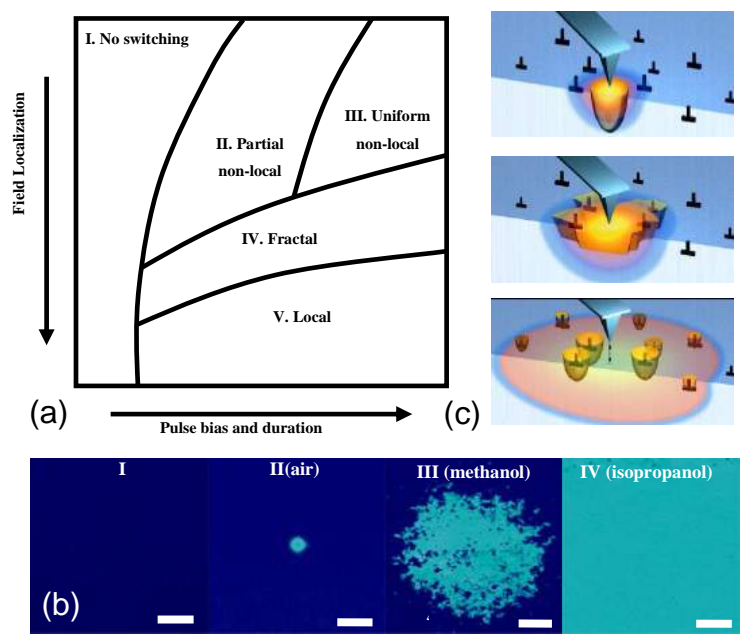

Figure 53. (a) Diagram of possible switching modes vs. electric field localization and dc bias magnitude and duration. (b) PFM phase images showing the resulting polarization map for regions I, III, IV, and V under different conditions (scale bar (I) 600 nm (II) 400 nm (III and IV) 2 µm. (c) Schematics of different switching regimes for local, fractal, and non-local cases. Adapted with permission from [606]. Copyright 2007, American Physical Society.

The development of liquid PFM has provided insights to the implementation of KPFM in liquid, in particular, the localization of ac and dc fields critical to the operation of the techniques in liquid. Challenges associated with the movement of mobile ions in solution and the presence of EDLs are present for both techniques. High frequency operation and the use of insulated conductive probes for liquid PFM[305, 306] may further localize applied fields and detected signals, lead to improved resolution, and are relevant for liquid KPFM as well. Implementation of metrological VM-AFM modes in liquid may further improve the quantification of surface deformations.[614] Clearly, implementation of PFM in liquid is considered preferred for several systems including biological materials, batteries, and super capacitors. Recently, significant progress has been made using electrochemical strain microscopy (ESM) for the study of ionic conductors and other materials.[80, 615-618] Like PFM, ESM measures the response of the cantilever to the voltage applied across the tip-sample junction, however, the signal formation mechanism is different. Both PFM and ESM have been implemented in various spectroscopic modes that track dynamic polarization switching and local reactivity.[80, 615-619] The further extension of these modes to liquid environments is a promising area of research.

## 6.4    Applications of classical KPFM in liquid

The goal of performing quantitative KPFM in liquid has been of interest to many research groups over several decades. The first major breakthrough in liquid KPFM was the implementation by Domanski *et al.*[45] of classical KPFM (operated in CL) in electrically insulating nonpolar solvents. In this seminal work, liquid KPFM was used to investigate the physisorption of decane molecules at a gold electrode interface, and its effect on the measured work function. By assuming both tip and sample surface were screened by the same amount of adsorbates and that the potential difference due to physisorbtion ($\Delta\Phi_{phys}$) was a constant, they were able to successfully calculate the effect of charge screening by the impurities (e.g. remnant water layers) ($\Delta\Phi_{H_2O}$) on the work function $\Delta\Phi_{H_2O} = \Delta\Phi_{ambient-decane} - \Delta\Phi_{phys} = 0.22 \pm 0.40$ eV demonstrating the importance of having clean surfaces in KPFM.[45] They further demonstrated the possibility of KPFM imaging on a patterned $SiO_x$/Au test sample, Figure 54(a). This allowed an investigation into the changes in work function of an Au substrate upon hexadecanethiol chemisorption.[45] More recently, using a similar approach, the surface potential of a pn patterned silicon sample was imaged in fluorocarbon liquid.[44] The veracity of classical KPFM in liquid has been verified using Kelvin probe force spectroscopy (KPFS)[44, 46]. We demonstrate such a measurement in Figure 54(b), over an electrochemically inert highly ordered pyrolytic graphite (HOPG) electrode in both ambient and nonpolar (decane) solvent. In both cases, the first harmonic electrostatic response, $A_\omega$, demonstrates linear dependence to $V_{dc}$V$_{dc}$, has a minimum $A_\omega$, and a resultant $180^o$ phase inversion. These characteristics are typical for a purely electrostatic response, or more generally when the force experienced by the system is governed purely by the time-independent Maxwell stress tensor. The observed shift in the CPD is likely a combination of the removal of the water layer from the tip and sample surfaces ($\Delta\Phi_{H_2O}$) and the interaction between the decane molecules and electrode (tip and sample) surface modifying the permittivity. From Figure 54(b), assuming that $\Delta\Phi_{phys}$ is similar for freshly cleaved HOPG as for gold, we find a similar value for $\Delta\Phi_{H_2O} = 0.18 \pm 0.01$ eV as reported by Domanski *et al.*,[45] demonstrating the possibility for well-controlled KPFM measurements in liquid, which are notoriously difficult in ambient conditions (see section 3.3.3).

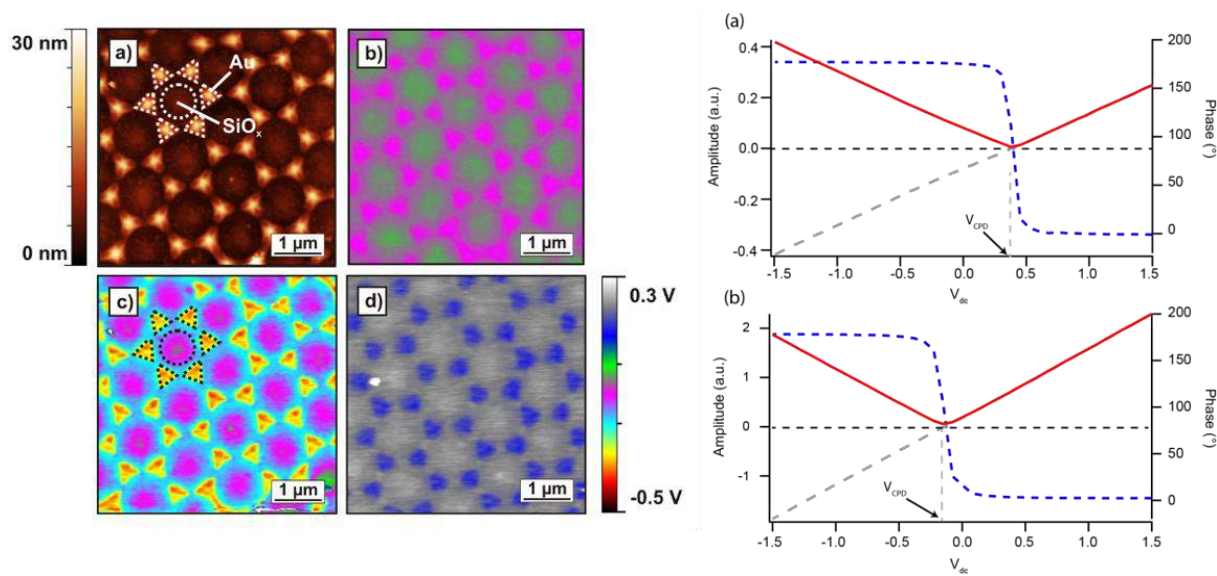

Figure 54. (a) (b) Bias dependence of the electrostatic force in nonpolar liquid. OLBS showing the oscillation amplitude (red line), phase (dashed blue line) and in-phase response (dashed grey line) in (a) ambient and (b) decane. OLBS was performed using a specified tip−sample separation of 200 nm, and voltage amplitude of 1 V with a frequency of 20 kHz applied to the tip. *Seeking permissions.*

In Figure 55 we present previously unpublished results which further highlight the possibility of performing liquid KPFM in non-polar solvents for energy materials, demonstrated on few and single layer graphene oxide (GrOx). Here, GrOx sheets were diluted in isopropanol and deposited on a freshly cleaved HOPG surface. KPFM was performed at a fixed height of 50 nm from the surface in under both ambient and decane environments. Unfortunately, the same GrOx structures could not be located for both images so measurements cannot be directly comparable. However, in both environments sharp KPFM potential variation was obtained. In ambient, the GrOx has a larger CPD (-422 ± 5 mV) than the underlying HOPG surface (-330 ± 3 mV). Upon immersion in decane, it is clear that the HOPG substrate no longer demonstrates a homogenous CPD with charge domains, which are not related to the surface topography, can be observed. Furthermore, it was found that images taken immediately after immersion in decane demonstrated a drift in the measured CPD. Figure 55(c,d) was taken ~ 5 mins after immersion in decane with each image taking ~18 mins to collect. The CPD of the graphene drifted linearly by ~65 mV during the first image. The underlying substrate, however, did not change significantly in this timeframe. The fact that the rate of change seemed to be material dependent suggests that a chemical reaction at the surface, not the tip, was taking place. Figure 55(e,f) shows a KPFM image collected ~80 mins after immersion in decane. Within this timeframe, the charged domains had reduced in size, the substrate had become more negative, while the GrOx had remained the potential reached during Figure 55(d). This suggests that multiple, sample and time dependent surface reactivity's are being probed, demonstrating the potential of KPFM for measuring dynamic chemical processes at the graphene oxide-liquid interface.

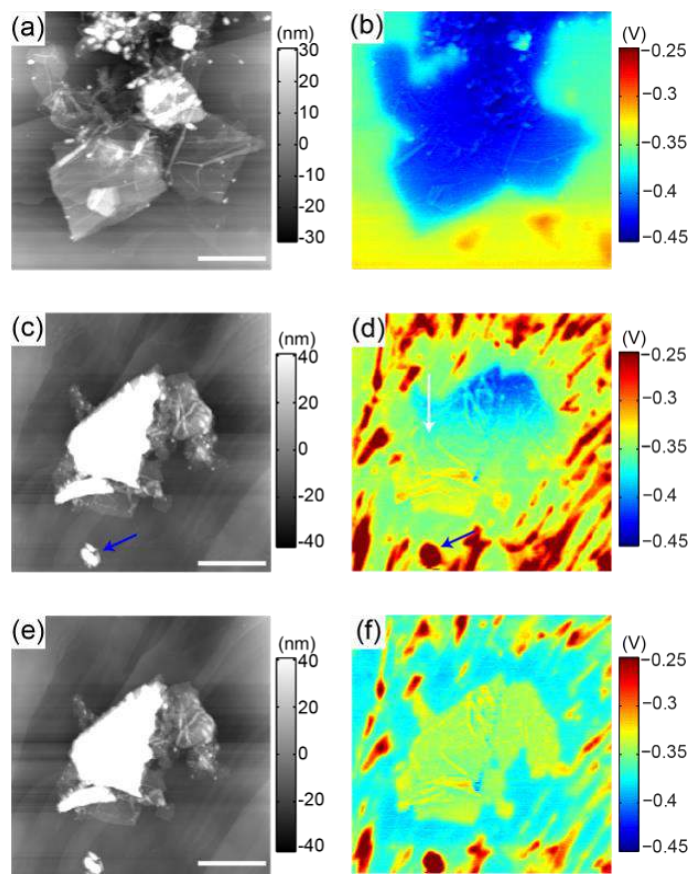

Figure 55. KPFM imaging of chemical reactions on GrOx in nonpolar liquid. GrOx (a) topography and (b) CPD measured in ambient at a lift height of 50 nm with $V_{ac}$=2 V with a frequency of 78.4 kHz. (c,e) topography and (d,f) CPD measured (c,d) ~5 mins and (e,f) ~80 mins after immersion in decane collected at a lift height of 50 nm with $V_{ac}$= 5 V with a frequency of 38.4 kHz. White arrow in (d) indicates scan direction (scale bar = 5 μm).

To summarize, operating classical KPFM in in non-polar solvents overcomes problems associated with the conductivity of polar solutions, including the delocalization of the applied fields which are largely absent. Furthermore, the lack of mobile charges in nonpolar liquids leads to behavior matching that of a lossless dielectric, a fundamental assumption underlying KPFM operation.[71, 74] Hence, under these conditions, classical KPFM can be successfully implemented and has shown promise in studying dynamic chemical processes at the solid-liquid interface.

## 6.5    Application of KPFM in non-polar liquids

Beyond non-polar liquids, several groups have been intensively exploring the possibility of using classical KPFM in polar liquids. Initially, this feasibility was tested by performing EFS in polar liquids (and comparing with air).[44, 46] An example of this is shown in Figure 56 where the $V_{dc}$ dependence of $A_\omega$ and $A_{2\omega}$ were probed under increasing bias ranges on Au surfaces in milliQ water.[46] Data sweeps were collected for a small (± 200 mV), medium (± 400 mV) and large (± 800 mV) bias range, consecutively. The measurement resulted in complex responses including hysteretic behavior and the presence of multiple maxima and minima similar in a bias dependent fashion, matching that reported by Umeda *et al.*[44] and Raiteri *et al.*[489] For small bias sweeps (± 200 mV), finite shifts in the magnitude of $A_\omega$ between negative to positive (green) and positive to negative (purple) bias sweeps were observed, likely resulting from a redistribution of ions. For medium bias sweeps (± 400 mV), hysteretic behavior was observed between sweeps. Large bias sweeps (± 800 mV) resulted in complex responses including hysteretic behavior and the presence of multiple maxima and minima. We note, that the general shape and magnitude of the response was heavily dependent on sweep rate, suggesting an underlying temporal dependence of the response. When using bias sweeps larger than 2 V, large changes in the AFM cantilever deflection signal occurred (not shown), often followed by visible bubble nucleation in the tip-sample gap (e.g., Figure 54(f)). Attempts at implementing KPFM in ionically-active liquids will often result in similar, if not more catastrophic, bubble formation by virtue of the absence of a defined minimum, which results in the application of large dc biases by the feedback loop.

These results demonstrate, that in polar solvents, even having minimal ion concentrations (e.g., deionized water) the absence of a unique minimum in $A_\omega$, and the presence of non-linear hysteresis and irreversible reactions at larger biases are fundamental barriers to the application of KPFM in milliQ water and other polar liquids. Noteworthy, these results demonstrate also highlight that the universal application of classical KPFM across all materials, all bias ranges and all solutions is not viable.

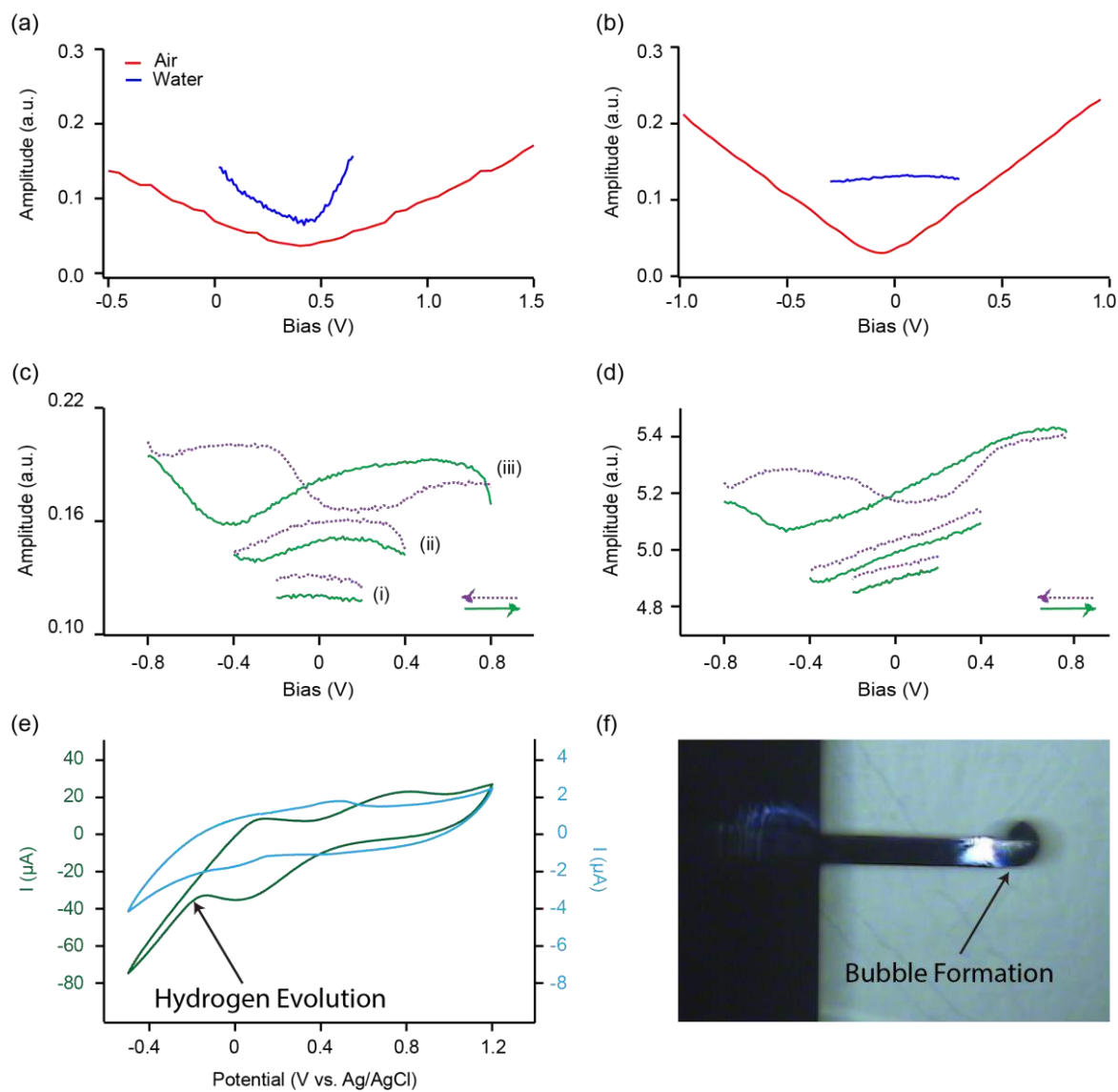

Figure 56. Investigation of KPFM in polar liquid. First harmonic amplitude in ambient (red) and milliQ water (blue) 50 nm above (a) HOPG and (b) Au surfaces as a function of the dc bias applied to the probe. (c) First and (d) second harmonic amplitude response recorded above an Au surface in the order of (i) small (± 200 mV), (ii) medium (± 400 mV), and (iii) large (± 800 mV) bias ranges. Measurements were performed using a sweep rate of 500 mV/s for HOPG and 100 mV/s for Au in ambient and 100 mV/s for both surfaces in milliQ water, while a 5 kHz signal with an amplitude of 0.5 V was applied to the probe. (e) Cyclic voltammetry measurements in milliQ water of HOPG (blue) and Au (green) electrodes. (f) Optical image of bubble formation at the tip in response to the application of 2 V (nominal cantilever length is 225 μm). *Seeking permissions.*

## 6.6    Applications of EFM in non-polar liquids

The application of EFM in liquid has been attempted by various groups over the past few decades with limited success.[530, 568] Typically approaches involved mapping static force vs distance curves to elucidate electrostatic force, FV maps, or static deflection measurements in lift mode such as fluid EFM. [530]Only recently have applications of VM EFM been demonstrated for quantitative measurements of electronic properties of materials in liquid.

To date most relevant applications of EFM have aimed to probe dielectric properties [52, 54, 55] which have recently been reviewed in brief by Kumar *et al.*[60] Dielectric polarization phenomena in electrolyte solutions is fundamental to the understanding of nanoscale properties of processes such as ion transport across biological membranes or metal corrosion. Much of the success in terms of probing such processes in liquid using SPM has emerged from the group of Gomilla, who recently presented an implementation of AM-EFM in liquid,[54, 55] enabling dielectric imaging with nanoscale spatial resolution. They demonstrated that nanoscale dielectric imaging was possible through the combination of high frequency ac excitation (>1 MHz) and a low frequency modulation (few kHz), overcoming detection bottlenecks. To validate their method, EFM was performed in aqueous electrolyte on a test structure (as shown in Figure 57) consisting of $SiO_2/Si$ microstructure. EFM measurements were repeated in 1 mM (Figure 57(b)) and 10 mM (Figure 57(b)), respectively. During the EFM measurement the applied frequency was decreased from 20 MHz to 0.1 MHz in order to investigate the influence of frequency on the image contrast. For 1 mM solution, the EFM dielectric images reproduce the periodic band structure, thus clearly demonstrating that the measurement is sensitive to local dielectric properties. This sensitivity to local dielectric properties diminishes at low frequencies. Contrast completely disappears below 100 kHz in 1 mM solution and below 1 MHz in 10 mM concentration.

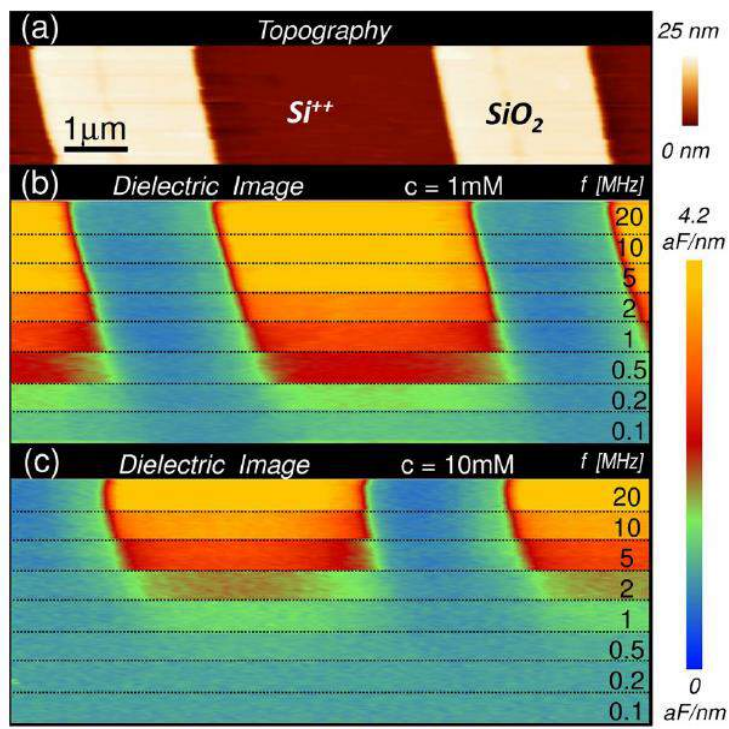

Figure 57. (a) Topography and ((b) and (c)) dielectric contrast images of 20 nm thin and 2.5 µm wide $Si^{++}/SiO_2$ microstripes obtained in electrolyte solutions. Dielectric images were obtained at constant height Z = 100 nm from the $Si^{++}$ baseline with ion concentrations c = 1 mM (b) and c = 10 mM (c) with v0 = 0.5 V. *Seeking permissions.*

Quantification of the dielectric constants of the materials was achieved using a previously developed procedure from the same group in air.[132, 133, 620, 621] Briefly, the procedure involves the use of numerical calculations to fit the experimental capacitive force vs distance curves obtained on a conducting substrate in order to calculate the tip radius and cone angle in liquid media. These values were used to further calculate the dielectric constant of the sample by fitting the numerical calculations to the experimental force curves obtained on the sample. This combination of experiment and finite element numerical calculations was subsequently used to determine dielectric constants of thin oxide films[52, 54] and lipid bilayers[52] in low molarity (< 10 mM) solutions.

In a similar approach, Kumar *et al.*[60] demonstrated the ability to characterize the Stern potential and Debye length of the EDL using EFM. This approach differs from that of Gomilla, as EFM is operated in the low frequency regime ensuring that EDL dynamics do not affect the tip-sample force. Furthermore, as the EDL shields the surface, measurements must be made close to the surface where the EDLs of the tip and sample overlap, but not so close that the EDL model is no longer accurate. This approach, however, was only demonstrated in single point z spectroscopy mode and its spatial resolution is expected to be limited as outlined by Umeda *et al.*[567] for excitations which are below the relaxation time. Furthermore, the Debye-Huckel approximation used is only valid for small applied voltages (a few $K_b T/e$).

## 6.7 Applications of open loop-KPFM in liquid

In recent years, significant work has been undertaken towards extending advanced OL forms of KPFM to liquid. Not content with operation in non-polar solvents alone, the first attempts harnessed the potential of operating the dc bias-free approaches (e.g., DH-KPFM) in conductive electrolytes. Kobayashi *et al.*[48] were the first to demonstrate such an approach for quantitative surface potential mapping in liquid, which they referred to as open loop electric potential (OLEP) microscopy.[48] They demonstrated that the local potential distribution at a solid-liquid interface could be imaged in a weak electrolyte solution (<3 mM).[48, 49] Later, Collins *et al.*[47] made a direct comparison between ambient and polar liquid (milliQ water) measurements using DH-KPFM by studying the influence of the imaging environment on the work function of graphene (Figure 58). It was found that the work function of the copper substrate changed upon immersion whilst the graphene surface remained unchanged, indicating that graphene acts as an effective corrosion inhibitor for the copper substrate. These results highlighted the potential of DH-KPFM for the quantitative investigation of the graphene-liquid interface, which may elucidate the role that synthesis processes, layer number, and defects have on the electrochemical behavior of graphene materials and devices.

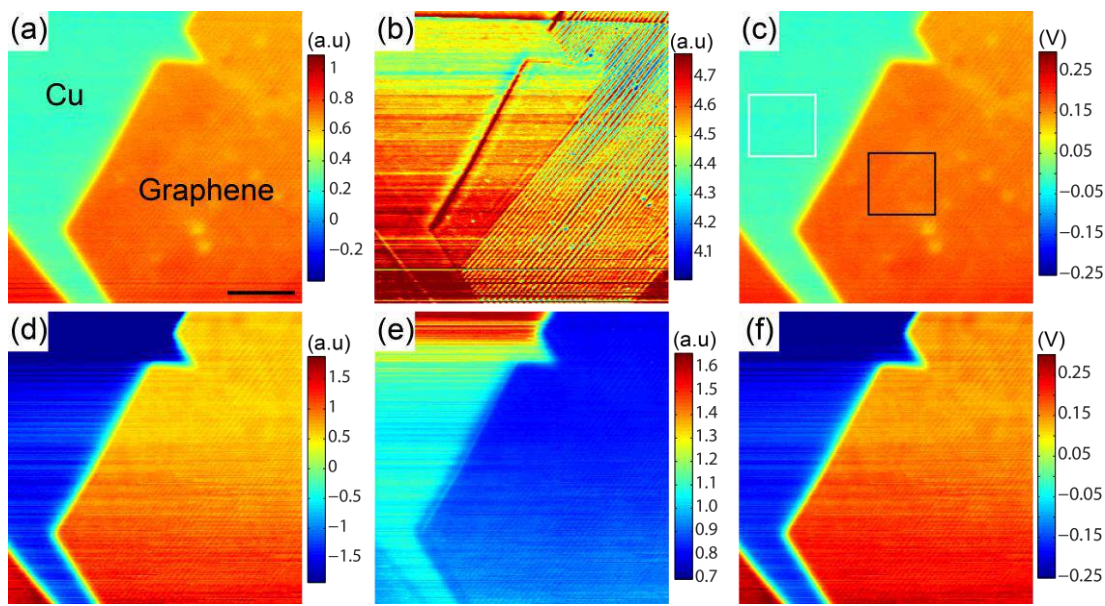

Figure 58. KPFM imaging of the graphene water interface. DH-KPFM images of single layer graphene on copper foil in (a-c) ambient and (d-f) milliQ water using a lift height of 50 nm (scale bar = 5 µm). (a, d) First harmonic mixed response images and (b, e) second harmonic amplitude images. (c, f) CPD. *Seeking permissions.*

In subsequent work, Kobayashi *et al.* developed a similar method called dual frequency (DF) OLEP-EFM.[50] In this method, two ac voltages having different frequencies are combined to produce mixing products, which are detected and used to calculate the electric potential. This mode has the advantage of high frequency modulation (twice as high as that in the conventional OLEP mode), however the underlying principles still rely on operating in a quasistatic equilibrium (i.e., the imaging mechanism is purely electrostatic)), which is difficult to determine experimentally. Nonetheless, they performed potential measurements of nanoparticles on a graphite surface in 1 mM and 10 mM NaCl solutions.[50] More recently, DF OLEP-EFM was used to investigate the nanoscale corrosion behavior of fine Cu wires and duplex stainless steel *in situ*.[56] Temporal variation in consecutive potential images show nanoscale dynamics and allow identification of corrosion sites, thus allowing real-time identification of local corrosion sites even when surface topography shows little change (see Figure 59). It is likely that such approaches will be useful for investigating reactions under surface oxide layers or highly corrosion-resistant materials, however, in terms of quantitative electrochemical potential measurements, these approaches are limited to low ionic concentrations where ion dynamics can be suppressed.

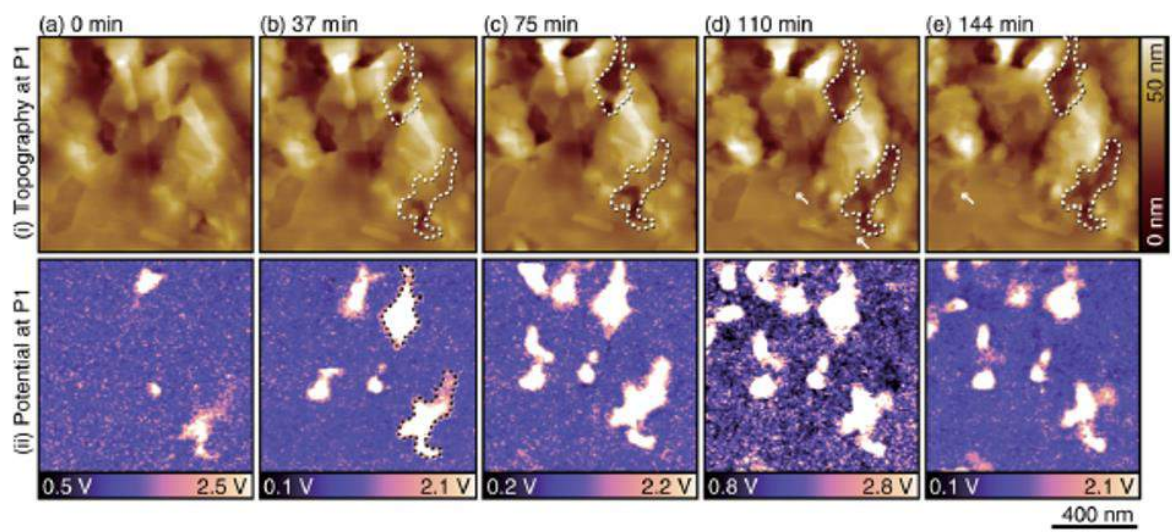

Figure 59. Temporal variation in surface potential of corrosion sites on duplex steal using DF OLEP- KPFM. *Seeking permissions.*

While measurements at the solid-water interface can be of importance to several fundamental areas of research,[622] measurements in higher molarity (> 100 mM) ion concentrations are attractive for most energy and biological systems. To explore the possibility of implementing KPFM in ionically-active liquids DH-KPFM measurements were performed in solutions containing increasing ion concentration (milliQ to 100 mM of NaCl). DH-KPFM was performed in single pass mode during the topography scan, which was collected with AM-AFM on the fundamental resonance frequency of the cantilever (35 kHz). An oscillation amplitude of 10 nm was used and the mean tip-sample position is taken as being 5 nm. Operating in single pass mode allowed increased sensitivity due to the proximity to the surface, however, it increases the difficulty in calibration of the cantilever transfer function, and hence quantifying the CPD due to the possibility of indirect crosstalk owing to shifts in the resonance peak. Therefore, a purely qualitative analysis on the first harmonic response was undertaken. In milliQ and 1 μM NaCl, Figure 60(a) and (b), clear contrast was observed between the Pt and Au electrodes which could be easily distinguished from each other. Note that the contrast observed is more localized than that observed in ambient or decane using CL-KPFM on the same samples (see Figure 60). Although is it difficult to make a direct comparison due to the different modes of operation (lift mode vs. single pass mode). This higher resolution can be expected because of the Debye screening length, ~419 nm in milliQ and ~96.2 nm in 1 μm NaCl, which effectively confines the interaction between probe and sample to the very tip apex and parts of the cone. In ambient or decane on the other hand, the measured response is a weighted mean of the interactions between the tip apex, cone and cantilever.

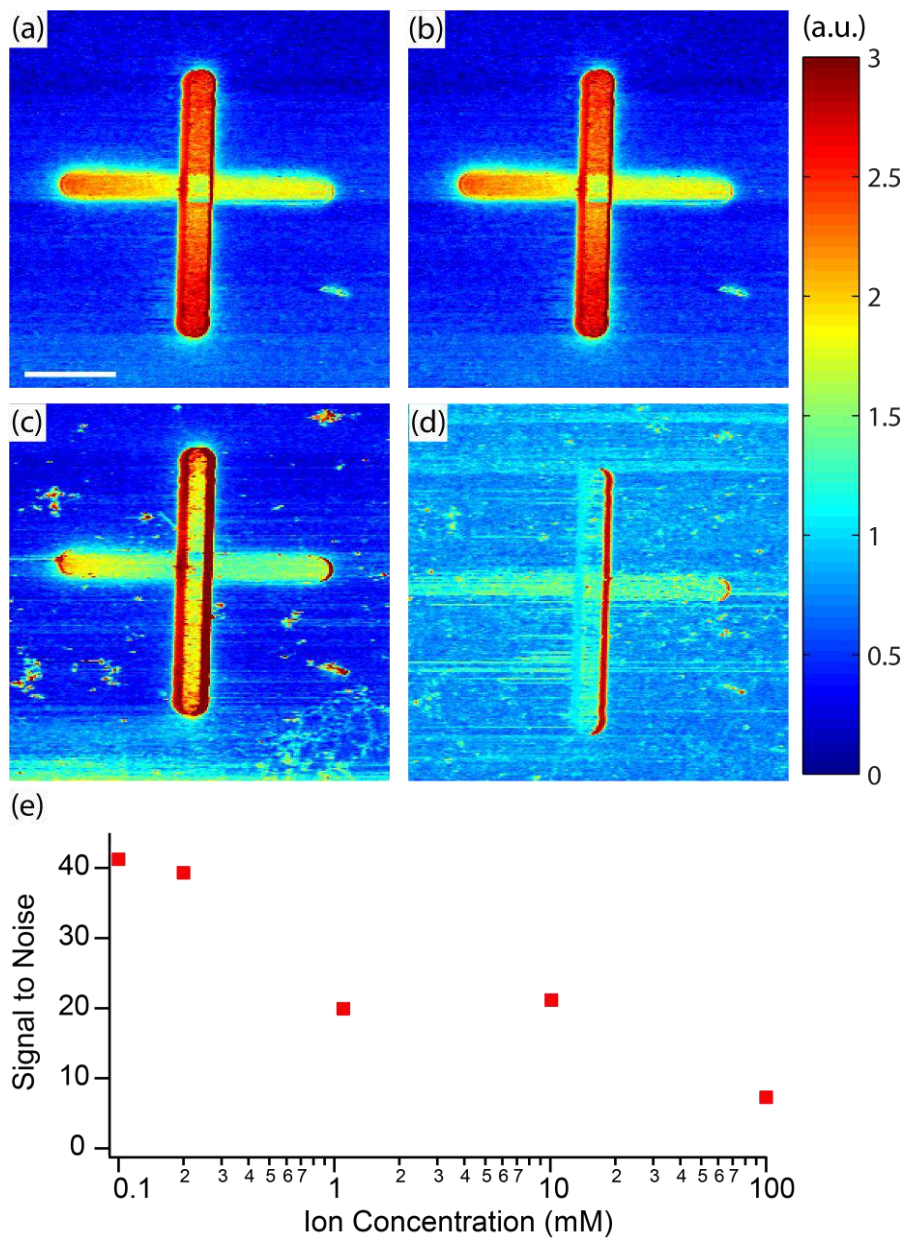

Figure 60. Breakdown of DH-KPFM in ionically-active liquid. First harmonic response of Pt/Au structure on a SiOx substrate measured in (a) milliQ (b) 1 μm (c) 1 mM and (d) 10 mM NaCl (scale bar =5 μm, normalized color scale =3 a.u.). (e) Plot of the signal to noise ratio as a function of increasing ion molarity. Images were recorded with in single pass mode with having a peak to peak amplitude of 10 nm. DH-KPFM was performed using a $V_{ac}$=500 mV and a frequency of 46 kHz.

As the molarity is increased further to 1 mM (λ=9.6 nm) and 10 mM (λ=3 nm), Figure 60(c) and (d), respectively, the contrast between the Pt and Au observed at lower molarities is lost and the signal to noise ratio deteriorates significantly, see Figure 60(e). These results are not surprising considering the complex ion dynamics and charge screening processes taking place in solution as a result of the applied bias.

## 6.8     Multidimensional methods

Despite the recent success in implementing KPFM type measurements in liquid, it is clear that their universal application to all systems is beyond current capabilities. Granted, at high frequencies (f >> $1/\tau_C$), EFM and OL-KPFM techniques (e.g., DH-KPFM) have been shown to provide useful information on system parameters (i.e., dielectric constant,[60] electrochemical potentials[47, 56, 623]), however this is not applicable to concentrated molarities (> 10 mM) in the frequency range addressable by AFM. In particular, for energy or bio applications the requirement of a high concentration of mobile ions (>> 10 mM) complicates the implementation of EFM or KPFM techniques. Indeed, in ionically-active liquid, ac voltage and dc bias applied between probe and sample result in electro-migration and diffusion of ions as previously described (see Section 5.1.3). Furthermore, steric effects and ultimately Faradaic reacts can be expected at typical voltages used in KPFM (100 mV – 3 V). In their current implementations, existing techniques (e.g., CL-KPFM, EFM, DH-KPFM) are not equipped to deal with the complex ion dynamics which result in a broad distribution of relaxation times and the associated transient relaxation of the resulting 'electrochemical' (e.g., electrostatic and osmotic) force in the presence of mobile ions. These techniques can only capture a snapshot of the complex ion dynamics taking place between tip and sample, precluding meaningful materials and system properties from being obtained, and necessitating the development of a suitable technique to study their influence systematically.

### 6.8.1    Requirement for multidimensional approach

Collins *et al.*[46, 51] outlined the need for multidimensional approaches to liquid KPFM, by considering the different relaxation processes expected in the presence of mobile ions. Figure 61, which shows the evolution of the concentration-dependent ( $\tau_D$ , $\tau_C$ ) and independent ( $\tau_L$ ) time constants, for relevant separations (i.e. tip-sample, cantilever-sample), as a function of the calculated relaxation times and ion concentration. In comparison with Figure 46(b), this phase diagram provides insight into the expected electrochemical processes taking place below the actuation frequency. The red box indicates the frequency space currently accessible using commercial AFM systems (dc – ~ 10 MHz). Within the frequency space addressable by AFM/KPFM there is a rich landscape of ion dynamics and electrochemical process taking place. In accordance with Figure 46(b), we can split this phase diagram into three different regimes. Notably, it may be possible to determine the electronic properties (e.g., CPD and dielectric constant) of the sample when operating in region III, where the EDL charging and ion diffusion processes occur at a time scale much slower than the measurement time (i.e., mobile ions are in quasistatic equilibrium). Again, this would agree with quantitative measurements of the dielectric properties[55] (and surface potential[47, 56, 623]) performed using EFM (or DH-KPFM) under quasistatic conditions. While measurements of CPD in higher (> 100 mM) ion concentrations are attractive for many energy and biological systems, they become impractical in the presence of ion dynamics (region II) and Faradaic processes at longer time scales (region I). However, for energy or bio applications it is not always possible to supress ion dynamics, which occur at extremely fast response times (>> 10 MHz) in concentrated solutions (> 10 mM). These time scales are outside the frequency space addressable with existing KPFM approaches, as limited by cantilever oscillation frequency and bandwidth of photodetectors used on most commercial AFMs. Furthermore, for classical CL-KPFM, which requires application of dc bias (region I) and ac voltage (region II / III) the complex coupling of the time dependent electrochemical processes can explain the hysteretic behavior observed in Figure 56. At the same time, operating in high molarities, regions I and/or II, presents an opportunity to probe important electrochemical phenomena in the tip-sample junction. This necessitates the development of techniques which are capable of probing and visualizing the ion dynamics and electrochemical processes which are inevitably taking place between probe and sample in the frequency space addressable with AFM. We note that the picture becomes more complicated due to ion crowding and ultimately Faradaic reactions for non-polarizable electrodes at larger biases. Furthermore, in a typical VM-AFM measurement the complex probe geometry can causes large electric fields, broken symmetry, electro-osmotic flows, and ultimately electrochemical reactions which are absent. Clearly, complete models of the electrochemical process taking into consideration 2D and 3D geometry of tip and sample is preferred.

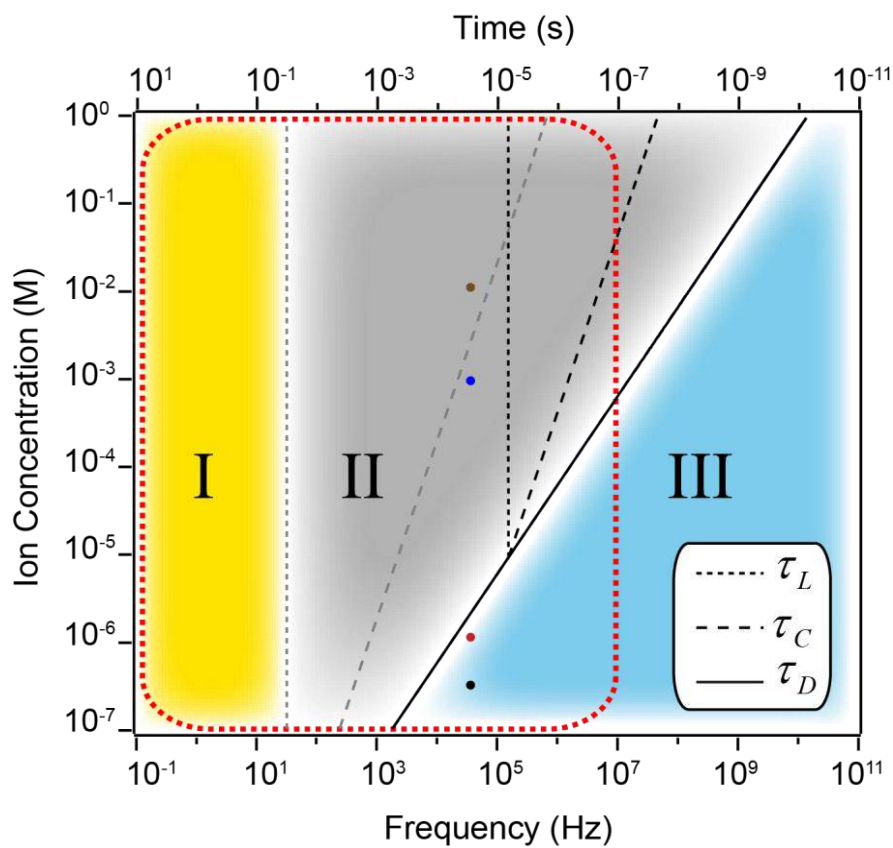

Figure 61. Charge dynamics and relaxation processes in VM-AFM. The characteristic timescales (Debye time, $\tau_D$, bulk diffusion, $\tau_L$, and charge relaxation, $\tau_C$) as a function of NaCl concentration for 200 nm (black) and 15 μm (grey) electrode separation representing tip- and cantilever-sample, respectively. The region enclosed by the red dashed line indicates the frequency space accessible by AFM. The shaded regions indicate the different electrochemical regimes where (I) Faradaic and (II) ion diffusion processes are expected to dominate the response and (III) where ion dynamics are largely absent (i.e., quasistatic equilibrium). The colored dots indicate the operational regime of images in Figure 60 (a) black (b) red (c) blue and (d) brown were performed. *Seeking permissions.*

### 6.8.2  Electrochemical force microscopy

Electrochemical force microscopy (EcFM)[51] is a generalized approach for electrochemical measurements in liquids containing mobile ions. EcFM combines the force sensitivity of AFM with the ability to probe the bias- and time-dependence of electrochemical dynamics at the probe sample junction and spatially across boundaries. In EcFM measurements, the tip is positioned a specified distance from the surface, and electrochemical processes in the probe sample junction are activated using a combination of single frequency ac voltages and dc bias waveforms, as shown in Figure 62. Upon positioning the tip at the appropriate distance from the surface (typically 10 − 200 nm), a user defined dwell period maintains the probe sample separation while the electrochemical measurement is performed and a trigger from the AFM controller initiates the EcFM measurement. In EcFM the probe is electrically modulated with a high frequency ac voltage used to detect the dynamic cantilever response using LIAs while the system is perturbed by dc bias waveforms applied to the probe. The data can be presented as an EcFM spectra representing the voltage- and time-dependent response for a single location.

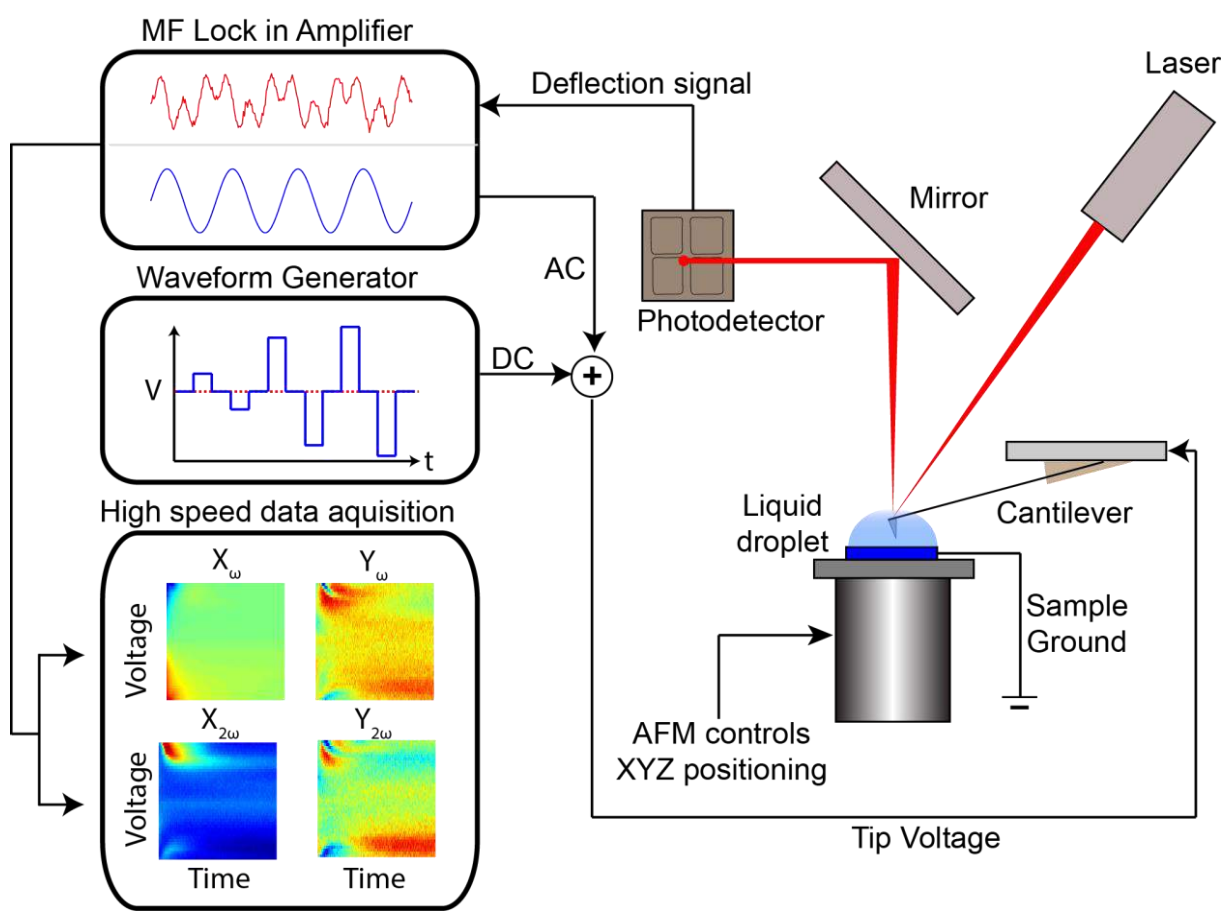

Figure 62. Schematic of the experimental setup for EcFM.

For all data the in-phase ($X$) and out-of-phase ($Y$) components of the tip response were recorded for each harmonic and later used to construct the amplitude and phase. Similar to macroscopic electrochemical measurements, the outer dc bias waveform can take a variety of forms, here, we primarily work with pulsed techniques for separation of electrokinetic phenomena.[624] For this we adopt a first order reversal curve (FORC) [625] approach using a set of bipolar pulses (typically $50 - 150$ ms) of increasing magnitude. Previously, this FORC approach was demonstrated to be highly effective in exploring local bias-induced electrochemical phenomena at the solid gas interface that are reversible for small biases and irreversible at high biases.[615, 619]

An example EcFM data set is shown in Figure 63, where a bipolar square waveform has been used. The data is collected both during the bias application (*bias-on* state) and following the bias application (*bias-off* state), as the magnitude of the bias pulses is increased linearly with time. Figure 63(a) shows a section of the dc bias waveform applied to the probe with the corresponding response recorded in air. Little to no relaxation of the electrostatic force is observed, with response following the applied bias, and therefore satisfying the second principle of KPFM, a time-invariant electrostatic response. Again, this is expected for a purely electrostatic response, or more generally when the force experienced by the system is governed purely by the time-independent electrostatic pressure.

The data can be presented as an EcFM spectra representing the bias- and time-dependent response for a single location, e.g., Figure 63(b). In nonpolar liquids, such as decane, Figure 63(d, e), a similar response is observed to that in air, where again the electrostatic force follows the applied dc bias and is time-invariant. In milliQ water, however, a very different response mechanism is observed. As can be seen in Figure 63(g, h), a large increase in the response is observed at the instant the bias pulse is applied, which relaxes within 5 ms to ~ 36 % of the peak value. Similar information can be obtained from the bias-off state. For both air and decane the bias-off state was constant and time-invariant for the entire bias range, Figure 63(c, f). For milliQ water (Figure 63(i)), however, transients were detected for all biases and the largest response (including for bias-on states) was seen after the application of positive bias pulses greater than +600 mV. The time-independent EcFM response for decane is in agreement with that previously reported.[45] In ionically-active liquids, however, the large peak and subsequent relaxation of the cantilever response observed is associated with the various relaxation times discussed previously, and as such presents a unique opportunity to probe the screening mechanisms in situ.

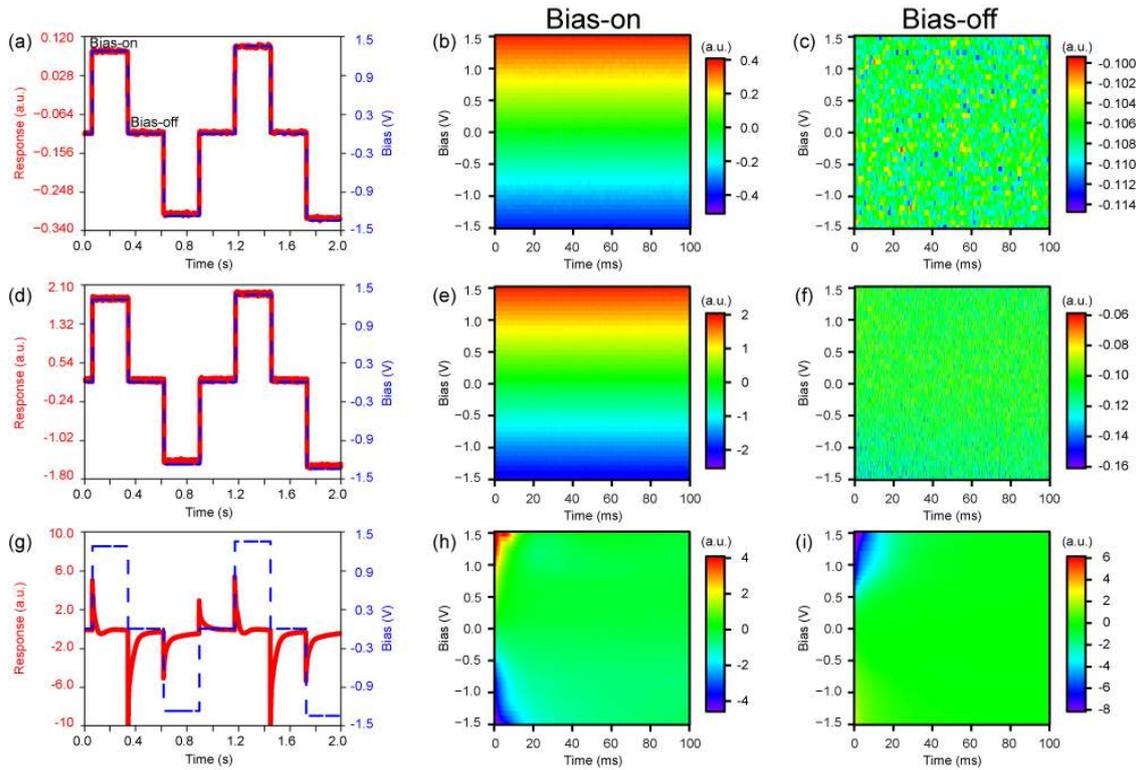

Figure 63. Single point EcFM in air, ionically-active and nonpolar liquid. EcFM mixed response collected 200 nm above a grounded Au electrode in (a, b, c) air, (d, e, f) decane and (g, h, i) milliQ water. (a, c, e) Temporal response (solid red line) of the EcFM response to the applied dc bias waveform (dashed blue line). Single-point EcFM response spectra showing bias-on (b, e, h) and bias-off (c, f, i) states. Measurements were performed with $V_{ac} = 0.5$ V [25 kHz] applied to the probe. *Seeking permissions.*

Figure 64(a, b) show the mixed response collected in decane and milliQ water. Here, the response is plotted as a function of tip bias, while the color scale represents the time at which the response was probed. In decane, the response follows a linear bias dependence and no deviation from linearity was observed for all times probed. Under these conditions, EcFM "converges" to KPFM in the sense that EcFM can be used to determine the CPD by linear fitting as is done in OLBS to find the bias at which the response is minimized. For decane, Figure 64(c), the CPD was determined for each time slice and it can be seen that the CPD (85 ± 2) mV was constant within experimental error. The slope of the line fit gives a measure of $C_Z'$, which changes only slightly during the measurement likely a result of small changes in tip-sample geometry due to drift in the tip-sample separation during the measurement.

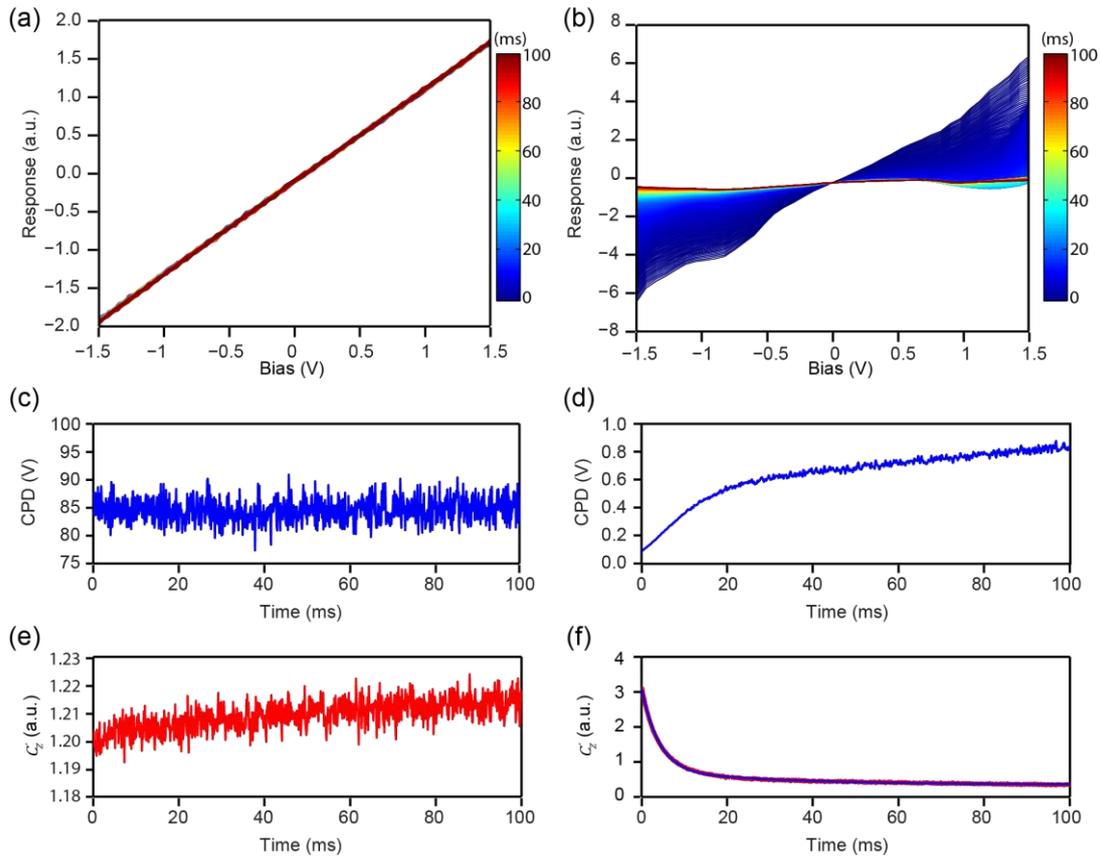

Figure 64. Dynamic CPD measurement using EcFM. EcFM measurements [bias-on] in (a, c, e) decane and (b, d, f) milliQ water recorded 200 nm above a grounded Au electrode. (a, b) EcFM bias-resolved mixed response with time represented on the color scale. (c, d) The measured CPD and (e, f) $C_Z'$ determined from linear fitting of mixed response. Measurements were performed with $V_{ac}$ = 0.5 V [25 kHz] applied to the probe. *Seeking permissions.*

In milliQ water, when the measurement timescale $<< \tau_C$, a measured value for the CPD can be obtained in a manner similar to that used for decane. The measured CPD was found to be $(112 \pm 14)$ mV from the first 2 ms of data recorded for the Au electrode in milliQ water. However, gradual changes of the measured CPD in milliQ water (Figure 64(d)) can be observed, which varies by greater than 700 mV within the 100 ms measurement as a result of screening by the EDLs between probe and sample. Note that at longer timescales, the zero response crossing becomes ill defined and thus the validity of the CPD measurement at these timescales is questionable. This is also reflected in the transient behavior in the $C_Z'$, which approaches zero at longer times as shown in Figure 64(f). It is likely that this decay is related to the charging of the EDL through the electrolyte resistance, as described in section 6.5.1. For charging of a simple $RC$ circuit it is anticipated that this relaxation will demonstrate a simple exponential response mechanism. In fact, it was found that the transient response of $C_Z'$ was well described by a double exponent fit (blue line in Figure 3(f)), where the fitting equation was defined as; $f(x) = y_0 + A_1 e^{-t/\tau_1} + A_2 e^{-t/\tau_2}$, having relaxation times of $\tau_1 = 4.2$ ms and $\tau_2 = 31.7$ ms, respectively. This time dependence in milliQ water, which is absent in the nonpolar solvent, decane, prohibits the implementation of CL-KPFM in ionically-active liquids, and validates the adoption of multidimensional bias and time resolved techniques, as demonstrated here.

Unlike KPFM, EcFM contains important information pertaining to charge screening mechanisms as evidenced in the transient relaxation of $C_Z'$, in Figure 64(d). To investigate this further, using the theory set out in section 5.1, we modelled the charging of polarized capacitor plates separated by 200 nm (tip-sample) and 15 μm (cantilever-sample) in response to a 250 mV bias pulse in milliQ water. As demonstrated in Figure 65(a) and (c), for an electrode separation of 200 nm, the charging of the capacitor is almost instantaneous with the applied bias. In this case the liquid behaves close to a lossless dielectric, Figure 65 (a), having a linear potential in the electrode junction and no significant screening from ions in solution. Here, the dynamics of the system are controlled by the Debye time ($\tau_D \approx 131$ μs,) and the total force (per unit area) is almost purely electrostatic with the osmotic component being negligibly small, Figure 65(e). In comparison, for the cantilever sample separation Figure 65(b) and (d) a much slower charging time is observed, governed by diffusion of ions from the bulk into the EDL ($\tau_L \approx 31.8$ ms). In this case, the net force (per unit area) is strongly influenced by the osmotic pressure, Figure 65(f).

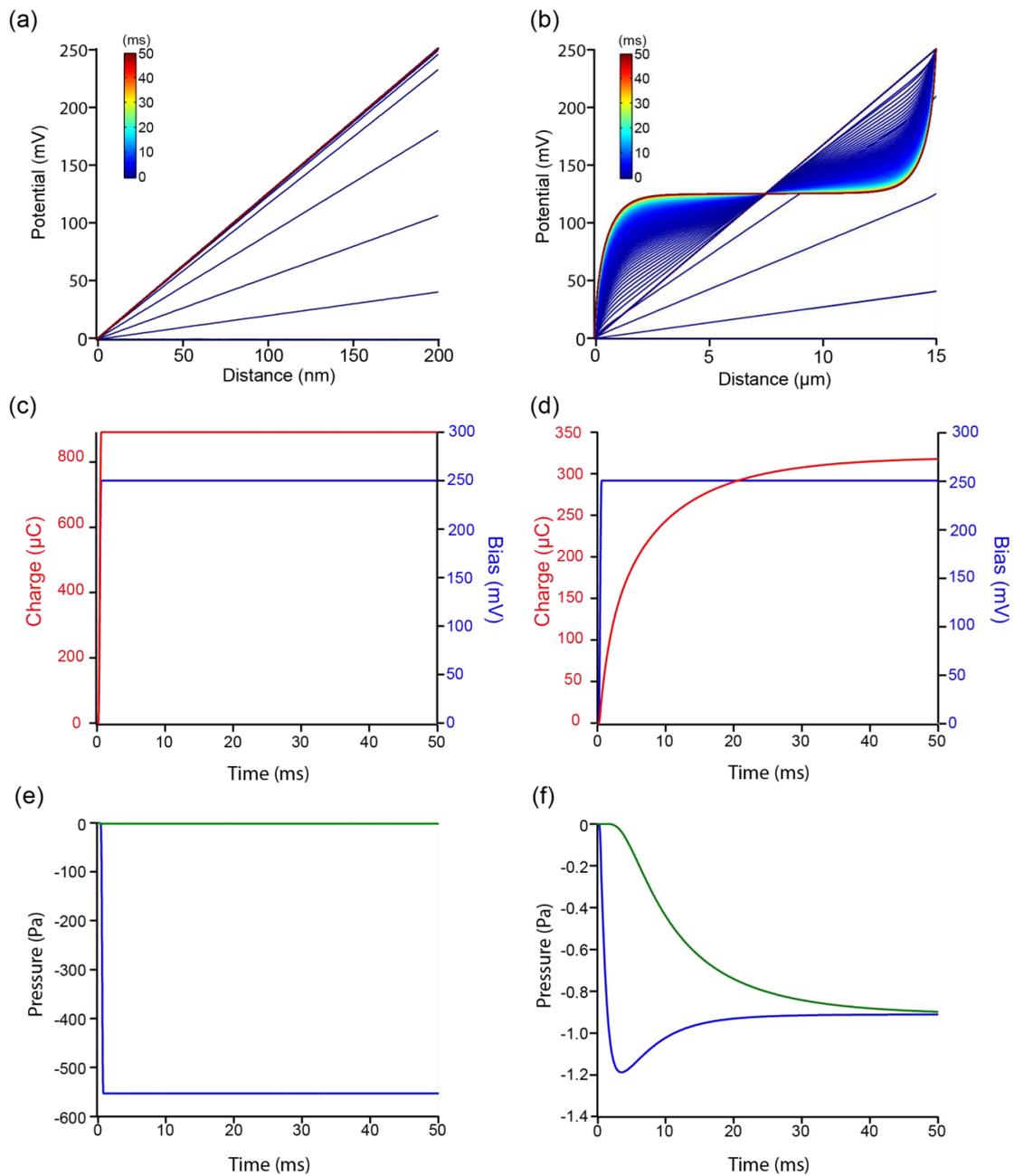

Figure 65. Charging of a capacitor terminal plate. Results of numerical modelling showing the evolution of the (a,b) potential (c,d) charge and (e,f) total pressure of a parallel plate capacitor in response to a 250 mV bias in milliQ water for (a,c,e) 200 nm and (b,d,f) 15 μm separations. (c,d) The bias step is the blue line and charge is shown in red. (e,f) Total pressure at the terminal plate in shown in blue and osmotic pressure at the midplane is shown in green.

In Figure 66 we compare modelled data in Figure 65(d) with experimental data in response to single 225 mV step bias. Note that EcFM data was recorded with bias steps of 75 mV, thus direct comparison with data recorded at 250 mV was not possible. In agreement with the results observed for the transient relaxation of $C_Z'$, in Figure 64(d), both the modelled data and experimental data were found to be best described by a double exponential fit. The modelled data was demonstrated to have relaxation times of (2.03 ± 0.01) ms and (10.48 ± 0.01) ms, comparable with the experimental data which yields relaxation time constants of (2.66 ± 0.02) ms and (22.55 ± 0.27) ms. The requirement for double exponential in both experimental and modelled data demonstrates a two-step charging process of the EDL. Indeed a nonmonotonic charging profile has been predicted theoretical for larger voltages (>> 25 mV).[9] The faster response time is in very close agreement with the expected charging time of the EDL between cantilever and sample determined analytically ($\tau_C$ is 2.0 ms). The second process is on a much longer timescale (modelled = (10.48 ± 0.01) ms, experimentally = (22.55 ± 0.27) ms) than this charging time and must be related to the slowest relaxation process, which is ion diffusion. Quantitatively there is a deviation from modelled and experimental data for the magnitude of this slower charging process, however, this is not surprising considering the complex geometry of the AFM tip compared to the 1D model, furthermore, the difficulty in determining the actual tip potential in solution, which can only be calibrated by incorporating a reference electrode into the setup. However, both modelled and experimental data are in agreement in terms of predicting a faster bulk diffusion time than determined analytically ($\tau_L$ is 31.8 ms).

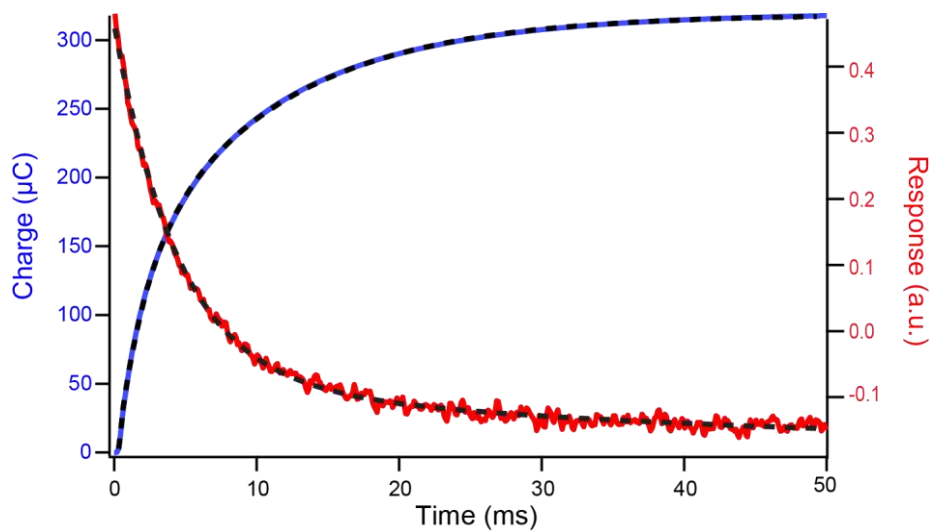

Figure 66. Comparison of calculated charge and EcFM mixed response relaxation. The modelling of terminal plate charging following a 250 mV bias in milliQ water for a 15 μm electrode separation yields timescales ((2.03 ± 0.01) ms and (10.48 ± 0.01) ms) which agree with the EcFM response to a 225 mV bias ((2.66 ± 0.02) ms and (22.55 ± 0.27) ms) in milliQ water. The time constants were determined from a double exponential fit (dashed black line). *Seeking permissions.*

When the numerical modelling was performed for different bias steps, for small biases (< 10 mV) the modelled EDL at a terminal plate demonstrated a purely exponentially decaying response, whereas for larger biases (> 100 mV) a double exponent was required to describe the modelled data, Figure 67, in agreement with experiment. The differences observed between low and high biases are likely a consequence of ion depletion close to the diffuse layers during the EDL screening process. This phenomena was previously described theoretically by Bazant *et al.*, [9]who demonstrated that when the potential becomes larger than $k_B T/e$, neutral salt uptake from the bulk, into the EDL results in a narrow region of ion depletion just outside the diffuse charge layers.[9] This neutral salt uptake continues up to the $\tau_C$ time, at which point simple diffusion of the neutral-bulk electrolyte fills in depleted zones until a uniform equilibrium state is reached.[9] To our knowledge this is the first time it has been demonstrated experimentally, particularly using a force based approach such as demonstrated here.

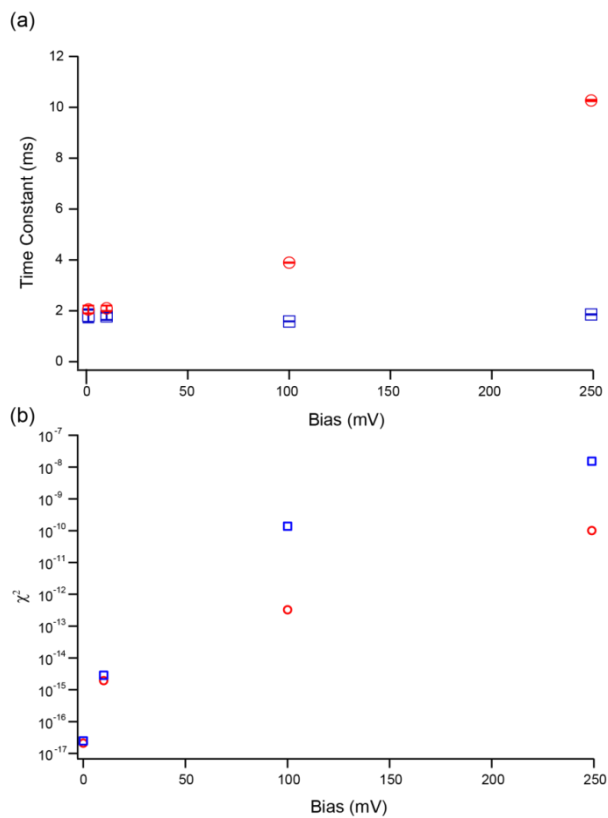

Figure 67. Deviation from exponential charging as a function of bias. (a) Time constants determined from double exponent fitting of the calculated terminal charging are shown for 1, 10, 100, and 250 mV. First (blue squares) and second (red circles) time constants coincide for smaller biases (≤ 10 mV) and diverge for larger biases (≥ 100 mV). (b) $\chi^2$ values for single (blue squares) and double (red circles) exponent fitting as a function of bias.

To explore the dependence of ion concentration on the electrochemical response, EcFM was used to study an Au electrode as a function of $K_2SO_4$ concentration. Figure 68 shows the first harmonic EcFM amplitude response vs. bias. Measurements are performed at 200 nm from the Au surface in milliQ water and 1 mM, 10 mM, and 100 mM $K_2SO_4$. In both milliQ water (Figure 68(a)) and 1 mM $K_2SO_4$ Figure 68 (b)) a bias-dependent response with a minima can be observed in the fast response regime, which quickly relaxes (< 2 ms). In 1 mM $K_2SO_4$, the transient relaxation time was determined in the same fashion as described for Figure 68(d) to be (391 ± 90) μs and (1.2 ms ± 0.1) ms, both of which are within the experimental detection limit (limited by the lock in amplifier time constant, 100 μs). In contrast, for both 10 mM (Figure 68(c)) and 100 mM $K_2SO_4$ (Figure 68(d)), no significant transient relaxation of the bias dependence in the chosen window (-375 mV − +250 mV) is observed. The charge screening is expected to occur faster as the concentration increases (12.9 μs for 10 mM 1:1 electrolyte, which is well below the current detection limit). To effectively probe charge screening dynamics in higher molarity electrolytes, sub-microsecond detection, as discussed in section 4.3, could be incorporated into the EcFM setup.

Interestingly, at larger biases in high molarity solutions, we begin to observe a strongly nonlinear response. These processes are too slow (> 75 ms) for ion diffusion across the cantilever-sample separation in response to charge screening, and are attributed to electrochemical reactions taking place at the solid-liquid interfaces. The activation bias of these processes for negative tip bias begins at ~ –500 mV for both 10 mM and 100 mM, however, larger amplitudes are observed for 100 mM. Further ion concentration-dependent processes are observed at positive biases (> 1.1 V) for 100 mM but not for 10 mM solutions. The concentration-dependence of this 'gap', between the apparent onset of bias-dependent processes warrants further investigation and could be used to map the activation potential of electrochemical reactions at the solid-liquid interface. This could have significant implications as it would allow quantitative mapping of local, material dependent, electrochemical reactivity.

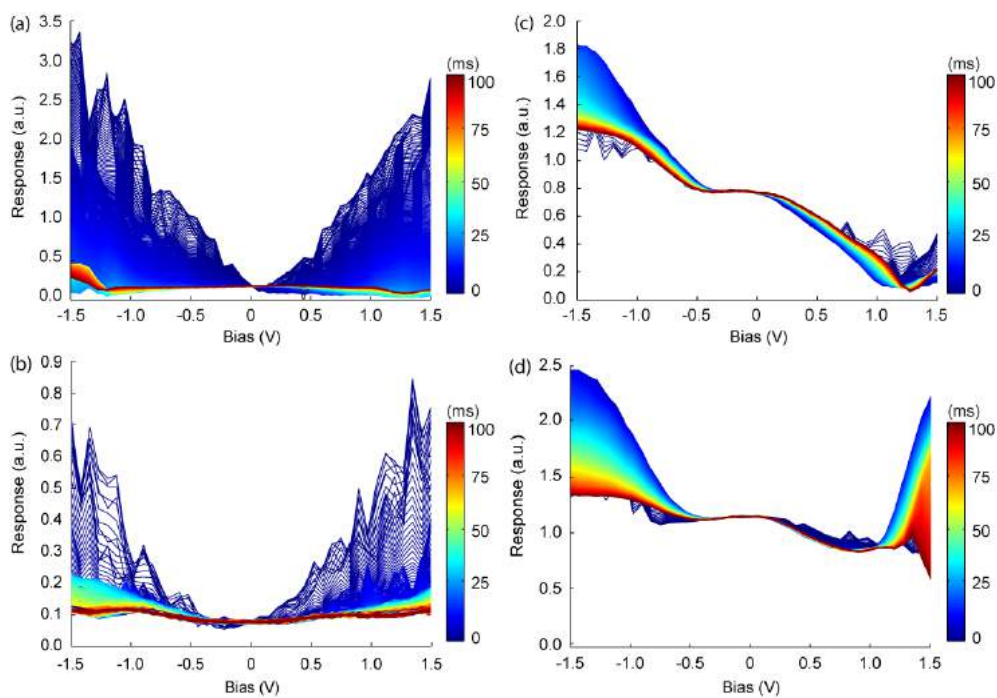

Figure 68. EcFM data recorded on Au as a function of ion concentration. The first harmonic EcFM amplitude collected 200 nm above an Au electrode in solutions of increasing salt concentration: (a) milliQ water and (b) 1 mM, (c) 10 mM, and (d) 100 mM K$_2$SO$_4$. Measurements were performed from low to high concentrations using

the same cantilever. Measurements were performed with $V_{ac} = 1$ V [17 kHz] applied to the probe. *Seeking permissions.*

### 6.8.3   EcFM imaging

Shown in Figure 69 is an example of 2D EcFM imaging in milliQ water across Au electrodes deposited on a $SiO_2$ layer. Here, 2D EcFM bias- and time-dependent spectroscopic response was recorded in a 30 × 30 grid across the electrodes (Figure 69(a)), giving rise to a 4D data set (*x*, *y*, bias, time). A quantitative and systematic analysis of the spatial variability of the bias and time dependence of the EcFM response requires a complete physical model of the signal formation mechanisms including Faradaic reactions at larger biases. However, the complexity of the response mechanism in EcFM currently precludes the establishment of a suitable physical model, as is the case for many other AFM techniques (e.g., electrochemical strain microscopy[80]). In order to demonstrate the spatial variability present in the data, we have plotted the first and second harmonic EcFM amplitude response recorded 5 ms after the onset of the +1.5 V and −1.5 V bias pulses (Figure 69(c)-(f)). Average cross sections for each image, from the region indicated in Figure 69(a), are shown in Figure 69 (b). The observed spatial variation between Au and $SiO_2$ is present for all biases and times. For both first and second harmonics, the EcFM response shows a bias-polarity dependence for $SiO_2$. This suggests that the effective carrier density of the $SiO_2$ surface has changed between the application of positive and negative biases. For Au, the first harmonic EcFM response is polarity-dependent, whereas the second harmonic response for Au is the same for both ±1.5 V. The presence of such spatially-dependent contrast demonstrates that the EcFM signal is localized and dependent on the electrochemical properties of the material below the tip.

Noteworthy, so far, EcFM studies have only focused on capture of the first two harmonics using LIAs; however the nonlinear response is expected to result in numerous other higher harmonics which may contain useful information. Furthermore, the time resolution of this approach is determined by the LIA time constant. Clearly, both of these bottlenecks could be overcome by the adoption of G-mode platform.

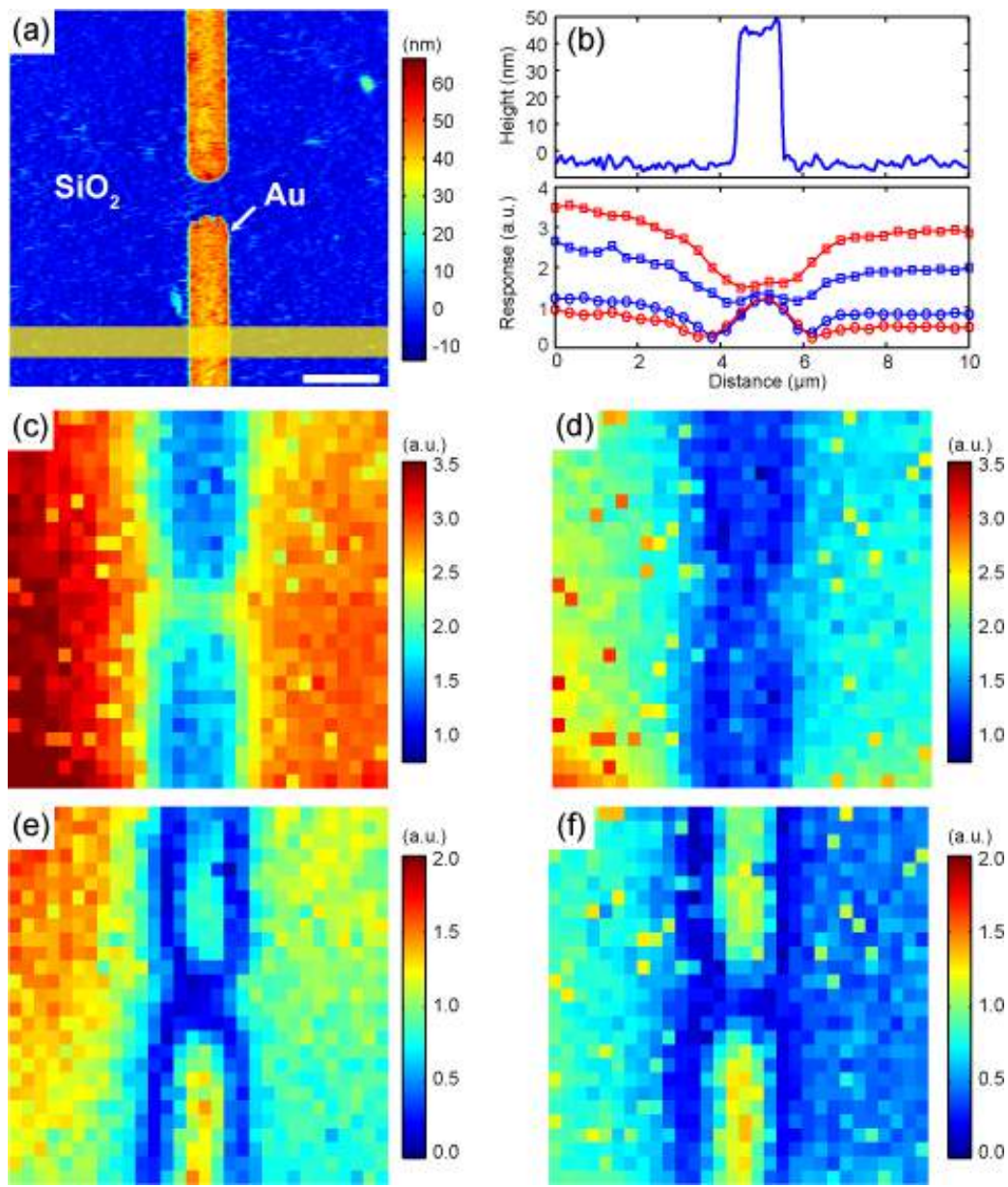

Figure 69. Spatial variability of EcFM data recorded across Au/SiO$_2$ boundaries in milliQ water. (a) Topography image of Au electrodes on a SiO$_2$ substrate recorded using AM AFM (scale bar = 2 μm, image size = 10 × 10 μm). (b) Average cross sectional EcFM data recorded 200 nm above the surface and determined from the area marked with a yellow box in (a) for first (squares) and second (circles) harmonic EcFM amplitude response during the application of +1.5 V (red) and −1.5 V (blue) bias pulses. Spatial variability maps from a grid of 30 × 30 2D EcFM measurements of the (c,d) first and (e,f) second harmonic EcFM amplitude response at 5 ms after the onset of the bias-on state for (c,e) +1.5 V and (d,f) −1.5 V, respectively. Measurements were performed with $V_{ac}$= 1 V [19 kHz] applied to the probe. *Seeking permissions.*

# 7   Future outlook and perspectives

In the past 25 years, KPFM has become a gold standard characterization approach for nanoscale for probing electrostatic/electrochemical properties on the nanoscale at solid surfaces, and is a focus of active research in SPM as well as having a big impact on a broad areas of material science. Recent years have seen multiple attempts to extend KPFM potential imaging to liquid environments – for applications ranging from energy science to biology to corrosion – with varying degrees of success. Throughout this review article, these approaches have been outlined and a background into the theoretical implications of performing KPFM in presence of mobile ions has been provided. Almost all of the techniques demonstrated to date break down when an appreciable concentration of mobile ions (> a few mM) are included in the system. From the body of work reviewed in this manuscript it is clear that electrokinetic and electrochemical phenomena in liquids have inherent complex time and bias dynamics that cannot be fully captured using traditional KPFM approaches.

The two major reasons behind this have been identified. First, KPFM is based on the principle that the junction between tip/cantilever-sample behaves like a lossless dielectric (e.g., vacuum) and cannot account for ion mobility or electrochemical reactions in liquid, which is detrimental to all previous attempts at liquid KPFM. Second, in liquid, the concept of a static CPD value is no longer valid and becomes a dynamic parameter, changing in time with ion concentration, electro-osmotic flow, and irreversible electrochemical reactions. As such, any attempt at traditional KPFM in liquid will need to deal with the influence of such processes. Even non-traditional OL modes of KPFM whose underlying principles are based on a lossless dielectric will breakdown, unless, there excitation frequency positions them in a quasistatic regime where ionic motion to the excitation is suppressed. Below this transition frequency, the description of the electronic properties of the solid-liquid interface by a single parameter is invalid, thereby explaining the limited success in achieving KPFM in liquid thus far. This requires an appropriate technique to study the complex dynamic bias- and time-dependent response of a VM-SPM probe, a gap currently satisfied by EcFM.

However, the combination of shielded probes [87, 626] and advanced dynamic modes holds the promise of probing the bias- and frequency-dependent details of electrostatic and electrochemical interactions in liquids at the physical limits of resolution. Crucial for this will be broad implementation of artificial intelligence to sample the regions of parameter and real space to avoid surface damage and non-equilibrium processes.[627]

**Acknowledgements**

We acknowledge the discussions, references to previous works, collaborations, support and friendship of all our colleagues in the ferroelectrics and scanning probe communities, without whom the progress indicated in this review would never have eventuated. We would also like to specifically thank Stephen Jesse, Evgheni Stelcov, Alexander Tselev and Stefan Weber for their fruitful discussions in preparation of this work. This research was conducted and partially supported (LC, SVK) at the Center for Nanophase Materials Sciences, which is a US DOE office of Science User Facility.